%% file: main.tex
\documentclass[useAMS,usenatbib,onecolumn]{mnras}

\usepackage{graphicx,url}
\usepackage[ngerman,english]{babel}
\usepackage{numprint,color}
\usepackage{amsmath,amssymb,verbatim}
\usepackage{textcomp}
\usepackage{xspace}
\usepackage[left]{lineno}
\usepackage{booktabs}
\usepackage{subcaption}
\usepackage[normalem]{ulem} 
\usepackage{relsize}
\usepackage{url}

\newcommand{\avg}[1]{\left< #1 \right>} 
\newcommand{\ud}{\mathrm{d}}
\newcommand{\m}{_\mathrm{m}}
\newcommand{\p}{_\mathrm{a}}
\newcommand{\f}{_\mathrm{f}}

\def \degree   {$^{\circ}$\xspace}
\def \degC     {$^{\circ}$C\xspace}

\definecolor{amethyst}{rgb}{0.6, 0.4, 0.8}
\definecolor{green}{rgb}{0.55, 0.71, 0.0}
\definecolor{apricot}{rgb}{0.98, 0.81, 0.69}
\definecolor{auburn}{rgb}{0.43, 0.21,0.1}
\definecolor{babyblueeyes}{rgb}{0.63, 0.79, 0.95}
\definecolor{bittersweet}{rgb}{1.0, 0.44, 0.37}
\definecolor{blue(munsell)}{rgb}{0.0, 0.5, 0.69}
\definecolor{oceanboatblue}{rgb}{0.0, 0.47, 0.75}
\definecolor{brightmaroon}{rgb}{0.76, 0.13, 0.28}

\newcommand{\collaborationreview}[1]{\textcolor{black}{#1}}

\renewcommand{\t}[1]{\mathrm{#1}} 
\npdecimalsign{.}
\newcommand{\np}{\numprint}

\newcommand\ddfrac[2]{\ensuremath{\frac{\displaystyle #1}{\displaystyle #2}}}

\newcommand{\orcid}[1]{{\href{http://orcid.org/#1}{%
\openin1 img/orcid-ID.png \ifeof1
\else%
\hskip2pt\includegraphics[width=9pt]{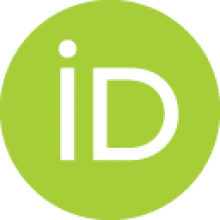}\fi}}{}}%

\title[Characterizing the atmosphere above the MAGIC site using LIDAR data]{Characterizing the aerosol atmosphere above the Observatorio del Roque de los Muchachos by analyzing seven years of  data taken with an GaAsP HPD-readout,  absolutely calibrated elastic LIDAR 
}

\author[C. Fruck, M. Gaug, et al.]{Christian Fruck$^1$\thanks{E-mail: christian.fruck@mpp.mpg.de}\orcid{0000-0001-5880-7518}, Markus Gaug$^2$\thanks{E-mail: markus.gaug@uab.cat}\orcid{0000-0001-8442-7877}, Alexander Hahn$^1$\thanks{E-mail: ahahn@mpp.mpg.de}\orcid{0000-0003-0827-5642}, Victor Acciari$^3$\orcid{0000-0001-8307-2007}, \newauthor  J\"urgen Besenrieder$^1$, 
Dijana Dominis Prester$^4$\orcid{0000-0002-9880-5039}, Daniela Dorner$^5$\orcid{0000-0001-8823-479X},\newauthor  David Fink$^1$, 
Llu\'is Font$^{2}$\orcid{0000-0003-2109-5961},
Sa\v{s}a Mi\'canovi\'c$^4$\orcid{0000-0002-0076-3134}, Razmik Mirzoyan$^1$\orcid{0000-0003-0163-7233},\newauthor
Dominik M\"uller$^1$,
Lovro Pavleti\'c$^4$\orcid{0000-0002-9926-0405}, Felix Schmuckermaier$^1$\orcid{0000-0003-2089-0277}, 
Martin Will$^1$\orcid{0000-0002-7504-2083}
\\
$^{1}$Max-Planck-Institut f\"ur Physik, 80805 M\"unchen, Germany \\
$^{2}$Departament de F\'isica, Universitat Aut\`onoma de Barcelona and CERES-IEEC, 08193 Bellaterra, Spain \\ 
$^{3}$  Institut de F\'isica d'Altes Energies, 08193 Bellaterra, Spain \\
$^{4}$University of Rijeka, Department of Physics, 51000 Rijeka, Croatia \\
$^{5}$Universit\"at W\"urzburg, D-97074 W\"urzburg, Germany.
}

\date{Accepted XXX. Received YYY; in original form ZZZ}

\pubyear{2022}

\begin{document}

\label{firstpage}
\pagerange{\pageref{firstpage}--\pageref{lastpage}}
\maketitle

\begin{abstract}
 We present a new elastic LIDAR concept, based on a bi-axially mounted Nd:YAG laser and a telescope with HPD readout, combined with fast FADC signal digitization and offline pulse analysis. The LIDAR return signals have been extensively quality checked and absolutely calibrated. We analyze seven years of quasi-continuous LIDAR data taken during those nights when the MAGIC \collaborationreview{telescopes} were operating. Characterization of the nocturnal ground layer yields zenith and azimuth angle dependent aerosol extinction scale heights for clear nights. We derive aerosol transmission statistics for light emitted from various altitudes throughout the year and separated by seasons. We find further seasonal dependencies of cloud base and top altitudes, but none for the LIDAR ratios of clouds. Finally, the night sky background light is characterized using the LIDAR photon backgrounds.
\end{abstract}


\input{introduction}
\input{lidar}

\input{signalinversion}
\input{molecular}
\input{calibration}
\input{cloudanalysis}
\input{sixyears}
\input{nsb}
\input{discussion}

\input{conclusions}

\section*{Acknowledgements}

This work would have been impossible without the support of our colleagues from the MAGIC collaboration. We are especially grateful to the many shifters who have helped to debug and improve the LIDAR system. We would also like to thank the Instituto de Astrof\'{\i}sica de Canarias for the excellent working conditions at the Observatorio del Roque de los Muchachos in La Palma.

We thank Emilio Cuevas-Agullo and Bahaiddin Damiri for their effort in establishing and maintaining the AERONET station and data at the Iza\~na site.

The financial support of the Spanish grant PID2019-107847RB-C42, funded by MCIN/AEI/~10.13039/501100011033, the German BMBF and MPG, and by the Croatian Science Foundation (HrZZ) Project IP-2016-06-9782 and the University of Rijeka Project uniri-prirod-18-48 is gratefully acknowledged.

The VIIRS/NPP Lunar BRDF-Adjusted Nighttime Lights Yearly L3 Global 15~arcsecond Linear Lat Lon Grid dataset was acquired from the Level-1 and Atmosphere Archive \& Distribution
System (LAADS) Distributed Active Archive Center (DAAC), located at the Goddard
Space Flight Center in Greenbelt, Maryland, USA.
 


\label{lastpage}

\bibliographystyle{elsarticle-harv}
\bibliography{biblio}

\end{document}

%% file: introduction.tex
\section{Introduction}

The atmosphere above the Canary Islands, and especially the western islands Tenerife and La~Palma, has been extensively characterized over the past 50 years~\citep{Kiepenheuer:1972,McInnes:1974,murdin1985,Brandt:1985,stickland1987,whittet1987,menendez1992,jimenez1998,Mahoney:1998,maring2000,torres2002,rodriguez2004,alonsoperez2007,basart2009,lombardi2009,delgado2010,Vernin:2011,rodriguez2013,cuevas2013,Varela:2014,Laken:2016,Vogiatzis:2018,Hidalgo:2021}. Due to the combination of large-scale atmospheric circulation on the descending branch of the Hadley cell~\citep[see e.g.][]{palmen1969,rodriguez2004}, and the ``Trade'' or ``Alisios'' winds coming from the Azores high area, a stable and strong temperature inversion layer (TIL) appears~\citep{font1956,huetz1969}, whose top is typically found \collaborationreview{at} heights 
around 1200\,m a.s.l.\ in summer and 1800\,m a.s.l.\ in winter~\citep{torres2002,Carrillo2016}. Whenever the temperature inversion is able to separate two well-defined regimes\collaborationreview{,} the moist \textit{marine boundary layer} (MBL) and above it, the dry \textit{free troposphere} (FT), the phenomenon is called an ``Alisio'' inversion. This happens about 80\% of the time~\citep{torres2002}. Under these conditions, the FT can be characterized as ``ultra-clean'', i.\,e., the concentration of particulate matter with an aerodynamic diameter of 10\,\textmu m or less (PM10) is lower than 10\,\textmu g\,m$^{-3}$.        The clean atmospheric conditions are one of the reasons why the Canarian observatories ``Observatorio del Teide (OT)'', located at $\sim$2400\,m a.s.l., and ``Observatorio del Roque de los Muchachos (ORM)'', located between $\sim$2100\,m and $\sim$2400\,m a.s.l., are known to belong to the best astronomical sites world-wide.

During the summer months, the TIL becomes weaker, and Saharan dust intrusions (the so-called ``calima'' phenomenon) may occur. The boundary layer may move then above the observatory, and higher aerosol densities are observed~\citep{lombardi2008}. Dust intrusions at the ORM are typically 3--4 days long, but also shorter intrusions of two days or, more rarely, long-lasting intrusion periods of 5--6 days are possible~\citep{lombardi2008}. Dust intrusions may occur throughout the whole year, but are generally confined to the period from June to October, with less intense and much less frequent outbreaks in February to April~\citep{Laken:2016}. Dust particles are characterized as \textit{marine}, \textit{desert} or \textit{anthropogenic aerosols}, depending on their \collaborationreview{origin}, which may be the Atlantic ocean, the Sahara desert, or Europe, respectively~\citep{rodriguez2004,rodrigues-desertdust}.

Aerosols at or above the observatories have been measured so far either by direct dust counters and analyzers on ground~\citep{kandler2007,lombardi2008,rodriguez2009,delgado2010,lombardi2010}, or on airplanes~\citep{collins2000,quinn2005}, or using integral light extinction measurements from the Sun~\citep{andrews2011,Laken:2016}, or from reference stars~\citep{murdin1985,whittet1987,stickland1987,MunozTunon:1997wu,jimenez1998,garciagil2010}. Some attempt has been made to use satellite data for aerosol content monitoring at the observatories~\citep{varela2008}, however with limited success so far. A LIDAR measurement campaign was carried out during several different characteristic situations~\citep{Sicard:2010}. All studies reveal predominance of relatively large particle sizes ($D\,>\,1$\,\textmu m) and hence gray extinction 
during ``calima'' conditions. Nevertheless, particles are then typically non-spherical and show a great variety of  shapes~\citep{kandler2007}.

Calima dust can most often be considered turbulently mixed with altitudes reaching up to 5\,km above the ORM, with most probable altitudes being found about 2\,km above ground~\citep{lombardi2008}. However, also complicated altitude profiles are possible, with several distinct layers reaching up to 8400\,m a.s.l.~\citep{Sicard:2010}.

We present here seven years of LIDAR measurements \collaborationreview{from 2013 till 2020,} taken at the ``Major Atmospheric Gamma-ray Imaging Cherenkov (MAGIC)'' telescopes~\citep{magic} site, located at the ORM ($28.78^\circ$~N, $17.89^\circ$~W, 2200\,m a.s.l.). The LIDAR was operated during night time, jointly with the MAGIC telescopes and pointing in nearly the same direction as the \collaborationreview{astrophysical} sources observed by MAGIC. The coverage of MAGIC observations with LIDAR data is above 85\%, apart from the commissioning of the LIDAR during the first two years of its regular operation and periods of upgrades of the LIDAR system. The MAGIC telescopes themselves are observing every night (apart from 3--5 days around full moon every month), unless the relative humidity surpasses 90\%, or the speed of wind gusts exceed 40\,km/h.

Several additional monitoring instruments are used at the MAGIC site~\citep{fonticrc,2014arXiv1403.5083G,Fruck:2015,will:atmohead2016}. A commercial weather station and All-Sky camera are mounted on the roof of the MAGIC Counting House in close proximity to the LIDAR dome. In addition, an infrared pyrometer is mounted on the dish structure of one of the MAGIC telescopes, pointing in the same direction as the telescope~\citep{will:atmohead2016}.

Every two seconds, the weather station is read out to evaluate if the current conditions are within the predefined safety limits for operations. The control programs of MAGIC check the values and act in case a safety limit is violated, putting the cameras and the telescopes in a safe state. The LIDAR control program checks the values as well and closes the dome of the LIDAR in case the wind or humidity values exceed \collaborationreview{those} safety limits. For later analysis, the weather station data are archived every two minutes. 

%% file: lidar.tex
\section{The MAGIC micro-Joule  LIDAR system with HPD-based photon detection}
\label{sec:lidar_hardware}


\begin{table}
  \begin{center}
    \caption{Components of the LIDAR assembly before and after the major upgrade.
    \label{tab:hardware_components}}
    \begin{tabular}{c|c|c}
    \toprule
    Component & Period 1   & Period 2   \\
              & 2013--2016 & since 2017 \\
    \midrule
    Telescope & \multicolumn{2}{c}{Welded aluminum} \\
    Mirror & \multicolumn{2}{c}{Diamond-milled Al, 60\,cm diameter, 150\,cm focal length} \\
    Mount & \multicolumn{2}{c}{ASTELCO NTM-500 high precision equatorial robotic mount, 80\,kg load capacity} \\
     & \multicolumn{2}{c}{Tracking accuracy: 5$^{\prime\prime}$} \\
     & max. acceleration: 10$^\circ$/s$^{2}$ & max. acceleration: 20$^\circ$/s$^{2}$\\
    Mount controller & \multicolumn{2}{c}{ASCOM interface, driver control through LabView} \\
    \midrule
    Laser & Teem Photonics STG-03E-1x0, 532\,nm & Horus Lasers HLX-G-F001-11101, 532\,nm \\
    Pulse energy, Stability (rms)  & 5\,\textmu J, 3\% &   25\,\textmu J, 3\% 
    \\ 
    \collaborationreview{Pulse} duration (FHWM) &  0.5\,ns & 1\,ns \\
    Beam divergence  & 10$\pm$2 mrad  & 12 mrad 
    \\
    Beam quality factor (M2) &   1.3   &  1.2  \\
    Beam expander & 10\,$\times$, Thorlabs BE10M-A & 20\,$\times$, Thorlabs GBE20-A \\
    Number of shots per data point, frequency & 50\,k, 500\,Hz every 5\,min & 25\,k, 250\,Hz every 3\,min \\
    Baffle tube & -- & 1\,m, 5\,cm \diameter, carbon fiber, 4 baffles \\
    \midrule\addlinespace[0.1cm] 
    Light detector & \multicolumn{2}{c}{Hamamatsu R9792U-40 HPD, QE\,$\approx$\,50\% at 532\,nm} \\
    Detector module & Custom built PCB, Al case & Custom PCB, 3D-printed housing, Al case \\
    Amplification & $2.6\,\times\,5\times 10^2$\,V\,A$^{-1}$ & $\sim 390$\,V\,A$^{-1}$ (differential, TI LMH6554) \\
    Photon pulse FWHM & 10\,ns (2.1\,ns without  OpAmp) & 2.3\,ns \\
    HV supply (7 kV, photocathode) &  external (NIM module) & EMCO C series,  external water-tight supply box \\
    HV supply (400 V, AD) & Inside detector module & iseg APS, inside HV supply box \\
    HV control & \multicolumn{2}{c}{MCP Microchip DACs 4822 and 4922 series, 12-bit resolution, SPI controlled} \\
    Temperature Compensation  &  Analogue   & Software \\ 
    FADC digitizer & Spectrum MI.2030, 8\,bit, 200\,MS/s & Spectrum M4i.4450-x8, 14\,bit, 500\,MS/s \\
    \bottomrule
    \end{tabular}
  \end{center}
\end{table}

\begin{figure}
  \begin{center}
  \includegraphics[width=0.8\columnwidth]{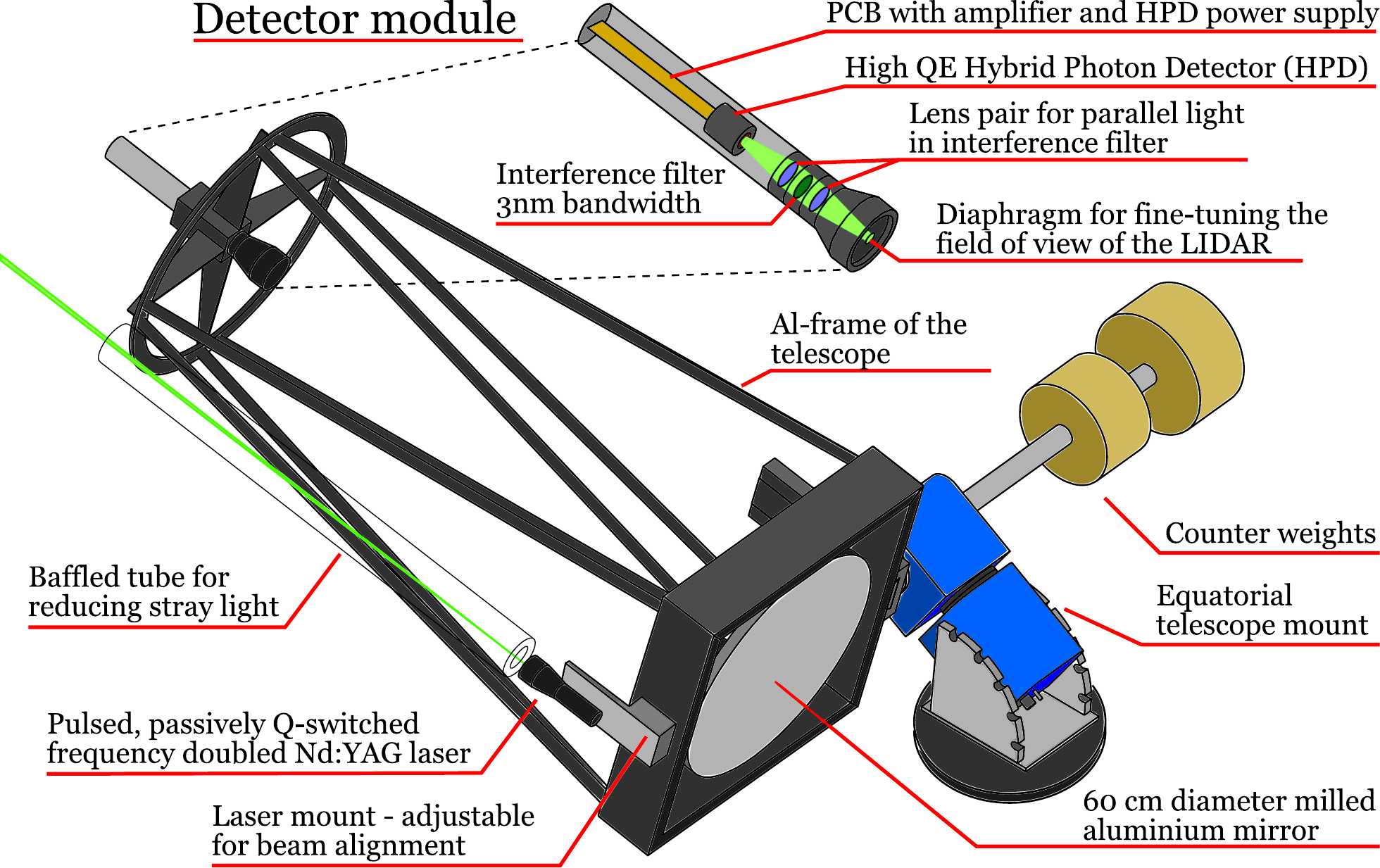}
  \caption[Hardware components of the MAGIC LIDAR system]{Hardware components of the MAGIC LIDAR system. Several individual components were modified or upgraded, but the overall scheme has remained the same.
  \label{fig:lidar_hardware}}
  \end{center}
\end{figure} 

\begin{figure}
  \begin{center}
  \includegraphics[width=0.68\columnwidth]{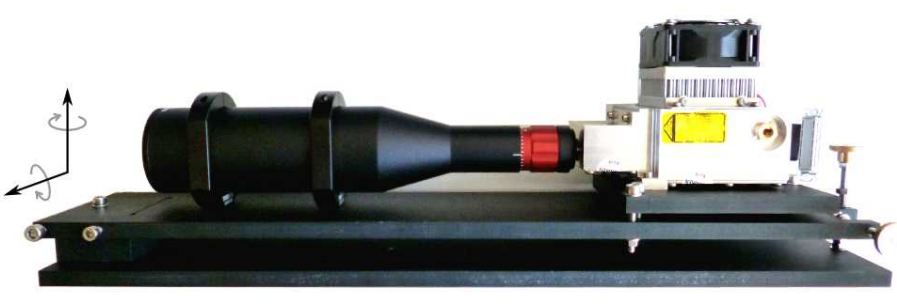}
  \includegraphics[width=0.462\columnwidth]{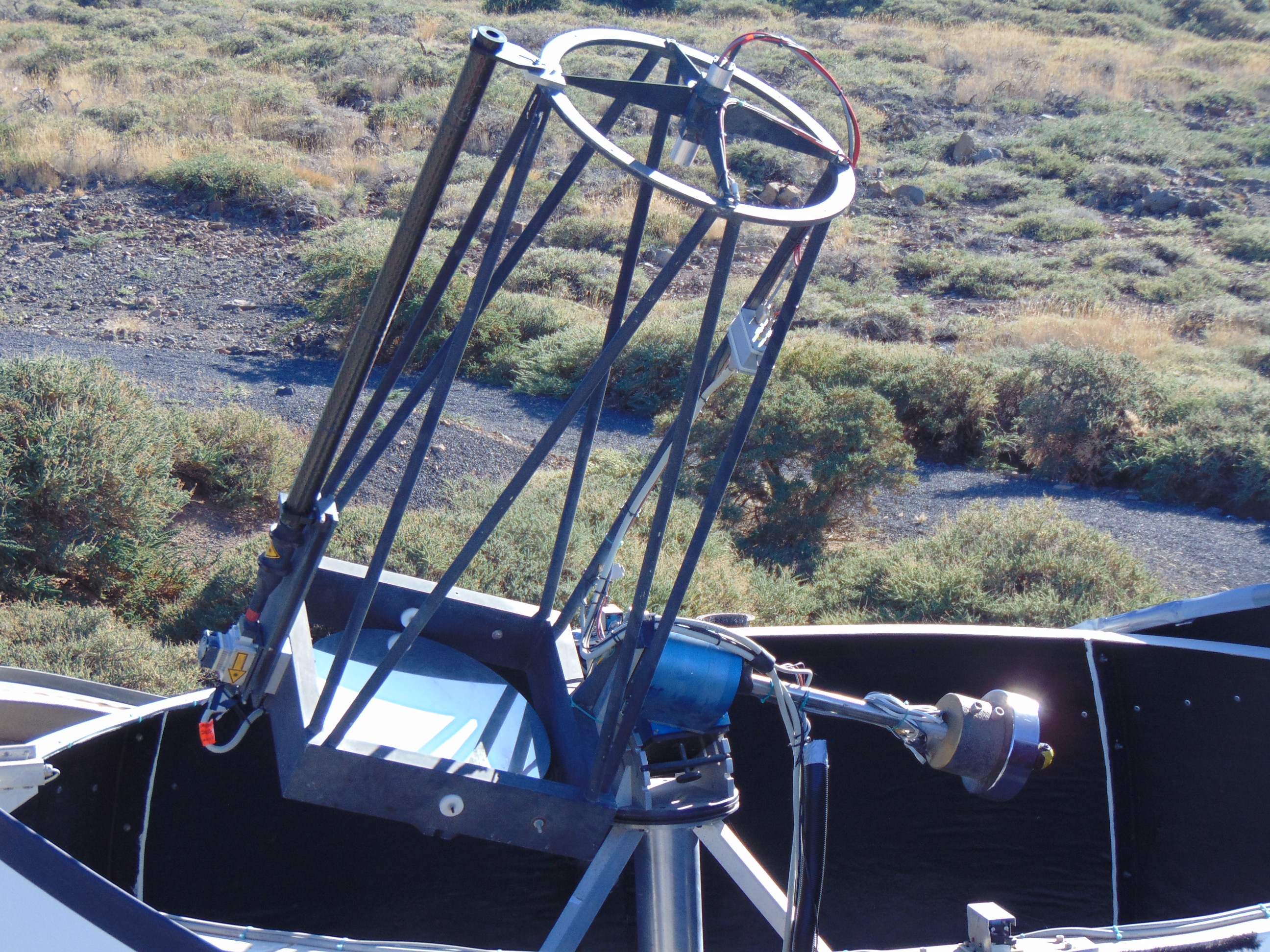}
    \includegraphics[width=0.52\columnwidth]{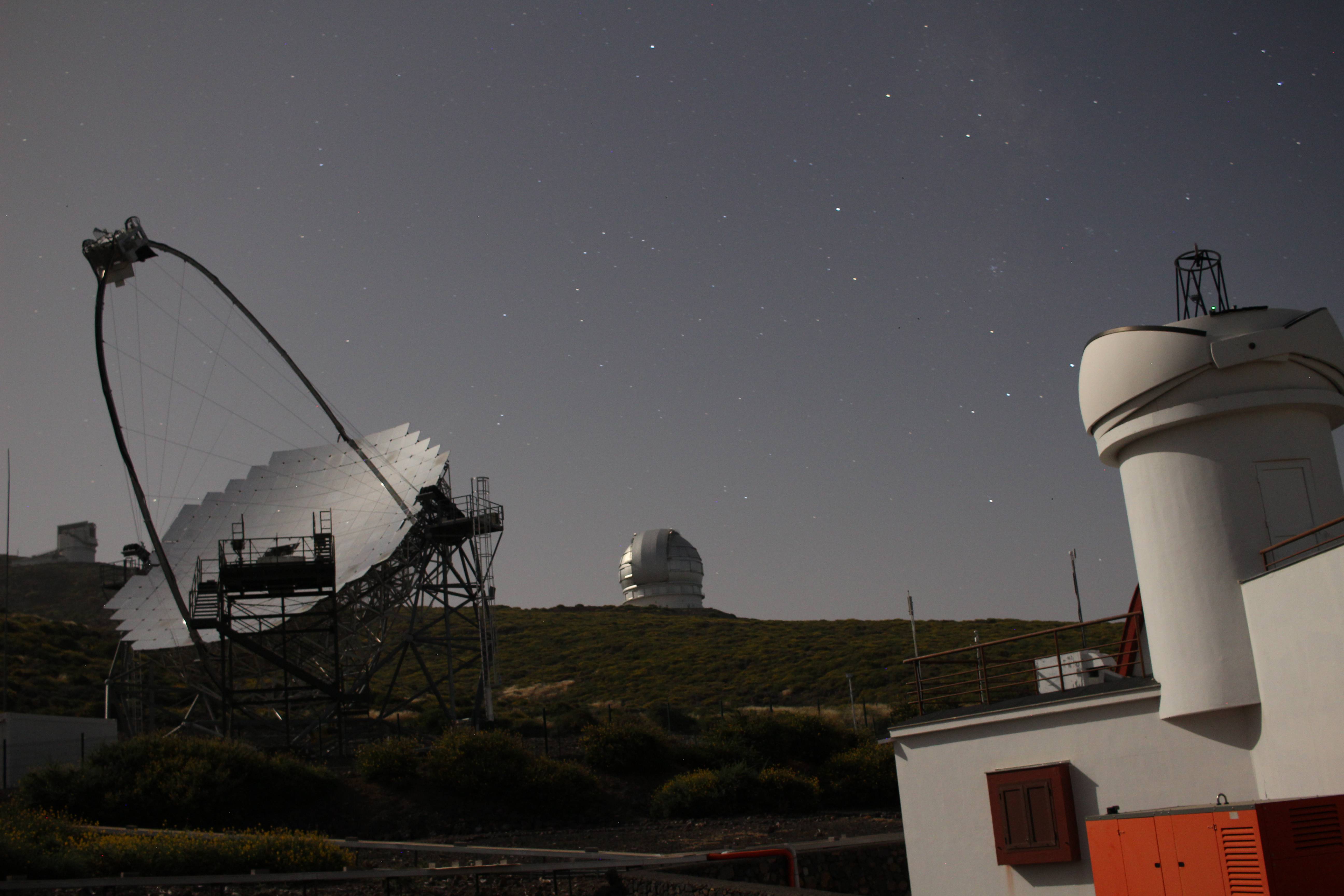}
  \caption[Laser setup]{Top: Upgraded laser setup: the laser on the right is connected by cable to the laser controller in the LIDAR control room, the output beam of the laser (silver colour box with a ventilator on top) is fed into the beam expander on the left. Both are mounted on a special plate assembly that can be easily adjusted in all directions to properly align the laser with the telescope and detector head. Bottom left: a picture of the LIDAR including structure, laser, mirror and light detector. On its left side, the  baffle tube made of carbon-fiber is coupled to the exit lens of the beam expander. Its function is to block the scattered light from the laser and expander, which could directly hit the light sensor module of the LIDAR. Bottom right: a picture of one of the two MAGIC telescopes visible together with the MAGIC counting house with on top the opened dome in which the black structure of the LIDAR telescope can be seen. 
  \label{fig:laser_new}}
  \end{center}
\end{figure}

\begin{figure}
  \begin{center}
  \begin{subfigure}{.36\textwidth}
    \centering
    \begin{subfigure}{\textwidth}
      \centering
      \includegraphics[width=\columnwidth]{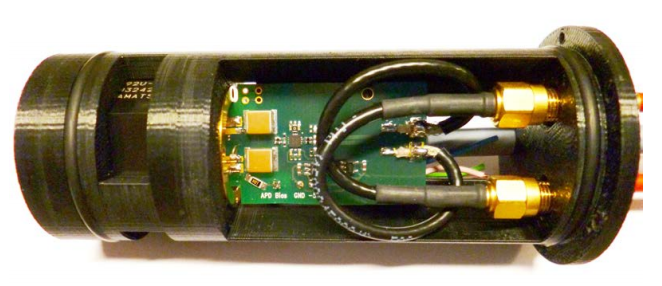}
    \end{subfigure}
    \begin{subfigure}{\textwidth}
      \centering
      \includegraphics[width=\columnwidth]{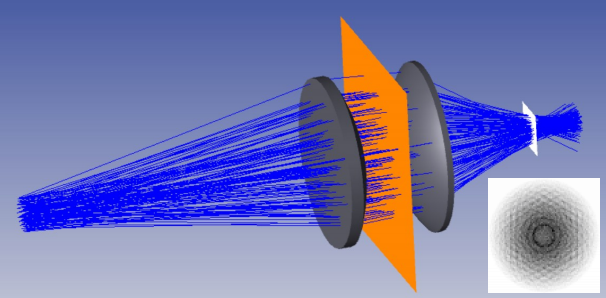}
    \end{subfigure}
  \end{subfigure}
  \hspace{1em}
  \begin{subfigure}{.59\textwidth}
    \centering
    \includegraphics[width=\columnwidth]{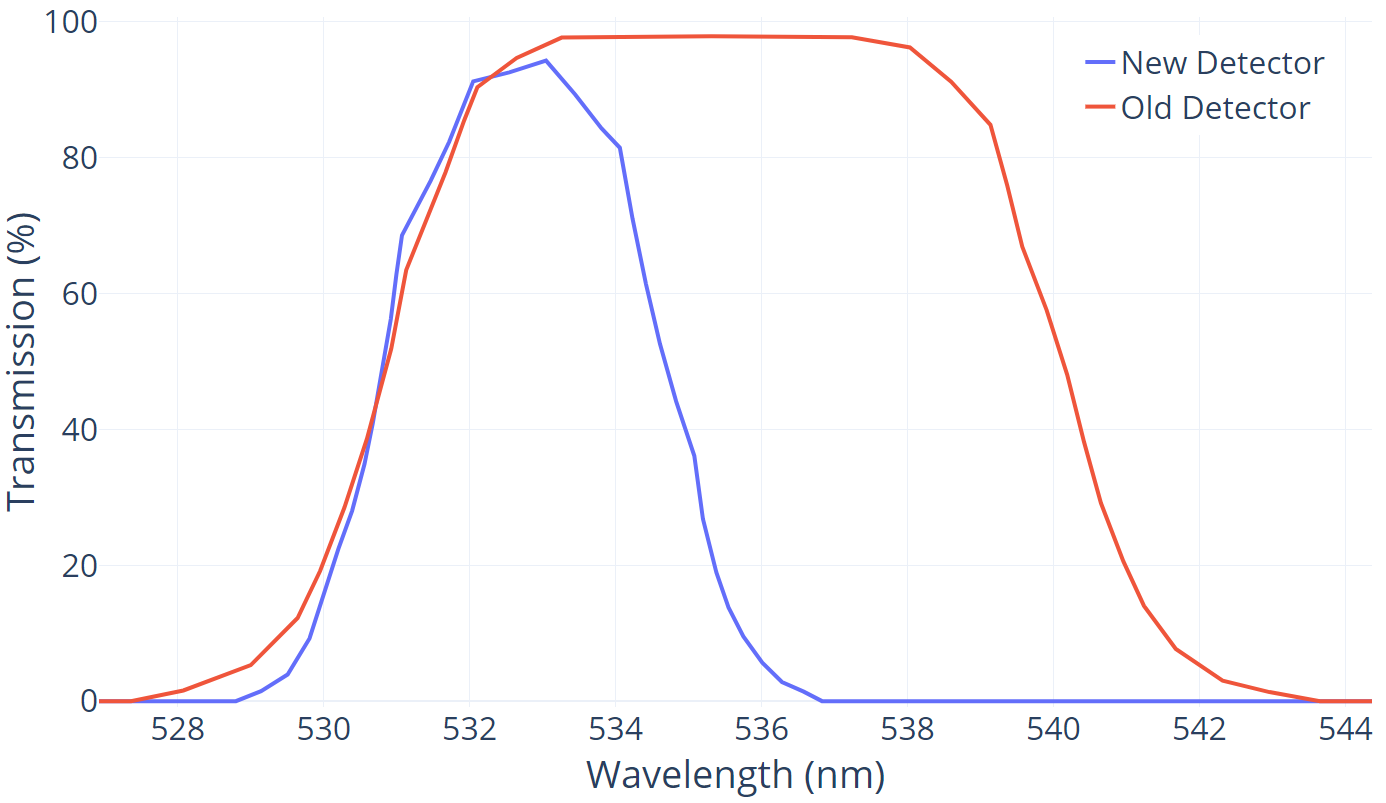}
  \end{subfigure}
  \caption[Detector assembly]{The upgraded detector (top left) houses custom-made electronics and the HPD light sensor in a 3D-printed case that is inserted into a watertight aluminum tube. To increase the effectiveness of the 3\,nm interference filter mounted before the HPD, two lenses make the light passing the filter parallel and image the entrance aperture of the diagram onto  the HPD  (bottom left sketch). On the right, the transmission of the improved lens assembly of the new detector assembly (blue)  is compared with the old one (red). The much narrower peak further reduces the amount of background light (mostly Light of the Night Sky (LoNS)) that enters the HPD, while preserving, as much as possible, the transmission of the back-scattered laser light at 532\,nm.
  \label{fig:detector_new}}
  \end{center}
\end{figure}

\begin{figure}
  \begin{center}
  \includegraphics[width=0.42\columnwidth]{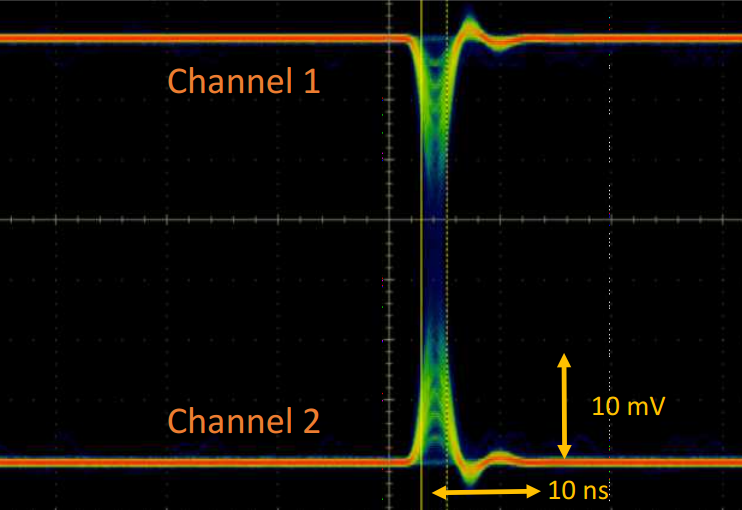}
  \hspace{1em}
  \includegraphics[width=0.48\columnwidth]{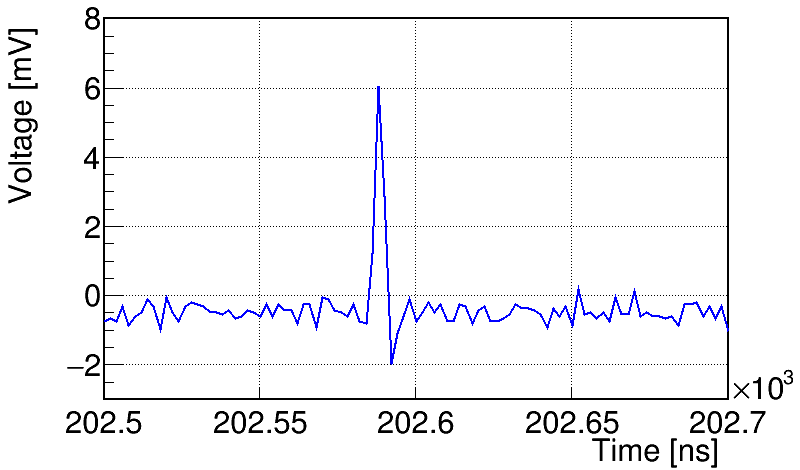}
  \caption[Differential amplifier]{Left: the upgraded detector includes a differential amplifier, producing two signals of opposite polarity. In this oscilloscope measurement, an HV of $-$8  \,kV was applied with a bias voltage of 422\,V, resulting in a detector gain of $\sim$~150\,000. The channels show very good coincidence, a FWHM of the photon peaks of 2.3\,ns. It is possible to resolve up to 6 photo-electrons. For actual measurements, the difference of the two channels is used, which has the advantage of amplifying the signal while suppressing the common-mode noise introduced in the data cables by the surrounding electronics, hence greatly improving the signal-to-noise ratio. Right: Waveform of a single photo-electron recorded with the FADC-card.
  \label{fig:detector_performance}}
  \end{center}
\end{figure}

The LIDAR is installed inside a protective dome above the LIDAR control room on the roof of the MAGIC control building, which houses computing and electronics for the MAGIC telescopes. The main purpose of the LIDAR is simultaneous monitoring of the atmospheric extinction profile of the observed field-of-view of the MAGIC \collaborationreview{telescopes} during the nightly observations. We use a micro-Joule LIDAR system operating at \collaborationreview{a wavelength of} 532\,nm (see Table~\ref{tab:hardware_components}). The wavelength of the Cherenkov light observed by MAGIC lies \collaborationreview{mostly} in the blue region, peaking at 320\,nm, with a median wavelength close to 470\,nm (at 15\degree zenith angle). The third harmonic of an Nd:YAG laser at 355\,nm would be the most appropriate wavelength to use for a LIDAR system, however,
\collaborationreview{however the need for a safer and more practical handling and adjustment} of a visible laser beam favoured the use of the second harmonic at 532\,nm,
\collaborationreview{in combination with the use of an HPD with GaAsP photocathode, which provides an unprecedented quantum efficiency of $>50$\% for the wavelength range 500-600~nm. }

This choice entails the need for minor wavelength corrections and assumptions about the local \AA ngstr\"om coefficient~\citep{angstrom:1929} when the derived atmospheric extinction is used to simulate that of Cherenkov light.

A low pulse energy was chosen in order to minimize  possible interference with operations of MAGIC and other telescopes of the observatory, 
\collaborationreview{and to avoid any safety concerns, e.g., related with eye-safety.}
A rendering of the LIDAR with the main hardware components is shown in Fig.~\ref{fig:lidar_hardware}, more details on the individual components can be found in Table~\ref{tab:hardware_components}. Several parts of the system were gradually upgraded between 2014 and 2019, but the overall functionality remained the same and is described in the following paragraphs with the upgraded components described in brackets.

The LIDAR uses a pulsed, passively Q-switched, frequency-doubled Nd:YAG laser with 5\,\textmu J (25\,\textmu J) pulse energy operated at 300\,Hz (250\,Hz) repetition rate. The laser incorporates a photodiode used to externally trigger the readout. A 10\,$\times$ (20\,$\times$) beam expander is used to reduce the beam divergence to 
\collaborationreview{10\,mrad (12\,mrad)}
The laser setup is mounted on a custom-built assembly that allows for easy and precise alignment of the laser with the LIDAR telescope and the detector, see top panel of Fig.~\ref{fig:laser_new}. The beam expander is placed in front of the laser with a  minimal gap to efficiently reduce the beam divergence, but dust particles on its exit window can scatter a tiny fraction of the laser light out of the beam. That scattered light may find its way to the LIDAR detector, but does not saturate the detector and can be discarded later on using the information from its early arrival time. In addition, scattered light, either from combination of the laser plus beam expander or from the air in the near range, can initiate false triggers of the MAGIC telescope PMT cameras. Depending on the viewing geometry of the LIDAR and the MAGIC telescopes, the scattered laser light can significantly increase event rates and dead time of the MAGIC data acquisition. To mitigate this problem, two measures have been taken: the scattered light was reduced with the help of a baffle tube (see bottom panel in Fig.~\ref{fig:laser_new}), made of a 1\,m long carbon fiber tube (a spare from the construction of the MAGIC telescope dish structure) with a diameter of 5\,cm and four inserted plastic baffles along its length. More recently this tube was replaced with a shorter commercially available alternative. In addition, the LIDAR events in the MAGIC data stream are tagged using a digital signal from the LIDAR electronics to the MAGIC readout electronics, which allows for the removal of such events offline during data analysis. For a short time, a slow mode was \collaborationreview{tested} to limit the interference of the LIDAR data in the MAGIC data stream, reducing the frequency of the pulsed LIDAR laser down to 150\,Hz or even 100\,Hz. After a trial period where no additional benefit could be found, it was decided not to \collaborationreview{use} this mode any longer and solely rely on the trigger tag and the baffle tube.

On the receiver side, a 60\,cm diamond-milled, massive aluminum mirror with 150\,cm focal distance is used in an off-axis geometry with respect to the emitter. The detector optics have an aperture diaphragm set to $\sim$\,6\,mm diameter in the focal plane of the mirror. A small telescope based on a pair of lenses images the diaphragm onto the surface of the HPD and as such limits the accepted solid angles.  The wavelength selection is made by using a 3\,nm bandwidth interference (IF) filter that is located between the lenses. It helps to reduce the light of night sky (LoNS) by more than a factor of hundred. The increased effectiveness of the IF filter of the upgraded detector model can be seen in Fig.~\ref{fig:detector_new}. After the narrow band filter, a second lens maps the light onto the photo-cathode of a hybrid photo \collaborationreview{detector} (HPD) with over 50\% quantum efficiency (QE) at 532\,nm. HPDs are very well suited for use with LIDARs, apart from single p.e.\ resolution and fast response, HPDs also show very low after pulsing (\citet{Saito:PhD,Saito:2011yp}). The signal at the output of the HPD gets directly pre-amplified with a trans-impedance gain of $\sim$~1300\,V\,A$^{-1}$. The upgraded detector electronics include a differential amplifier ($\sim$~390\,V\,A$^{-1}$), producing two signals of opposite polarity that are transmitted to a dedicated readout PC and subtracted from each other. This further reduces the pick-up noise introduced largely by the drive electronics. Some measurements of the new detector module can be found in Fig.~\ref{fig:detector_performance}. The amplified signal is received and recorded by a computer equipped with an 8-bit (14-bit) 200\,MSamples/s (500\,MSamples/s) FADC PCI card, externally triggered by the photodiode of the laser. The voltage range of the FADC card is $\pm$500\,mV. It is operated in FIFO mode for continuous data transfer to the PC memory. More than 14\,GB of data are such recorded in less than 60 sec. The PC itself runs under Windows~XP (Windows~7) and communicates with the LIDAR subsystems through a 32 channel TTL I/O card. For each of the typically used 50\,000 (25\,000) laser shots, about 160\,\textmu s of waveform data are written to memory for onsite analysis.

The LIDAR is controlled through a graphical user interface written in LabVIEW. Complex or computationally intensive tasks are performed outside of LabVIEW by a series of small \collaborationreview{subroutines}
written in C{\tiny$^{++}$}. During standard operation the LIDAR receives the pointing coordinates of the MAGIC telescopes and follows these with an offset of $\sim5^\circ$, avoiding that the laser enters the  MAGIC \collaborationreview{telescopes}' field-of-view. Since October 2019, the LIDAR also avoids conflicts with the prototype Large-Sized Telescope \citep{the_cta-lst_project_status_2021} of the future Cherenkov Telescope Array Observatory (CTAO), located less than 100\,m in western direction. In this so-called "auto-mode", the LIDAR performs a full measurement cycle every four minutes.

\begin{figure}
	\centering
	\includegraphics[width=1\textwidth]{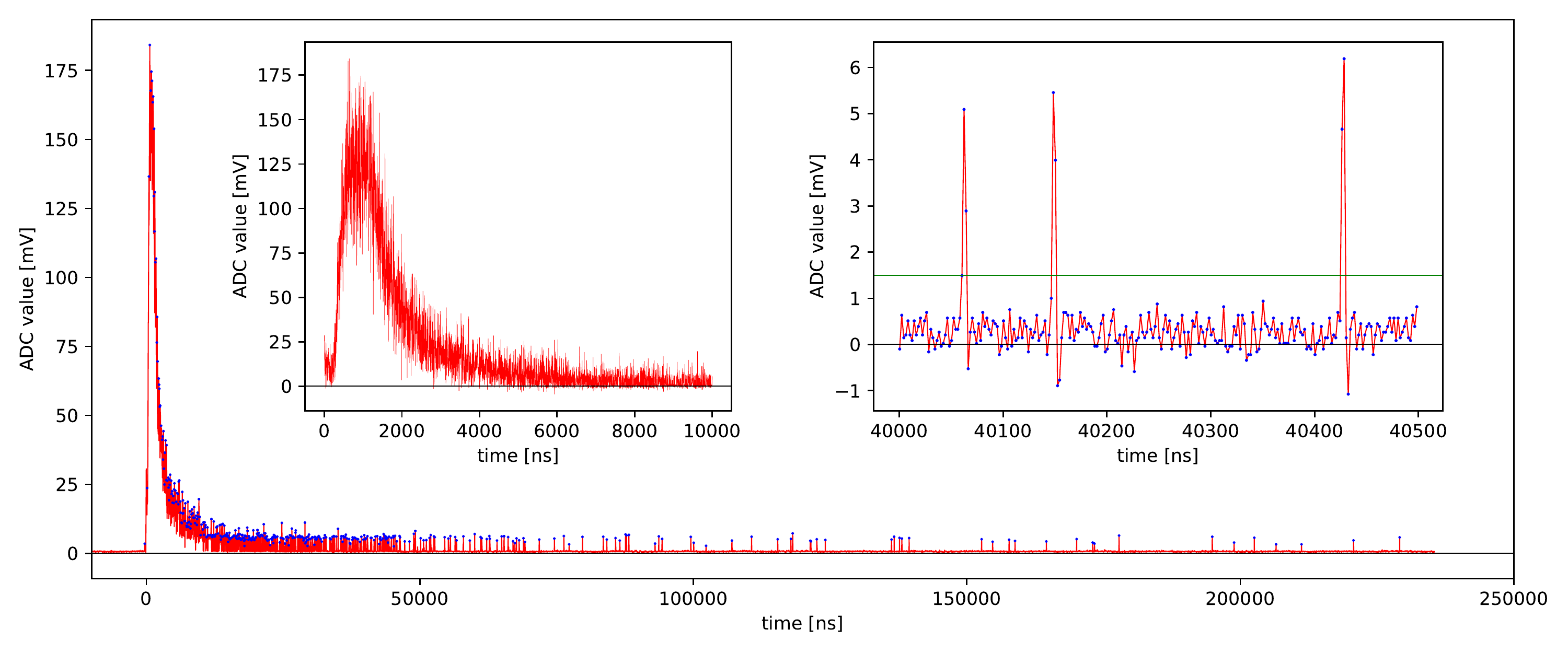}
	\caption{\label{fig:fadcsignals}
	Time profile of the back-reflected light measured by the LIDAR (Period 2, after the 2015 upgrade) from a single shot of the laser fired into the vertical atmosphere (main plot), and a zoom into the near range, low altitude region where the signal estimation  has to be made by charge integration (left inset plot). Peaks of the detected single photons are shown in the right inset plot, the horizontal line depicts the photon counting threshold $V_t$. The photon peaks have a similar charge integral, but due to the limited sampling rate, they appear at different heights, albeit still well above the threshold.}
\end{figure}

After a LIDAR shot sequence has finished, two different photon counting algorithms are applied to the raw data, which itself gets deleted immediately after, in order to save disk space. The first "photon counting" algorithm searches for single-photo-electron peaks. The second "analogue" algorithm integrates the waveform and divides by the mean charge of a single photo-electron, which itself is obtained from the distribution of single photo-electron charge integrals, from the same data sample. Single photo-electron waveform integration is possible due to the excellent charge resolution of the HPD used~\citep{Saito:PhD}, in combination with fast FADC pulse sampling. Later on, the single photo-electron counts are used from a given distance onward (see Table~\ref{tab:hardware_periods}), and the number of photons from charge integration is used for closer ranges, where the probability for photon pileup is not negligible anymore. The algorithm for counting single photons makes use of the the short exponentially decaying edge of the pulses, which are a function of the capacitance of the avalanche photodiode (APD) and the (low) input impedance of the trans-impedance amplifier, along with stray inductance and the bandwidth of the amplifier and the digitizer. This can be used to distinguish them from possible high-frequency noise, picked up after the amplifier. The algorithm works in two steps: \collaborationreview{first}, the whole FADC curve is scanned for events exceeding a certain threshold $V_\t{t}$. Then, the corresponding regions are scanned for the decaying edge of single-photon events by searching for such peaks that also exceed $V_\t{t}/2$ in the second bin after the peak and $V_\t{t}/4$ in the third bin~\citep[for further details, see][]{fruck:phd}. We have also implemented and tested an analysis based on digital filters, following the approach of~\citet{Albert:NIMPRSASDAE2008a}, but only marginal improvements could be obtained and the analysis based on simple thresholds was maintained for simplicity. The recorded waveform before the laser-induced signal is used for a baseline correction of the data. First, a mean pedestal baseline is calculated and then subtracted from each waveform sample of the shot sequence. Then, the region before the laser-induced pulse is searched for background photo-electron pulses, from which a background rate is calculated and subtracted from the photo-electron rates of the return signal. Examples for FADC sample data and an illustration of the single-photon counting algorithm can be found in Fig.~\ref{fig:fadcsignals}. Due to inelastic back-scattering of electrons off the avalanche diode, the achieved charge resolution is somewhat worse  than the theoretical expectations of $\sim 2.6$\%, estimated from the Poissonian fluctuations of the HPD bombardment gain $G\sim$1450~\citep[for detailed numbers, see][]{ambrosio}. Also the HPD quantum efficiency decreases by 12\% due to this effect. 

The Nov. 2015 upgrade of the HPD-control and readout electronics paid special attention to mitigate the strong temperature dependence of the avalanche photo-diode (APD) gain, part the HPD. To achieve this, a thermistor has been thermally
coupled to the APD and corrects the bias voltage. This setup is capable of compensating the main temperature dependence
of the APD gain, reducing it to an acceptable stability of
$\sim$0.3\,\%$/^\circ$C in the temperature range between 25\,$^\circ$C to 35\,$^\circ$C~\citep{Orito:HPD:2009, mueller:master}. In order to compensate for the residual temperature dependency of the single photo-electron charge, 
the FADC card is put into software-trigger mode and 1000 waveforms with background photons are acquired. In the data, single photo-electron events are then searched for with a threshold voltage of 1.5\,mV and the waveform integrated over three FADC time slices, before, on and after the peak. Moreover, a mean baseline level is calculated across a region starting 50\,ns before the found peak to 10\,ns before it. This baseline is subtracted from the integrated photo-electron waveform. The truncated mean of the resulting charge distribution is then evaluated over a range from 6 to 60\,pVs, and the APD gain iteratively adjusted to achieve a mean photo-electron charge of $\sim 16$\,pVs. 
This monitoring and feedback algorithm gets executed every 15 seconds, 
as long as no further data acquisition is being executed.

\begin{figure}
  \begin{center}
  \includegraphics[width=1.0\columnwidth]{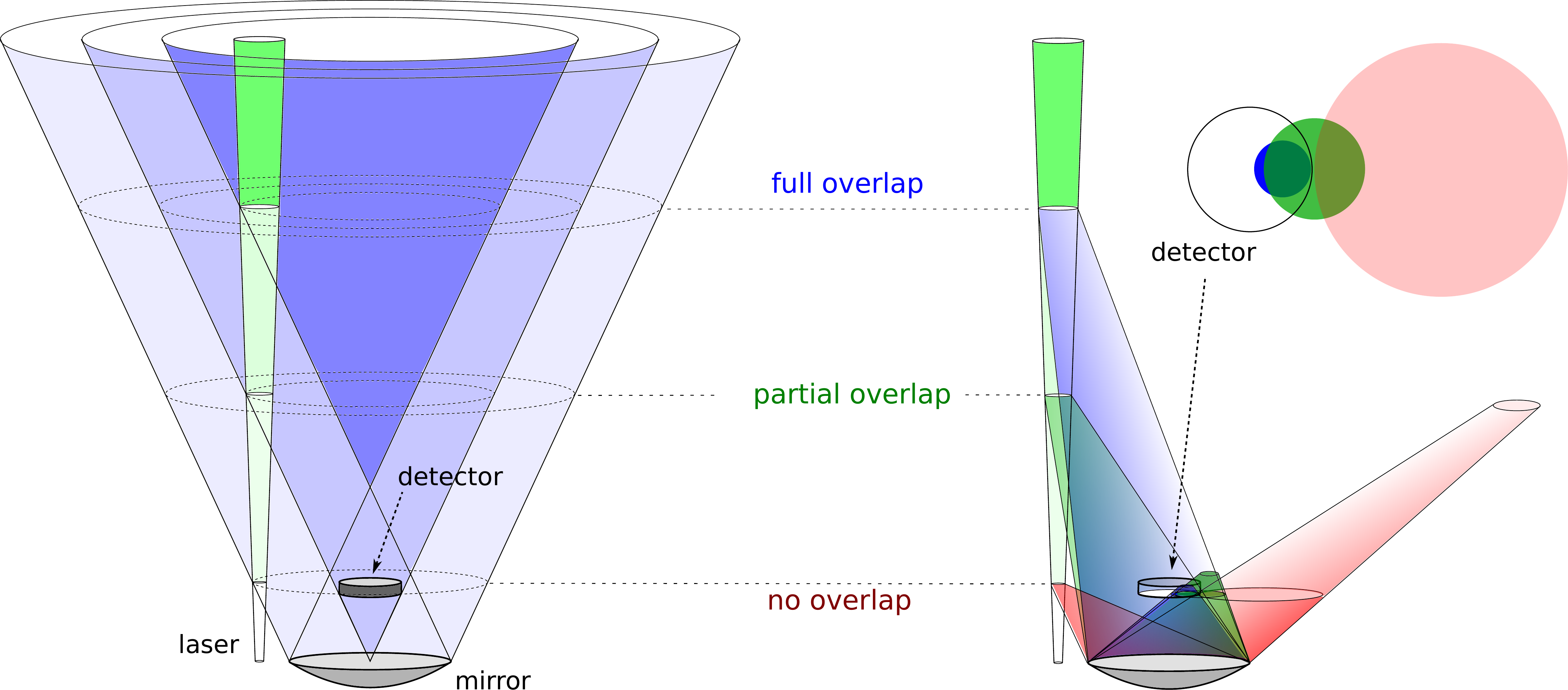}
  \end{center}
  \caption[Overlap of laser beam and detector FoV]{Overlap of the laser beam in green and the detector field of view in blue. Since they are mounted next to each other, there is a region close to the LIDAR where the backscattered light from the laser does not reach the detector (no overlap), or only part of it is seen (partial overlap). 
  \label{fig:overlap}}
\end{figure}

\begin{figure}
	\centering
	\includegraphics[width=0.5\textwidth]{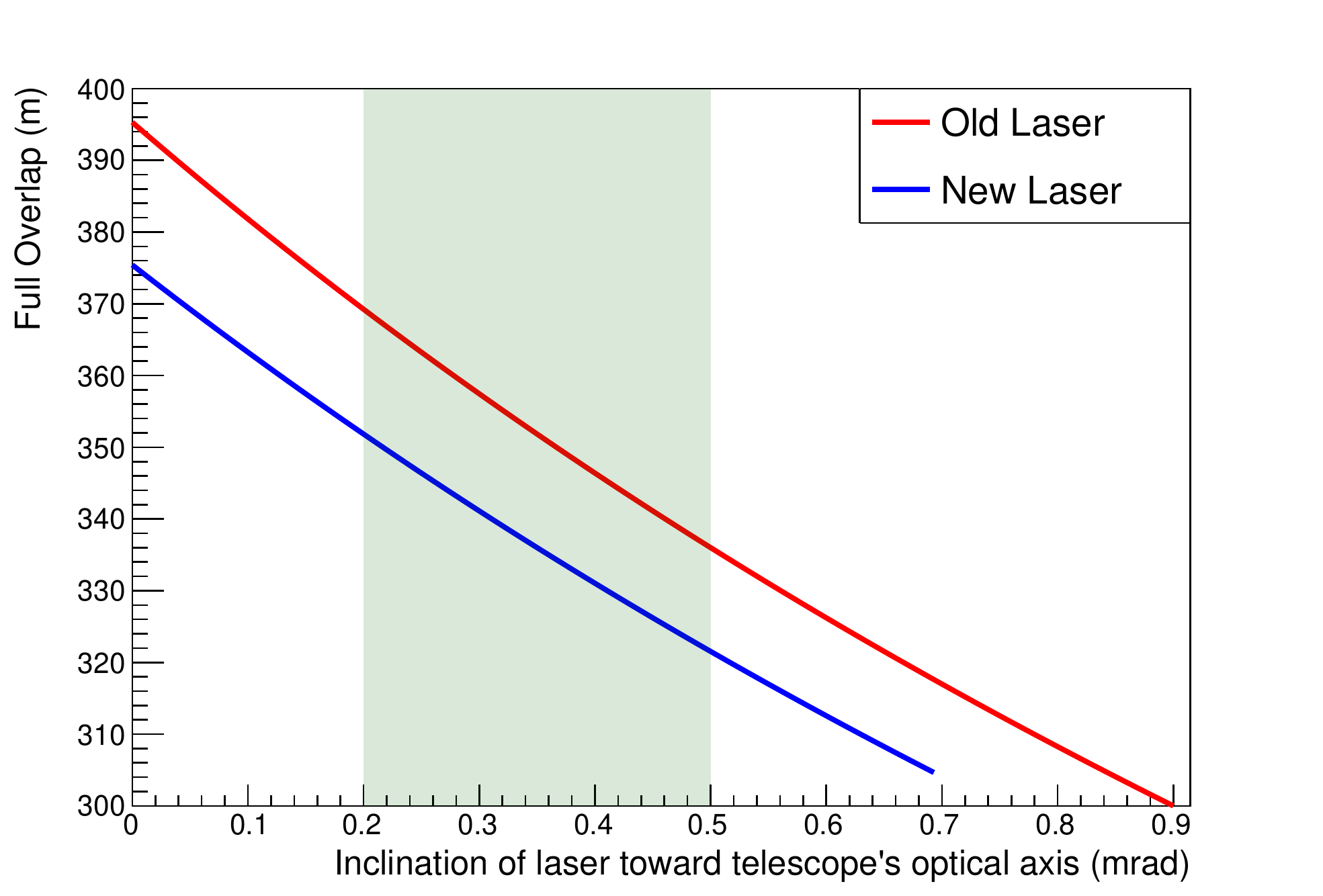}
	\caption[Expectations for full overlap]{Design expectations for full overlap distances for the MAGIC micro LIDAR system. The lines show the result of Eq.~A2 of~\citet{Biavati:2011}, depending on different inclination angles of the laser towards the optical axis of the telescope. The green shaded regions show the range of angles realized in our system.
	\label{fig:biavati}
    }
\end{figure}

Full geometric overlap between the laser beam and the fully focused detector field of view \collaborationreview{(see Fig.~\ref{fig:overlap})} is reached at a distance of about 350\,m from the telescopes (e.\,g., calculated with Eq.~A2 of~\citet{Biavati:2011}, see Fig.~\ref{fig:biavati}). These calculations are in good agreement with ray-tracing detector simulations~\citep{fruck:diploma}.

\begin{table}
    \centering
    \begin{tabular}{cc|ccccccp{5.0cm}}
   \toprule  
       Nr. & Time  & Sampling &  Shots  & Distance & $T_{0}$ at & Distance & Comments \\
        & range & rate     &                 & to p.e.  & slice      & to full  &          \\
       &       &          &                 & counting &            & overlap  &          \\
       &       & (MS/s)   & ($\times 10^3$) & (m)      & (1)        & (m)      &          \\
   \midrule
    A1   & 02/2013--08/2014 & 200      &  50 & 
3836 &   16.3   &  ~\llap{$\sim$}400         & Start data taking with commissioned system \\ \addlinespace[0.1cm]
  A2   & 09/2014--11/2015 & 200      &  50  & 
3836 &   16.3  &   ~\llap{$\sim$}360         & LIDAR lifted by 32\,cm, change signal estimator \\ \addlinespace[0.1cm]
    A3   & 12/2015--11/2016 & 500      &  25  &  
3836 &   16.3  &        ~\llap{$\sim$}360  &  New FADC card  \\
      \midrule
    B1   & 12/2016--03/2017 &  500        &  25  &  
4652 &  15.1 &       ~\llap{$\sim$}2000  &  New laser \\ \addlinespace[0.1cm]
    B2   & 04/2017--04/2017 & 500       & 25  &  
4652   &  15.8  &          ~\llap{$\sim$}340 &  Baffle tube installed  \\ \addlinespace[0.1cm]
    B3   & 05/2017--12/2018 & 500\rlap{$^\dagger$} & 25  & 
6977    & 31.1  & ~\llap{$\sim$}340 &  New raw data binning \\ \addlinespace[0.1cm]
    B4   & 03/2019--02/2020 & 500\rlap{$^\dagger$} & 25  &  
6977 & 31.1  & ~\llap{$\sim$}340  &  New readout, new power supply \\
    \bottomrule
    \end{tabular}
    \caption{Characteristics of the MAGIC LIDAR during different hardware periods defined in Table~\protect\ref{tab:hardware_components}. $^\dagger$\,Dynamic range is reduced subsequently in fixed intervals at larger distances, see text.
    }
    \label{tab:hardware_periods}
\end{table}

%% file: signalinversion.tex
\section{Adapted signal inversion algorithm}

Elastic LIDAR signal inversion is a long-standing problem, extensively discussed in the literature~\citep[see, e.\,g.,][for an excellent overview of the problem]{kovalev}. For clean, or only marginally turbid atmospheres, the molecular scattering contribution to the LIDAR return becomes important, if not predominant, and systematic uncertainties of the assumed molecular profile start to dominate the accuracy of the aerosol inversion products.

Analytical solutions for a two-component atmosphere have been proposed by~\citep{collins1966,fernald1972,platt1979,fernald1984,klett1985,browell1985,weinman1988,kovalev1993,kovalevmoosmueller1994,kovalev1995,kunz1996,McGraw:2010}, however none goes without previous assumptions about the aerosol extinction-to-backscatter efficiency (LIDAR ratio) throughout the retrieval range. 

Correcting science data of MAGIC for aerosols and clouds in its field-of-view is, however, based on two fundamental assumptions: 
\begin{enumerate}
  \item The nocturnal boundary layer is normally found below the typical emission height of Cherenkov light from gamma-ray showers observed by the MAGIC Telescopes (see, e.\,g., Fig.~10 of~\citet{HILLAS:JPGPP1990a}). Its structure hence does not need be resolved, only its overall transmission counts.
  \item In case of layers at higher altitudes, the vast majority can be considered thin compared to the typical longitudinal shower profiles which span several kilometers. Hence their internal structure \collaborationreview{does not need to} be resolved either, only the overall transmission of the cloud layer is required~\citep{fruck2013}.
\end{enumerate}\noindent

For these reasons, neither a good range resolution is necessary, nor retrieval of the backscatter coefficient. Quality criteria for a LIDAR used together with IACTs include instead: good transmission resolution for optically thin aerosol and cloud layers, for distance ranges from ground to at least 25\,km, the possibility to point the LIDAR to any direction in the sky, and acceptable accuracy of the transmission retrieval. 

Analysis of the LIDAR data typically starts with the single scattering LIDAR equation:
%
\begin{equation}
  N(r) =  N_0 C G(r) \ddfrac{A}{r^2} \beta(r) l \exp{\!\left(\!-\!2\!\int_0^r\!\! \alpha(r') \ud r' \right)}.
  \label{eq:lidar}
\end{equation}

%
where $N(r)$ is the 
number of observed photo-electrons during, 
a digitization length $l$, 
corresponding to distances between $r$ and $r+l$ of the back-scattered light. $A$ is the effective area of the system, $N_0$ the number of photons emitted per laser pulse, $C$ is the overall detection efficiency related to the experimental setup, $G(r)$ is the geometrical overlap factor, $\beta(r)$ is the local back-scatter coefficient and $\alpha(r)$ is local the extinction coefficient of the atmosphere. Both contain a molecular scattering part ($\beta\m(r)$ and $\alpha\m(r)$) and an aerosol scattering part ($\beta\p(r)$ and $\alpha\p(r)$); $\beta(r) = \beta\m(r) + \beta\p(r)$, $\alpha(r) = \alpha\m(r) + \alpha\p(r)$. Furthermore, the extinction-to-backscatter ratio or the \textit{LIDAR ratio} $S(r)=\alpha(r)/\beta(r)$ is introduced~\citep{takamura1987}, for aerosols ($S\p(r)$) and molecules ($S\m=8\pi/3$), respectively.

%


We will use inversion of the dual-component atmosphere, introduced by~\citep{fernald1984,klett1985,sasano1985,kovalevmoosmueller1994}, in its non-logarithmic form, as pointed out by~\citep{young1995}. In this formulation, the range-corrected LIDAR return is multiplied with a new function $Y(r)$
\begin{equation}
  Y (r) = S\p(r) \cdot \exp\left( -2\int_0^r \!\!\big( F(r') - 1 \big) \alpha\m(r') \ud r' \right) ~, 
\end{equation}
where $F(r) = S\p(r)/S\m$. The LIDAR equation transforms then into
\begin{eqnarray}
  Z(r) &=& \ln \left( N(r)\cdot r^2 \cdot Y(r) \right) \nonumber\\
  \ddfrac{\ud Z(r)}{\ud r} &=& \ddfrac{1}{y(r)}\cdot \ddfrac{\ud y(r)}{\ud r} - 2 y(r) \quad,
  \label{eq:sasano}
\end{eqnarray}
where the new variable $y(r)$ is defined as
\begin{equation}
  y (r) = \alpha\p(r) + F(r) \cdot \alpha\m(r) \quad.
\end{equation}
Eq.~\ref{eq:sasano} can be solved following the prescription of~\citep{klett1981}:
\begin{equation}
  \alpha\p(r) = -F(r)\alpha\m(r) + \ddfrac{e^{(Z(r)-Z(r\f))}}{y\f^{-1} + 2\int_r^{r\f} e^{(Z(r')-Z(r\f))} \ud r'}
\end{equation}

As we will see later, the LIDAR return can be calibrated almost always at reference points $r\f$ of practically no aerosols and pure Rayleigh scattering, in our case at few kilometers above ground, as will be shown later. At these points,
\begin{equation}
  y\f \simeq F(r\f) \cdot \alpha\m(r\f)\quad. 
\end{equation}

If the LIDAR operates at a zenith angle $\theta$, it is sometimes useful to express Eq.~\ref{eq:sasano} as a function of height $h = r / \xi$ with $\xi = 1/\cos\theta$ and $\ud h / \ud r = 1/\xi$:
\begin{equation}
  \ddfrac{\ud Z(h)}{\ud h} = \ddfrac{1}{y(h)}\cdot \ddfrac{\ud y(h)}{\ud h} - 2 \xi \cdot y(h) \quad,
\end{equation}
with the solution
\begin{equation}
  \alpha\p(h) = -F(h)\alpha\m(h) + \ddfrac{e^{(Z(h)-Z(h\f))}}{y\f^{-1} + 2\xi \int_h^{h\f} e^{(Z(h')-Z(h\f))} \ud h'} \label{eq:sasano2}
\end{equation}

%% file: molecular.tex
\section{Subtraction of the molecular profile}

Because of the extremely clean atmosphere found above the nocturnal boundary layer at the MAGIC site, we can calibrate the LIDAR return over large altitude ranges, where practically the entire signal is due to pure Rayleigh scattering on the air molecules.

The volume Rayleigh back-scatter cross section for unpolarized or circular polarized light is very well understood~\citep{bucholtz,mccartney} and can be parameterized in the following form:
\begin{align}
\beta\m(\lambda,h) 
           &= \ddfrac{6 \pi^2 \cdot (n_s^2-1)^2}{N_s \cdot \lambda^4 \cdot (n_s^2+2)^2}
           \cdot \left( \ddfrac{6+3\rho(h)}{6-7\rho(h)} \right) \cdot \ddfrac{P(h)}{P_s}\cdot \ddfrac{T_s}{T(h)} \nonumber\\
  &\cdot \ddfrac{3}{4}\cdot\left(\ddfrac{2+2\rho(h)}{2+\rho(h)}\right) \cdot \left( 1 + \ddfrac{1-\rho(h)}{1+\rho(h)} \right) 
           ~\mathrm{m^{-1}\,Sr^{-1}} \qquad , \label{eq:Rayleigh} 
\end{align}
where the first line shows the original cross-section formula, including the \textit{King correction}, due to the depolarization $\rho$ of the air molecules, and the correction for the different air densities at height $h$, measured through temperature $T(h)$ and pressure $P(h)$. The second line is the \textit{Chandrasekhar corrected} phase function~\citep{Chandrasekhar,mccartney}. Further, $n$ is the refraction index of air and $N_s$ the number density of molecules per unit volume, at standard conditions ($N_s =2.5469\cdot 10^{25}~\mathrm{m}^{-3}$~\citep{bodhaine} at $T_s = 288.15$\,K and $P_s = 101.325$\,kPa). According to~\citet{tomasi}, both $n$ and $\rho$ depend slightly on the wavelength of light, atmospheric pressure, temperature, relative humidity and the concentration of CO$_2$. Assuming dry air and $C_\mathrm{CO_2} = 400$\,ppmv (corresponding roughly to the year 2015), 
standard values $P_s$ and $T_s$, we obtain for the Nd:YAG laser at 532\,nm: $(n_s-1) = 2.779\cdot 10^{-4}$. The depolarization coefficient $\rho = 0.0283$ can be used for typical atmospheric conditions in the lower troposphere, and varies by less than 0.5\%. 

Using these numbers and the relation $9\cdot(n^2-1)^2/(n+2)^2 \approx 4\cdot(n-1)^2$ (more precise than 0.01\% for all our cases), the volume scattering cross section can be written as \citep[see also][]{GaugDoro:2018}:
\begin{align}
\beta\m (\lambda = 532 \mathrm{nm},h) & := \beta_0 \cdot \ddfrac{P(h)}{P_s}\cdot \ddfrac{T_s}{T(h)} \approx 1.545 \times 10^{-6} \cdot \ddfrac{P(h)}{P_s}\cdot \ddfrac{T_s}{T(h)}~\mathrm{m^{-1}\,Sr^{-1}} \quad , \label{eq:Rayleigh1} \\
\avg{\beta(h_\mathrm{LIDAR})}  & :=  \beta_0 \cdot \ddfrac{\avg{P(h_\mathrm{LIDAR})}}{P_s}\cdot \ddfrac{T_s}{\avg{T(h_\mathrm{LIDAR})}} \approx  1.228 \times 10^{-6}~\mathrm{m^{-1}\,Sr^{-1}} \quad , \label{eq:Rayleighavg}
\end{align}
where the standard Rayleigh backscatter coefficient at Sea level, $\beta_0$, and the average backscatter coefficient at the altitude of the LIDAR, $\avg{\beta(h_\mathrm{LIDAR})}$,  have been introduced, assuming an average atmospheric pressure of 788.2\,mbar and an average temperature of 8.89\degC. Eq.~\ref{eq:Rayleigh1} is precise to at least 0.5\%, with the main uncertainty stemming from the unknown water vapor content~\citep{tomasi}.

\begin{figure}
  \begin{center}
    \includegraphics[width=0.485\columnwidth]{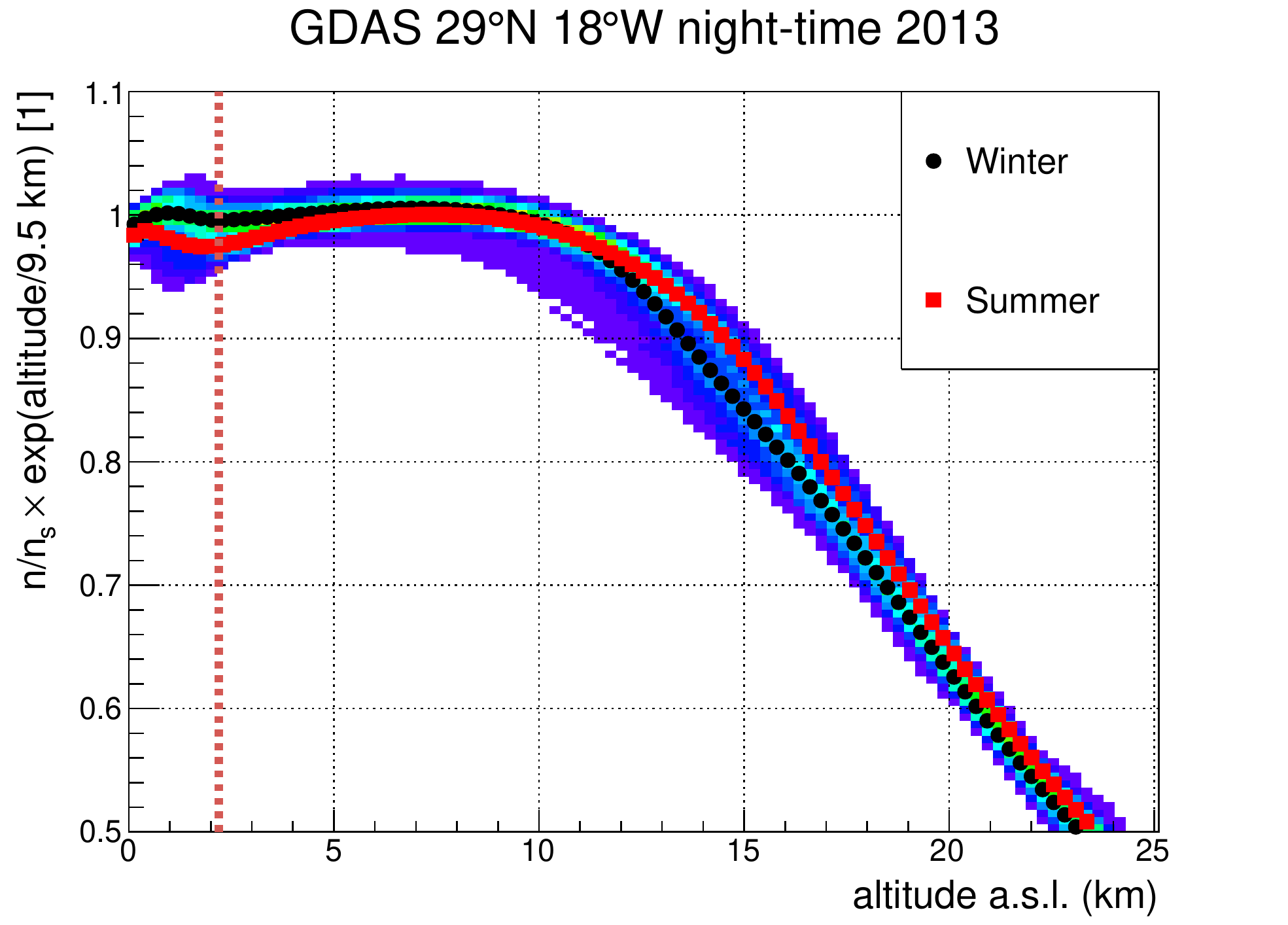}
    \includegraphics[width=0.485\columnwidth]{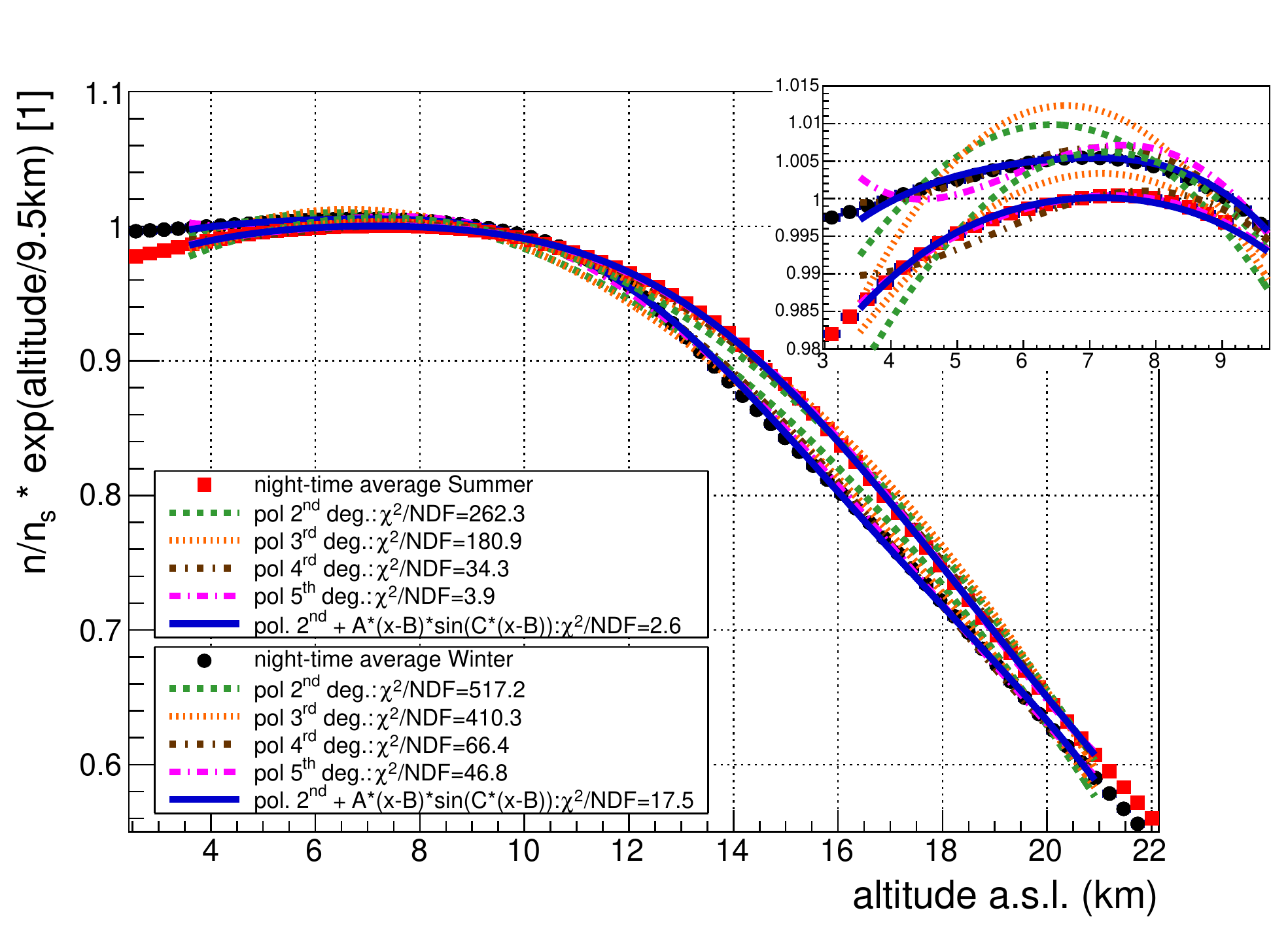}
  \end{center}
  \caption[GDAS height profiles for MAGIC]{Left: altitude profiles for the closest GDAS grid point to the MAGIC telescope during all nights of 2013. The discrete GDAS data points had been interpolated by a spline before. The winter and summer averages are shown as black circles and red squares, respectively. For better visibility of the modulation along the exponential decay, all data points have been multiplied by an exponential with an arbitrary scale height of 9.5\,km. The vertical dotted line indicates the altitude of the MAGIC site. Right: the winter and summer averages have been fitted to a series of polynomials and the parabola plus sinusoidal (Eq.~\protect\ref{eq:parsin}). The inlet shows a close-up to the first 9 kilometers, where an approximately linear dependency is expected.
  \label{fig:lnn}
  }
\end{figure}

\begin{figure}
  \begin{center}
    \includegraphics[width=0.5\columnwidth]{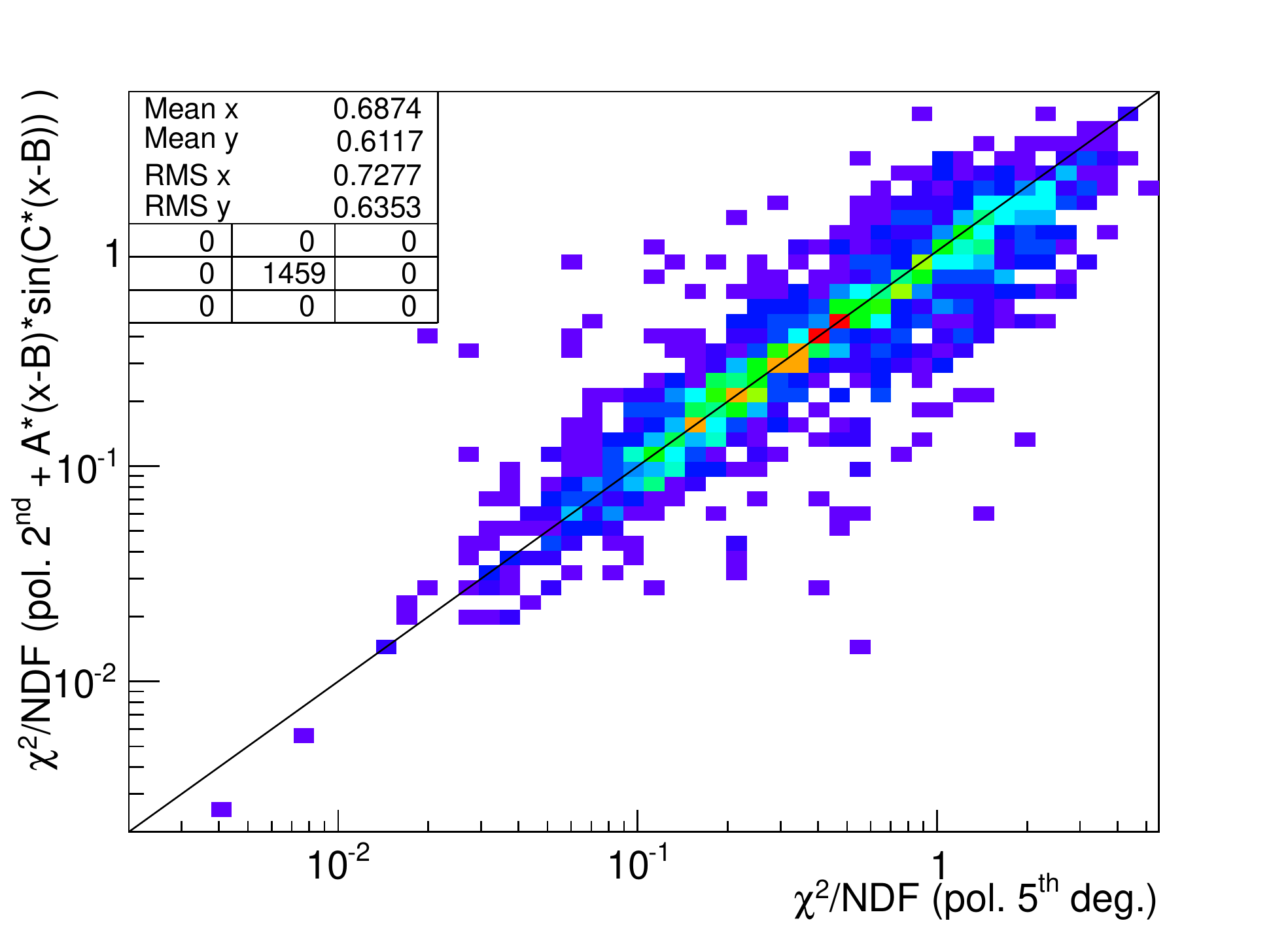}
  \end{center}
  \caption[Chi-squares from fits to GDAS height profiles]{Chi-square values from the fits to all night-time GDAS data of 2013. The horizontal axis shows the result from a polynomial of 5$^\mathrm{th}$ degree, while the vertical axis from the fit Eq.~\protect\ref{eq:parsin}. The statistics box shows that both the mean and the RMS are slightly \collaborationreview{better} for the fit Eq.~\protect\ref{eq:parsin}, and no values outside the displayed range have been obtained. \collaborationreview{The lower part statistics matrix shows the number of entries displayed (center) and the over- and underflow statistics (non-central values).}
  \label{fig:chi2gdas}}
\end{figure}

We assessed the molecular profiles above the MAGIC site through the Global Data Assimilation System (GDAS), which provides an atmospheric analysis four times per day, a 3, 6 and 9~hour forecast as well as an archive dating back to 2005~\citep{gdas}~\citep[see also][]{PAO2012-2}. The data are free and available online\,\footnote{\url{ftp://arlftp.arlhq.noaa.gov/pub/archives/gdas1/}}. For our needs, we extract temperature, geopotential height and relative humidity on 23 fixed pressure levels, ranging from 1000 to 20\,hPa. The geopotential heights have been converted to altitudes above ground using the prescription of~\citep{list}. The GDAS data are provided for a grid of integer latitude and longitude (in degrees), hence the closest grid point for the MAGIC telescope corresponds to 29$^\circ$\,N and 18$^\circ$\,W. The GDAS predictions have been ground-validated with the MAGIC weather station resulting in excellent agreement of the predicted pressure values and a small positive bias of two degrees for the temperatures~\citep{Gaug:2017site}, attributed to local ground cooling effects at night, which is not reproduced at the GDAS grid point, located over the Sea. A more accurate local modelling of the molecular profile involving the Weather Research and Forecasting (WRF)\footnote{\url{https://www.mmm.ucar.edu/weather-research-and-forecasting-model}} is currently investigated, but out of the scope of this article.  

The term $P(h)/P_s \cdot T_s/T(h) = n(h)/n_s$ can be approximated by an exponential decrease from ground up to the about 6\,km a.s.l., with a scale height of approximately $H_s \approx 9.5$\,km, except for the first two kilometers (below the MAGIC site), where the effect of the quasi-permanent temperature inversion, very characteristic for that geographic area, is present. In and above the tropopause, a different and more complex behaviour is observed. Fig.~\ref{fig:lnn} (top) shows
how $n(h)/n_s$ deviates from the pure exponential decrease during night-time in 2013, while Fig.~\ref{fig:lnn} (bottom) shows several attempts to fit it over a range from 3.5\,km to 21\,km~a.s.l. After several trials, we found that the following function can be used best to describe the behaviour of $n(h)/n_s$:
\begin{figure}
  \begin{center}
    \includegraphics[width=0.5\columnwidth]{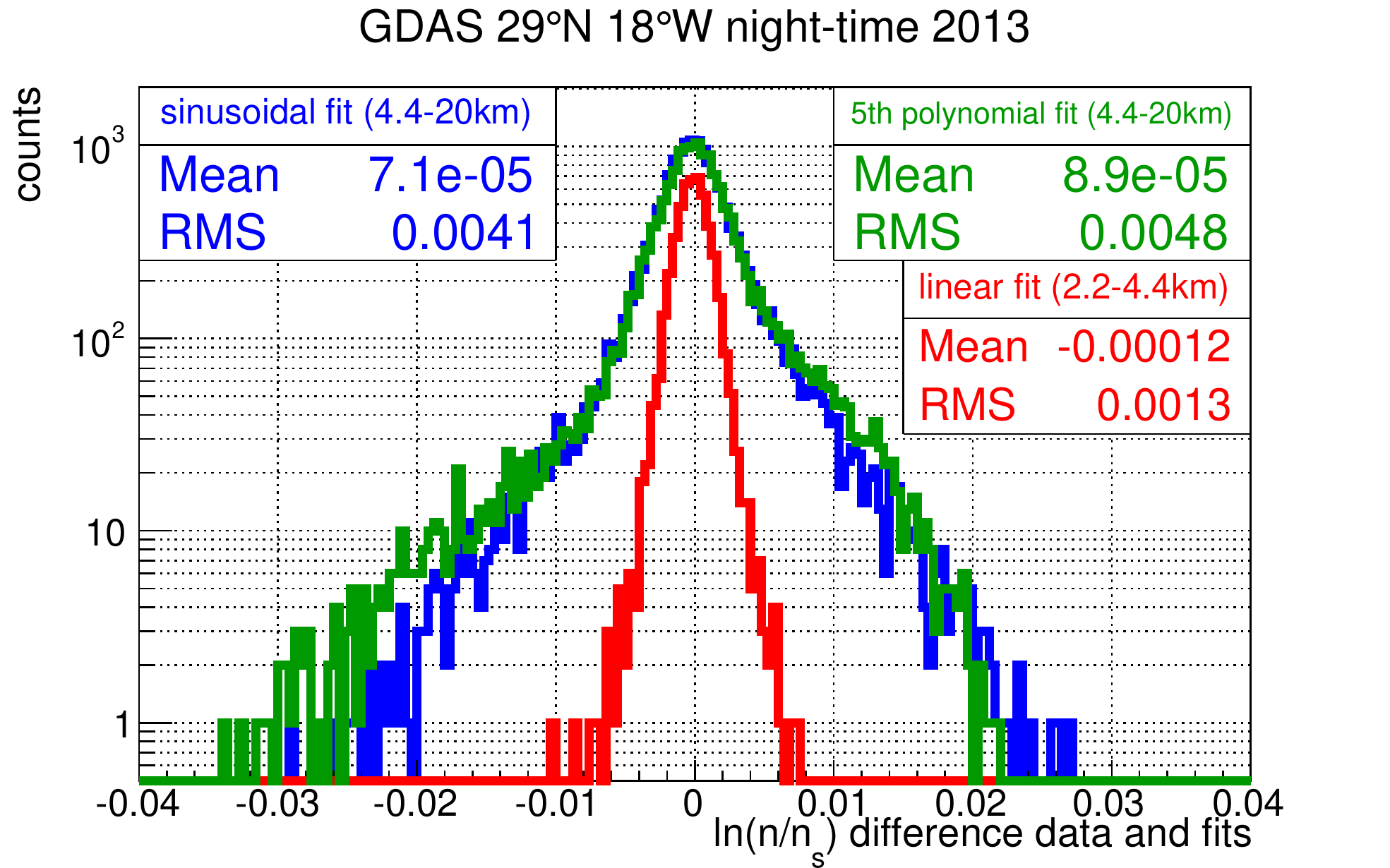}
  \end{center}
  \caption[Absolute differences between fits and GDAS data points]{Absolute differences between fits and GDAS data points. In red for a linear fit from 2.2 to 4.4\,km a.s.l., in green for polynomial of 5$^\mathrm{th}$ degree, fitted from 4.4 to 20\,km and in blue for the sinusoidal fit Eq.~\protect\ref{eq:parsin} in the same range. The statistics boxes shows that both the mean and the RMS are slightly superior for the fit Eq.~\protect\ref{eq:parsin}, no deviations larger than 3\% are observed.
  \label{fig:deviations}}
\end{figure}\noindent

\begin{eqnarray}
n(h)/n_s &=& e^{f(h)} \quad \textrm{with:} \nonumber\\
f(h) &=& A \cdot(h-B) \cdot \sin(C\cdot(h-B)) + \nonumber\\
         && {}~{}~{}~ + D + E\cdot h + F \cdot h^2  \quad,
\label{eq:parsin}
\end{eqnarray}\noindent
where the constants $A-F$ are free parameters. This fit gives superior results to polynomials up to 5$^\mathrm{th}$ degree, especially in the low-altitude range, where the linear behavior is always well reproduced. Especially in case of the winter and summer averages, the residual error of the fit \collaborationreview{always lies} well below 1\%. We tested the quality of the fit, Eq.~\ref{eq:parsin}, on all GDAS data points of one full year, and found a distribution of reduced chi-squares, shown in Fig.~\ref{fig:chi2gdas}. \collaborationreview{Low chi-square values} 
are only obtained if all fits are properly initialized, i.e. the start values of polynomials of a certain degree need to be initialized to the fit results from a polynomial of degree one order less, with the coefficient of the new order set to zero. In the case of Eq.~\ref{eq:parsin}, good start values are shown in Table~\ref{tab:parsin}. As we have not found any previous attempts to fit molecular density height profiles with the proposed equation in the literature, we show also our experience of useful parameter limits in Table~\ref{tab:parsin}. Apart from assuring that the fit converges at the correct minimum, the limits on the parameters $B$ and $C$ guarantee that the sinus in Eq.~\ref{eq:parsin} will not move to a parameter range, where the measurement points get interpolated by fast oscillations.

\begin{table}
\centering
\caption{Initialization values and limits for the parameters of fit Eq.~\protect\ref{eq:parsin}.
\label{tab:parsin}}
\begin{tabular}{c|c|c}
Parameter & Start value &  Limits \\
\hline
A   &  -0.003 & (-0.025,0.025) \\
B   &  1.5 &  (-30,15)  \\
C   &  0.4 &  (0.12, 0.72) \\
D   &  from pol. 2$^\mathrm{nd}$ &   none  \\
E   &  from pol. 2$^\mathrm{nd}$ &   none  \\
F   &  from pol. 2$^\mathrm{nd}$ &  (-0.0045,0.002) \\
\end{tabular}
\end{table}

For the case of pure Rayleigh scattering at atmospheric heights, which are practically free of aerosols, we can re-write the LIDAR equation (Eq.~\ref{eq:lidar}), using the previous parametrization:
\begin{align}
\ln\left(N\!(r) r^2\right)_\mathrm{mol.~part} = & {} \quad \ln\left(N_0\, A\, l\, \avg{\beta(h_\mathrm{LIDAR}}\right) + f\left(h\right) 
 - \ln\left(\ddfrac{\avg{P(h_\mathrm{LIDAR})}}{\avg{T(h_\mathrm{LIDAR})}}\cdot \ddfrac{T_s}{P_s}\right)  - 2 \int_0^{r}\! \left( S\m \, \beta_0 \, e^{f(r'\cos\theta)}\! +\! \alpha\p(r') \right) \, \ud r'   \\[0.15cm]
         {} = & {}~  C_0 -  \ddfrac{2}{\cos\theta} \textit{VAOD} + F(h) \label{eq:molecular_return}\\[0.2cm]
                   & \textrm{with:} \nonumber\\[0.2cm]
            h = & {}~ r \cdot \cos\theta + h_\textrm{LIDAR} \\[0.2cm]
          F(h)  = & {}~  f(h) - \ddfrac{2}{\cos\theta}\, S\m \, \beta_0 \! \int_{h_\textrm{LIDAR}}^{h}\!\!\!\! e^{f(h')} \,\ud h' - \ln\left(\ddfrac{\avg{P(h_\mathrm{LIDAR})}}{\avg{T(h_\mathrm{LIDAR})}}\cdot \ddfrac{T_s}{P_s}\right) \label{eq:lnnr}  \\[0.2cm]
          \textit{VAOD} =  &~ {} \int_{h_\textrm{LIDAR}}^{h_t}\!\!\!\! \alpha\p(h')\, \ud h'~   \label{eq:ht}
\end{align}\noindent
where the pointing zenith angle ($\theta$) of the LIDAR has been introduced, the altitude of the MAGIC LIDAR ($h_\textrm{LIDAR} = 2188.2$\,m~a.s.l.), and a transition height $h_t$, which is used to distinguish the altitudes possibly affected by the planetary boundary layer and the free troposphere above. Experience has shown that $h_t$ can vary mainly from below 1\,km up to about 4.5\,km~a.s.l.\ at La~Palma~\citep{lombardi2008}, and the free troposphere can be considered free of aerosols most of the time, except for the possible occurrence of clouds, and stratospheric volcanic debris~\citep{garciagil2010}. The Vertical Aerosol Optical Depth (VAOD) up to $h_t$ can be retrieved entirely from the LIDAR data, only if an absolute calibration of the system constant~\citep{Gao:2015}:
\begin{align}
    C_0 = \ln(N_0 A\, l\, \avg{\beta(h_\mathrm{LIDAR})}) 
\end{align} \noindent
is available. Above that altitude $h_t$, the function $F(h)$ describes the logarithm of the range-corrected signal correctly in the absence of clouds and can be used to fit $\left(2/\cos\theta\ln(T\p) + F(h)\right)$. The latter will be hitherto called the ``Rayleigh fit''. Below the region of the Rayleigh fit, Eq.~\ref{eq:sasano2} can be used to derive the aerosol extinction profile, apart from the first hundreds of meters, where the back-scatter signal $N(r)$ is affected by the region where full overlap of the fields-of-view is not yet reached, or the signal saturates the FADC readout.

\begin{figure}
  \centering
  \includegraphics[width=0.49\columnwidth,trim={0.9cm 0 1.8cm 0},clip]{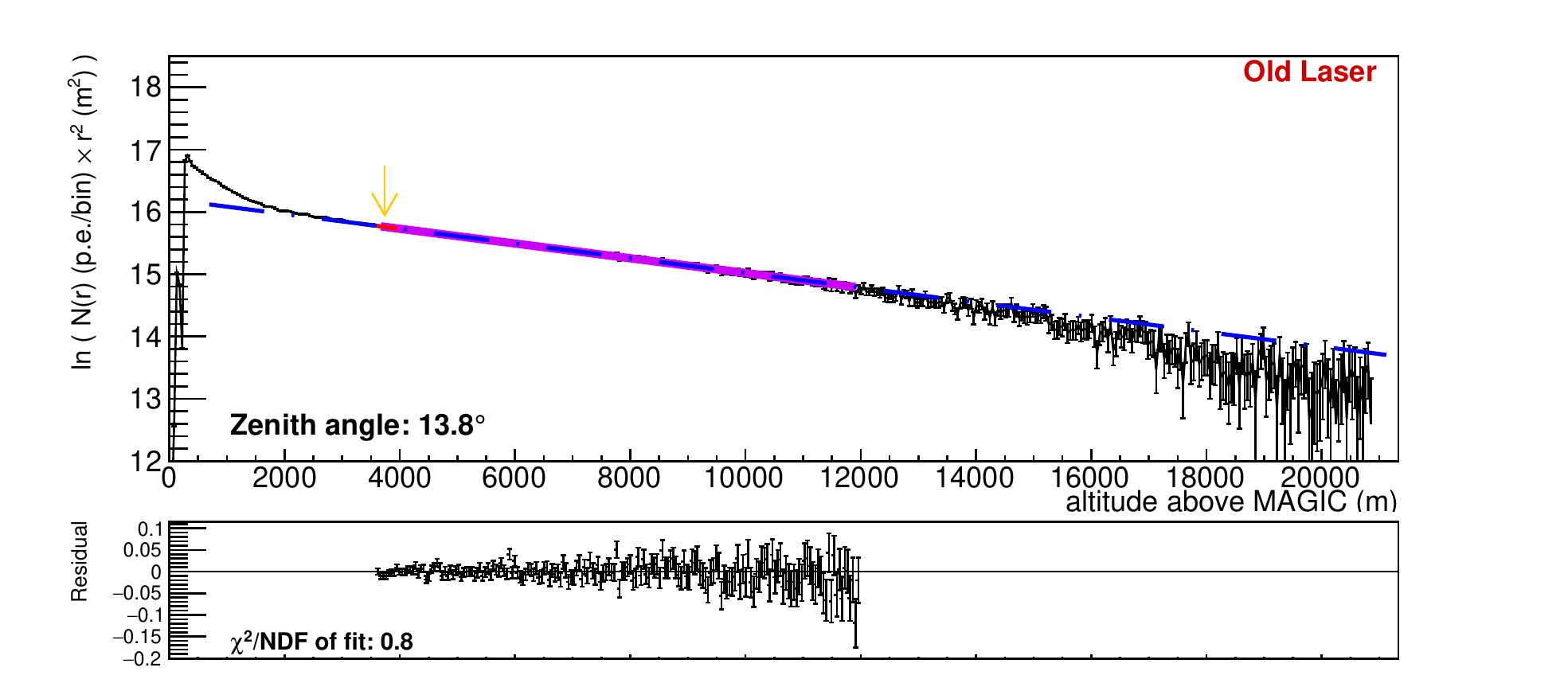}
   \includegraphics[width=0.49\columnwidth,trim={0.9cm 0 1.7cm 0},clip]{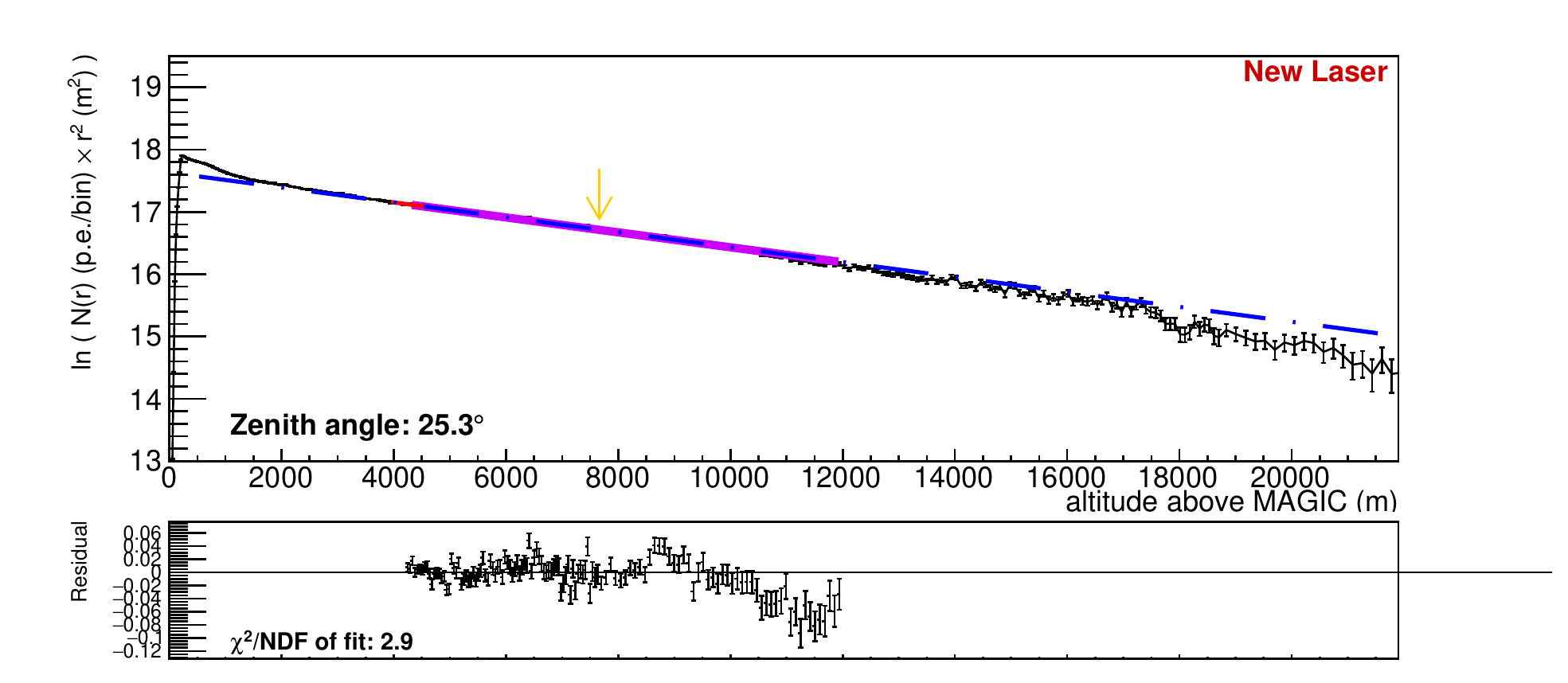}
 \includegraphics[width=0.49\columnwidth,trim={0.5cm 0 1.8cm 0},clip]{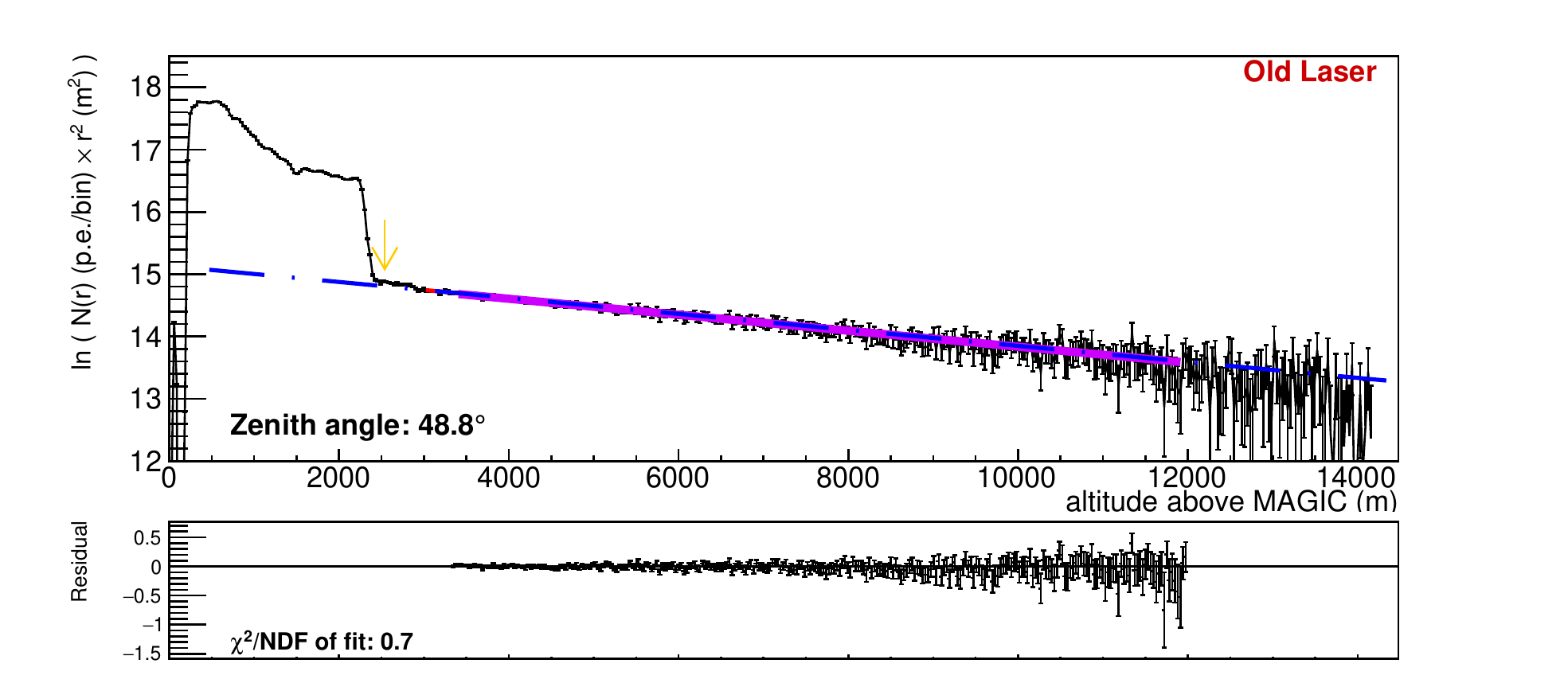}
  \includegraphics[width=0.49\columnwidth,trim={0.9cm 0 1.7cm 0},clip]{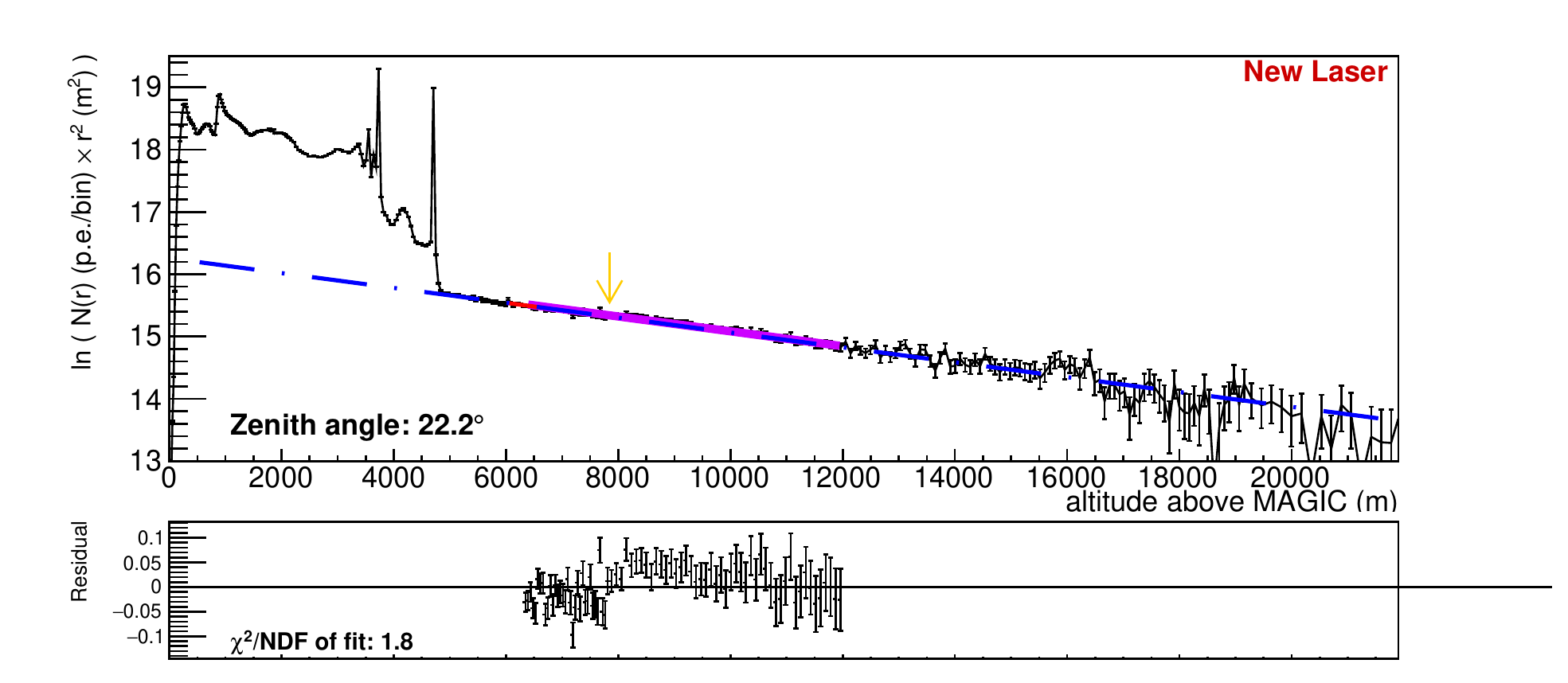}
  \includegraphics[width=0.49\columnwidth,trim={0.9cm 0 1.8cm 0},clip]{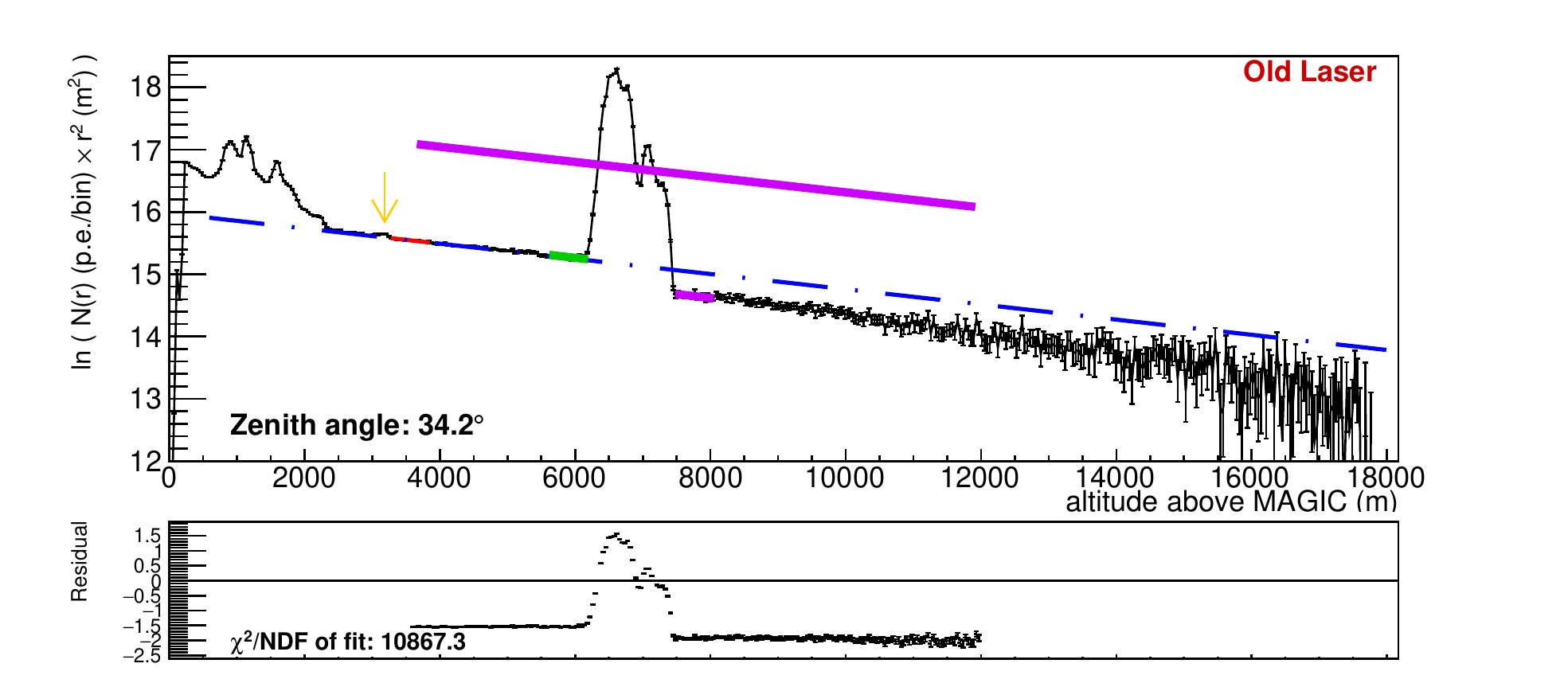}
  \includegraphics[width=0.49\columnwidth,trim={0.9cm 0 1.7cm 0},clip]{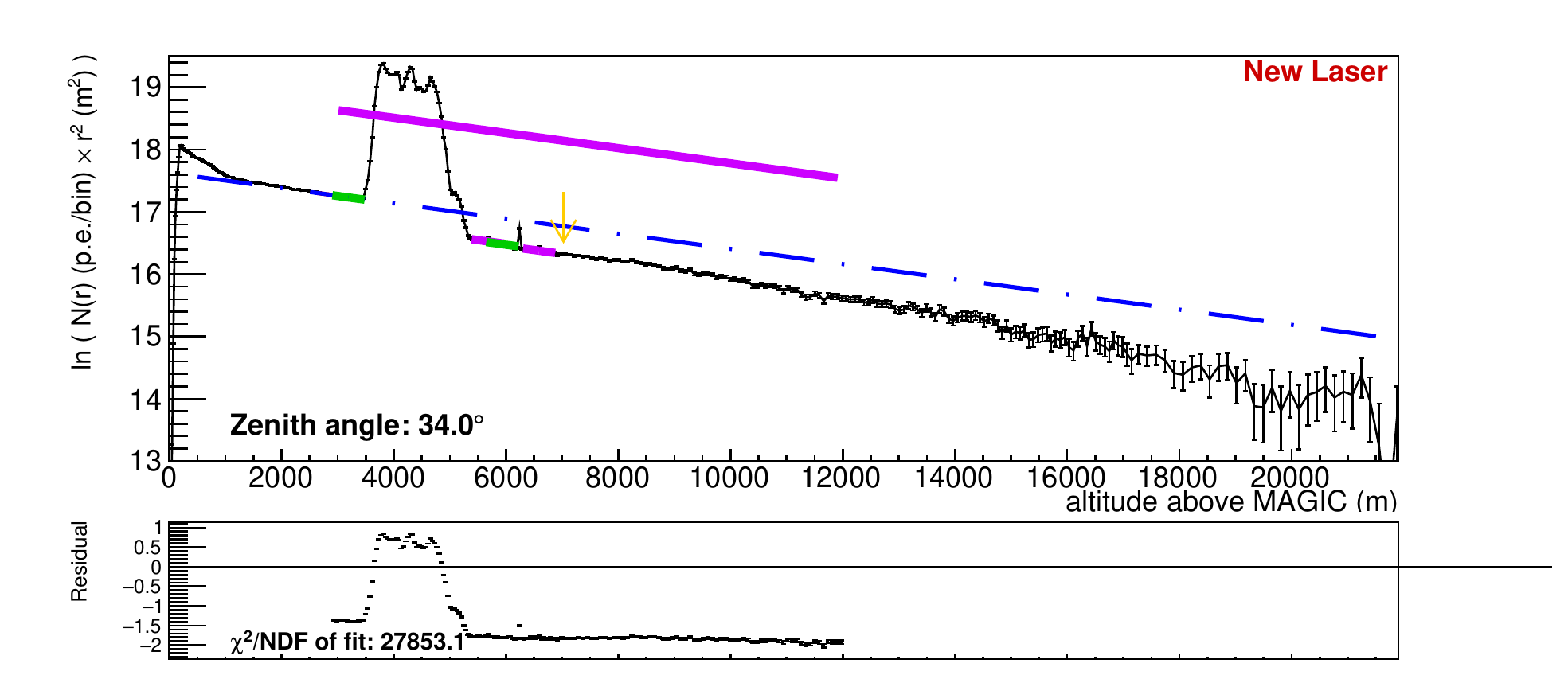}
  \caption[Example of Rayleigh fit]{Examples for "Rayleigh fits" during clear nights (top), nights with calima (center), and nights with clouds (bottom). On the left side, representative data sets from 2014, taken with the old laser (Period~1), are shown, on the right side from 2019 with the new laser (Period~2). The violet line shows the region fitted to function $F(h)$ (Eq.~\protect\ref{eq:molecular_return}) from the end of the boundary layer up to 12\,km above the MAGIC site, the blue dashed lines show the extension of $F(h)$ to cover the full signal range. 
  In the case of clouds, green lines lines show the fits to $F(h)$ right below and above the detected cloud. The small yellow arrow indicates the transition from the amplitude to the photo-electron counting regime. Below each signal plot, the residuals between data and $F(h)$ are displayed \collaborationreview{from the start of the free troposphere up to 12\,km a.g.l.} and the corresponding reduced $\chi^{2}$. The latter is used to determine whether a cloud search is initialized.
\collaborationreview{Only statistical uncertainties of the LIDAR data are shown in the residuals and have been used to calculated the reduced $\chi^{2}$s. Note that with the new system the number of laser shots used has been reduced by one half.}
  \label{fig:rayleighexample}}
\end{figure}

\begin{figure}
  \centering
  \includegraphics[width=0.69\columnwidth]{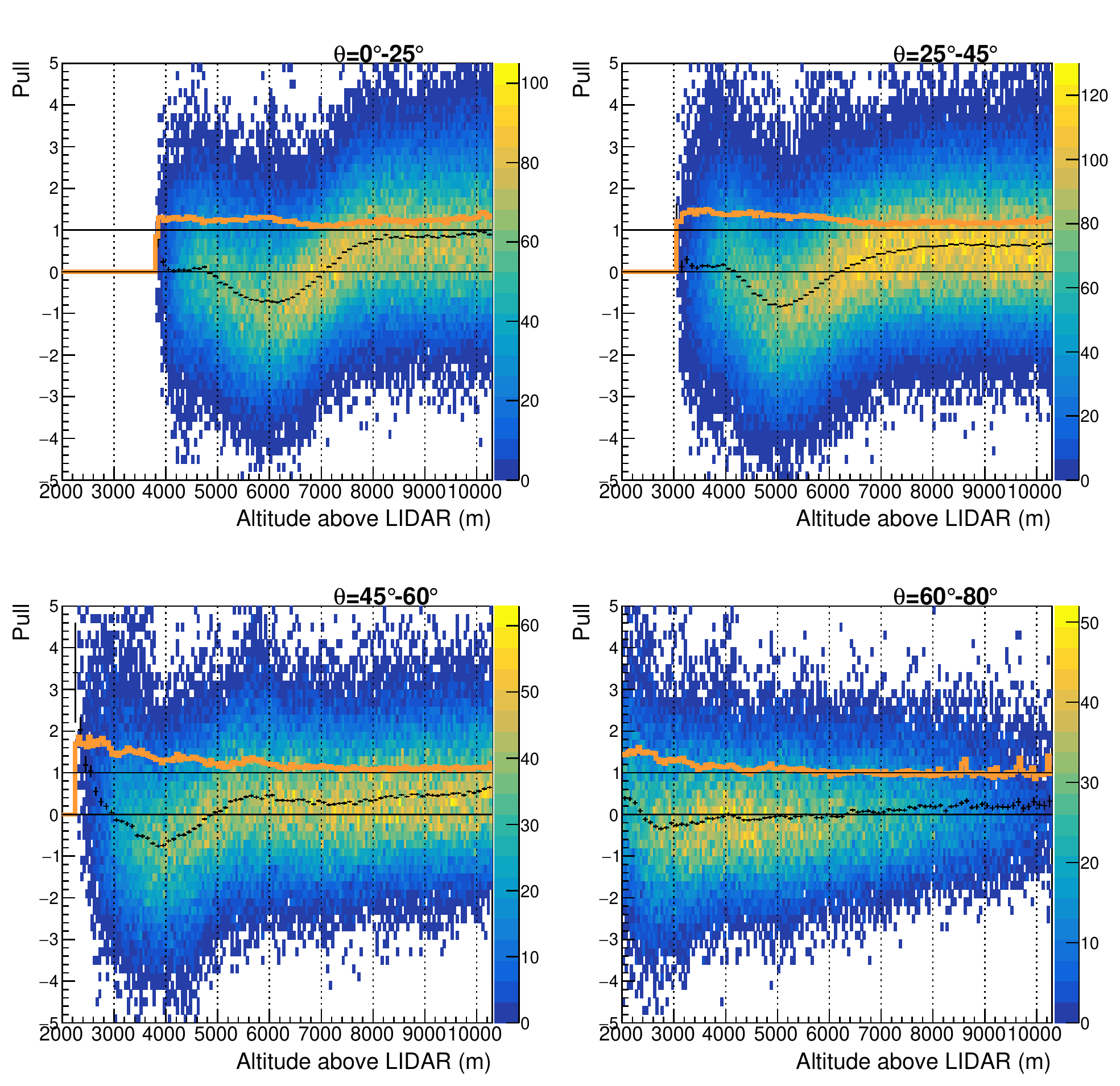}
  \caption[Distribution of pull-plots old laser]{Profiles of ``pull'' distributions $\left( \ln(N(r)r^2) - \ln(N(r)r^2)_\mathrm{mol.~part}  \right)/\Delta \ln(N(r)r^2)$ from Period~1  (old laser) photon counting data, for different zenith angle bins. The black points show the median and the orange histogram the standard deviation of the distribution for each altitude bin. Poissonian statistics has been assumed for $N(r)$.
  \label{fig:pull2_old}}
\end{figure}

\begin{figure}
  \centering
  \includegraphics[width=0.49\columnwidth]{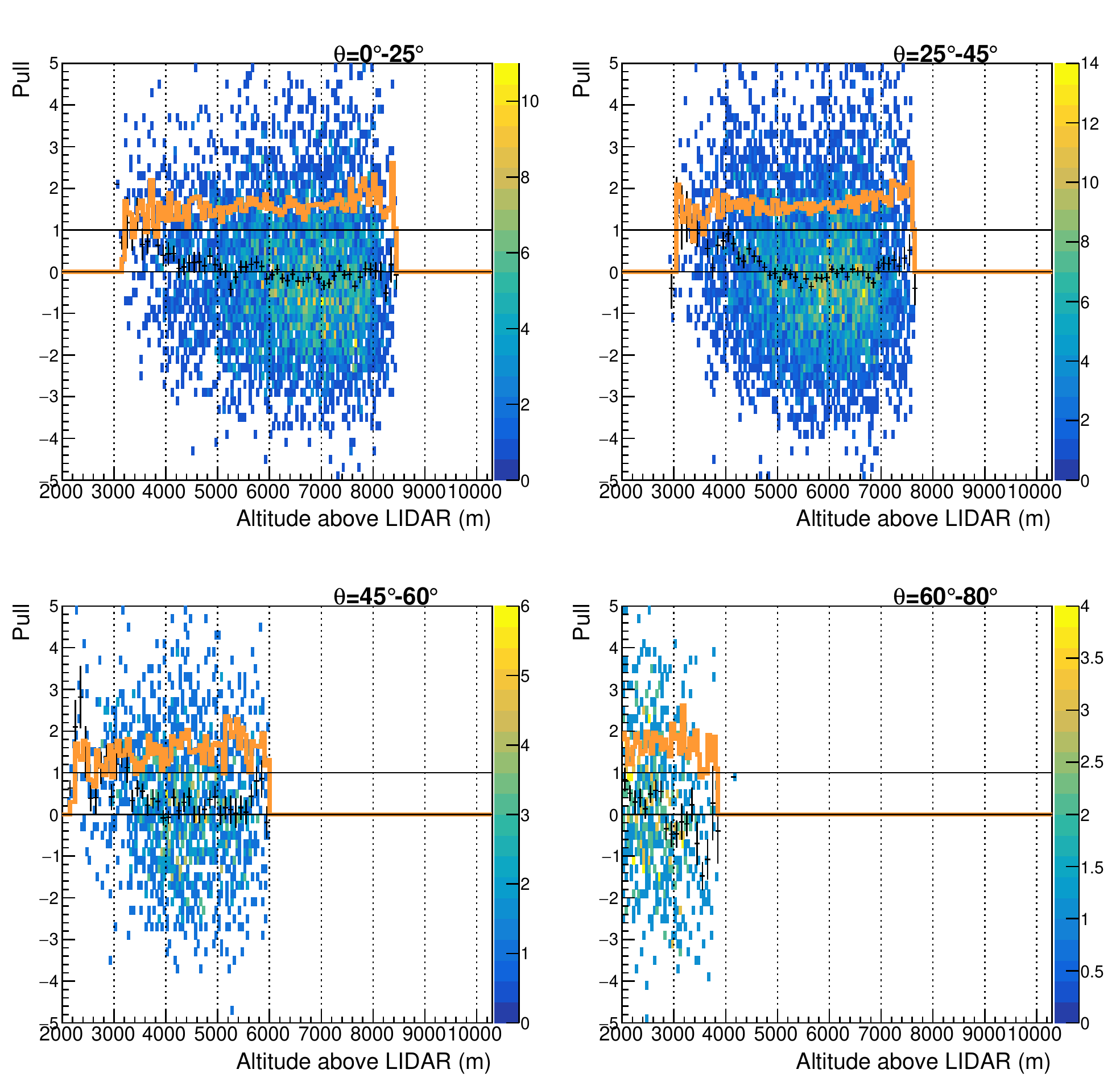}
  \includegraphics[width=0.49\columnwidth]{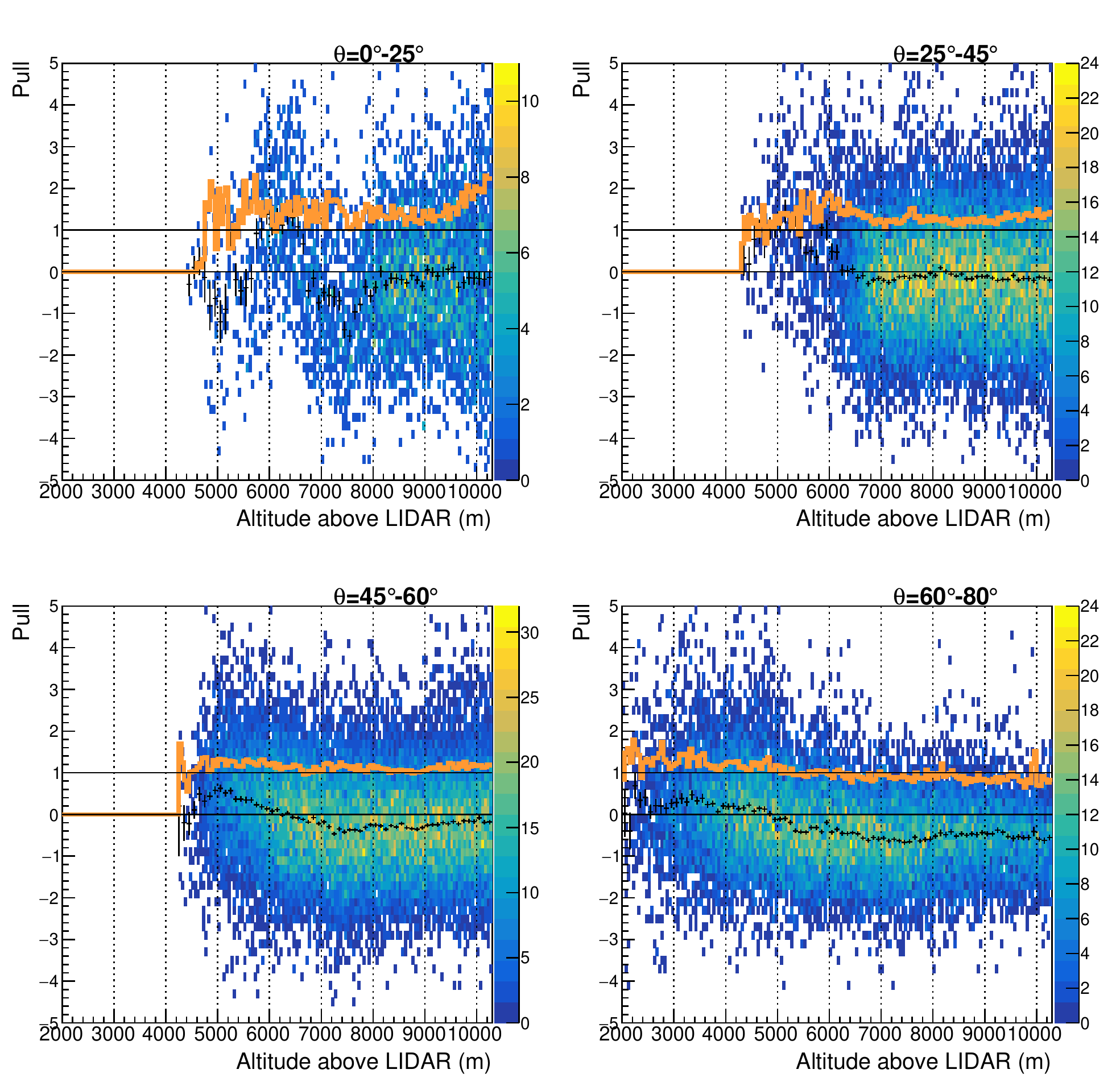}
  \caption[Distribution of pull-plots old laser]{Profiles of pull distributions $\left( \ln(N(r)r^2) -  \ln(N(r)r^2)_\mathrm{mol.~part}  \right)/\Delta \ln(N(r)r^2)$ from Period~2 (new laser) analogue (left four figures) and photon counting (right four figures) data, for different zenith angle bins. The black points show the median and the orange histogram the standard deviation of the distribution for each altitude bin. Poissonian statistics has been assumed for $N(r)$, except for the analog signal, where the signal variance has been multiplied by an additional excess noise factor of 1.2.
  \label{fig:pull_new}}
\end{figure}

Fig.~\ref{fig:rayleighexample} shows three typical examples of situations where the Rayleigh fit is successful, applied to different regions of the atmospheric profile, according to the atmospheric conditions. In order to check how well the fit function, Eq.~\ref{eq:molecular_return}, describes the LIDAR return for clear nights in the absence of clouds, we have studied "pull" distributions\footnote{Pull plots are normally defined as the difference between the estimated and the true value of a parameter (e.g., known in a test using simulations), divided by the estimated standard deviation.}, in this case defined as the residuals between the LIDAR return $\ln \left(N(r)\cdot r^2\right)$ and the molecular profile predictions  (Eq.~\ref{eq:molecular_return}), which themselves had been fitted with respect to $C_0$ across the full troposphere above the nocturnal boundary layer. The residuals were then divided by the estimated standard deviation of the signal  and filled into histograms binned in altitude. Fig.~\ref{fig:pull2_old} shows the distributions of these normalized residuals for the photon counting data of Period~1 (old laser, the analog signal does not reach the end of the boundary layer here), Fig.~\ref{fig:pull_new} shows both analogue as photon counting residuals for Period~2 (new laser).  One can observe that the old system shows a systematic bias towards negative values at distances between 5\,km and 7\,km from the LIDAR, and a positive bias above. Since the magnitude and location of these biases seem to occur always at a same distance to the LIDAR, independent of the observation zenith-angle,  we conclude that this must have been due to a residual under-shoot and subsequent over-shoot of the signal, produced by differentiation of the high-frequency part of the signal due to non-optimal capacitive coupling of the sensor with the amplification chain. Subtraction of the expected HPD after-pulse contribution to the signal \citep[see][]{Saito:PhD} did not modify significantly the biases observed in Figs.~\ref{fig:pull2_old} and~\ref{fig:pull_new}, and are hence not responsible for these. Note, however, that the magnitude of the biases is always smaller than the standard deviations of the distributions, which are hence still dominated by statistical fluctuations. This is important later on, when statistical criteria for cloud detection will be formulated. Residual mismatches of the true atmosphere with respect to  the molecular model, Eq.~\ref{eq:molecular_return}, are nevertheless observed, when the standard deviations of the pull distributions become larger than one. This is the case for altitudes up to about 6\,km for small zenith angles, and up to 3--4\,km for large zenith angle observations. We interpret these (small) excesses of fluctuations as residual aerosols, the contribution of which to the signal is, however, so small that it can only be detected after combining hundreds of data sets.  Such an excess fluctuation is also found in the new laser data, see Fig.~\ref{fig:pull_new}. Here, the observed biases \collaborationreview{are} considerably smaller than those observed with the old system and are visible only at very large observation zenith angles. We surmise that the curvature of the Earth, which has not been taken into account, might be responsible for such residual biases at low pointing altitudes.  

\noindent

\noindent

\noindent

%% file: calibration.tex
\section{Calibration of the Analogue vs. Photon Counting Signals}

The fact that the ORM often shows ultra-clean environmental conditions, leading to negligible aerosol content above the nocturnal boundary layer, allows us to fit backscatter return signals to selected data and study the behaviour of the analogue vs. photon counting signal calibration. 

The excellent time resolution of the combination of HPD and fast FADC card allow us to resolve and integrate onsite each individual photo-electron pulse and achieve an immediate charge calibration. Furthermore, the start of the photon counting region could then be chosen sufficiently far from the strong low-altitude signal range, such that pulse pileup can be almost completely excluded there. Unfortunately, a full FADC trace occupies about 1.6\,GB of disk space and needs to be deleted after the first onsite signal estimation.   
The reduced data contain only binned photon rates, which were then quality-checked offsite.  

\begin{figure}
  \centering
  \includegraphics[width=0.32\columnwidth,trim={0.2cm 0cm 0.2cm 0cm},clip]{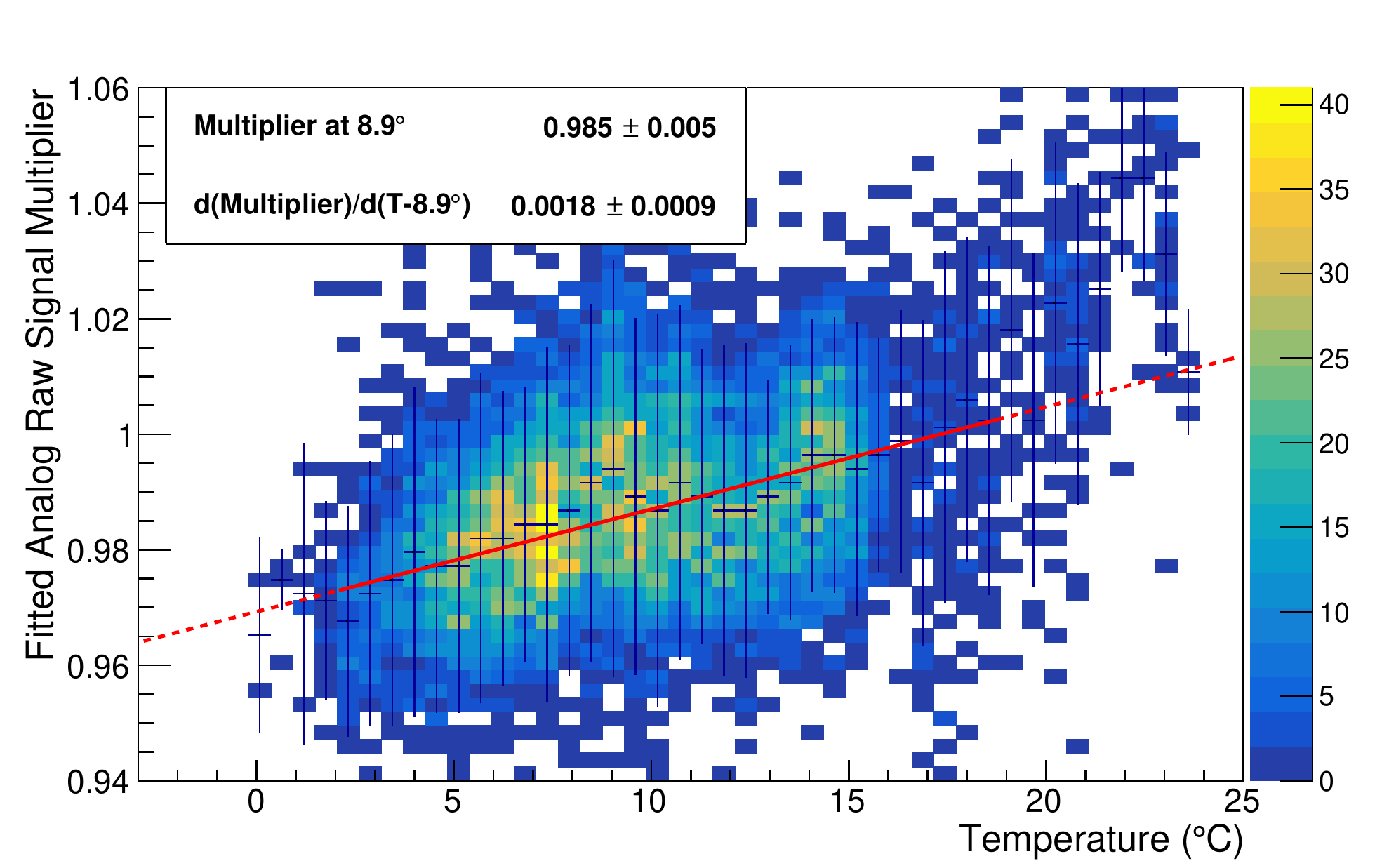}
  \includegraphics[width=0.32\columnwidth,trim={0.2cm 0cm 0.2cm 0cm},clip]{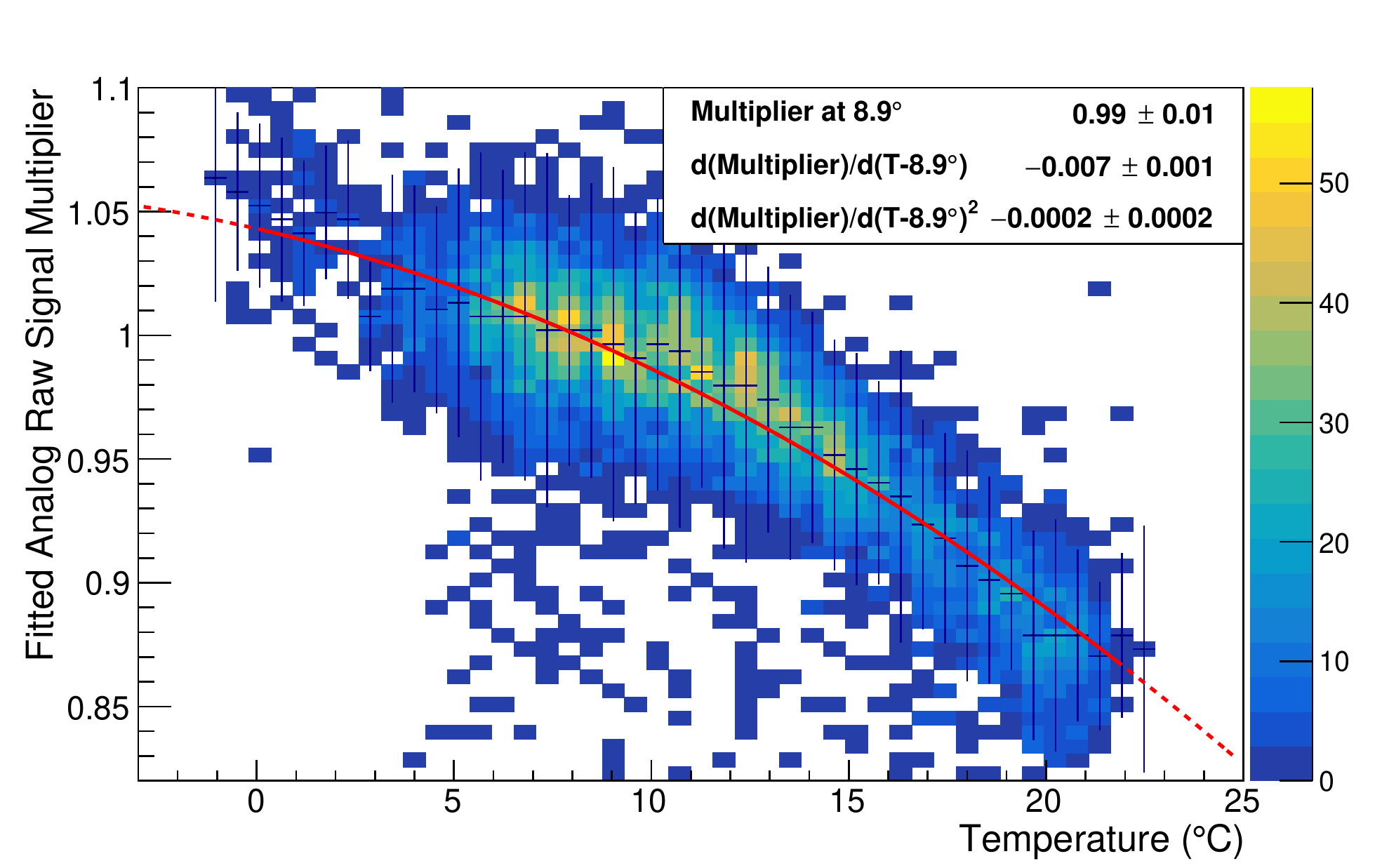}
  \includegraphics[width=0.32\columnwidth,trim={0.2cm 0cm 0.2cm 0cm},clip]{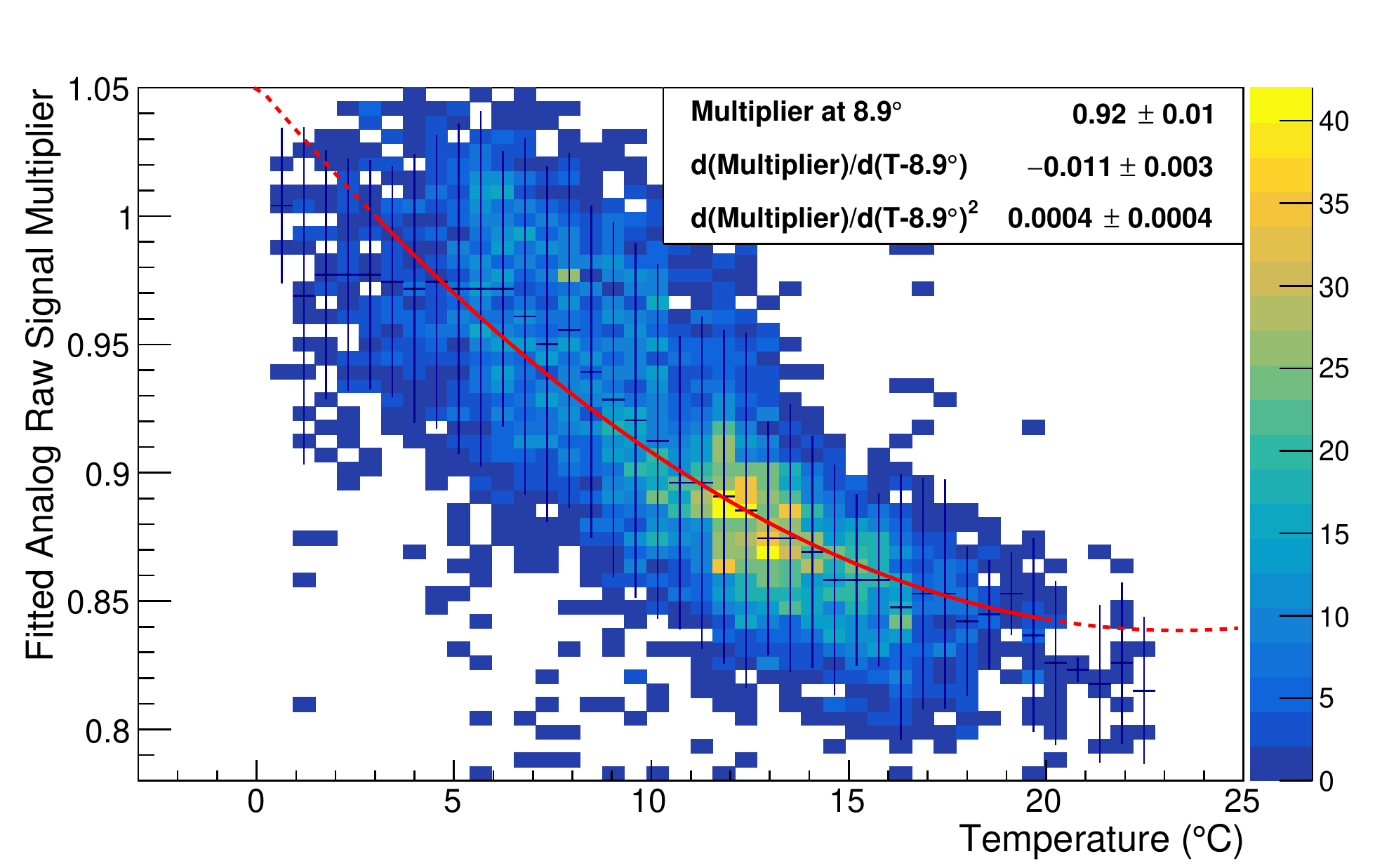}
  \caption[Fitted Analogue Raw Signal Multipliers]{Analogue signal multipliers, obtained from fits to selected pure molecular backscatter profiles, as a function of temperature for three different time periods. Left: Period~1, before installation of new FADC card, center: Period~1 after installation of the new FADC card, right: Period~2. The medians of the distribution of each temperature bin have been fitted to a 1$^\mathrm{st}$  (left) and 2$^\mathrm{nd}$ (center and right) order polynomial.  \label{fig:hwswitch_multipliers}}
\end{figure}

In order to investigate the accuracy of the onsite charge calibration, eight range bins before and eight range bins after the switch from analogue to photon counting have been fitted jointly to the expected pure molecular signal expectation (Eq.~\ref{eq:molecular_return}), and the fit quality assessed in terms of a reduced $\chi^{2}$. In an iterative approach, the analogue part of the raw signal (before background subtraction) was then multiplied with a signal multiplier, until the reduced  $\chi^{2}$ achieved a minimum. Figure~\ref{fig:hwswitch_multipliers} shows the dependency of the obtained raw signal multipliers vs. temperature for three different hardware periods. One can see that the dependency has changed over time, from a positive temperature trend to a negative one, after installation of the new 500\,MS/s FADC card in December 2015. The latter is particularly intriguing after the APD gain stabililization had been introduced in order to obtain stable single photo-electron charges (see Section~\ref{sec:lidar_hardware}). The Period~2 temperature dependencies hint to an \textit{over-correction} of APD gains, possibly  due to residual gain dependencies of the HPD dark count rates, or distortions of the pulse widths as the APD gain gets regulated. 

Further correlations with atmospheric pressure, humidity, photon background level, LIDAR pointing zenith angle, time and aerosol optical depth have been investigated, but none found to be significant. 

The analogue signals were then corrected according to the following criteria: 

\begin{enumerate}
\item if both parts of the signal, left and right from the switch, behave sufficiently molecular-like (i.e. show a low reduced $\chi^2$ of the fit to Eq.~\ref{eq:molecular_return}, the analogue raw signal multiplier providing the lowest reduced $\chi^2$ for the joint fit across the full 16 bins is used. This happens when no calima is present and no cloud is found at or around the switch from analogue signal estimation to photon counting. 
\item Otherwise, the polynomial fits shown in Fig.~\ref{fig:hwswitch_multipliers} are used, together with the measured temperature, to apply an average analogue signal multiplier correction. 
\end{enumerate}

\section{An Absolute Calibration of the LIDAR Return Signal}

The system constant, $C_0$, of a LIDAR includes contributions from the laser power, the geometric area of the mirror and its reflectance, shadows, the HPD photon detection 
efficiency, and the single photo-electron counting efficiency of the raw signal analysis. For the purpose of the system calibration, we do not include here the overlap factor in our definition of the system constant, but assume it to be equal to 1, and treat only the part of the data with full overlap and no saturation of the readout. It can be calculated from the measured quantities of the different components that $C_0 \approx 16.56 (18.87) \pm 0.15$ for the old (new) laser, respectively.

%
 
Its accuracy is, however,  hampered after degradation of the mirror and diaphragm, e.\,g., due to ageing and  dust deposit, and after hardware or software upgrades. Moreover, different periods in time  show sudden changes of the system constant, e.\,g., after installation of new hardware, after cleaning  the mirror, or changes in the onsite photo-electron counting algorithm. We have recorded, in total, 24~such hardware-intervention periods between the beginning of 2013 and March 2020. In addition, the behavior of the LIDAR has suffered eight unrecorded sudden changes of the system constant\footnote{Cleaning of the primary mirror and the lens in the focal point of the telescope was part of the operational personnel training program during several years, which was not properly logged.}. Additionally, the system constant may show weak dependencies on other parameters, such as the ambient temperature and humidity.

To correctly account for these changes, and to improve accuracy of the system constant, we have developed a new method, which allows for a more precise monitoring of the system constant. It consists of an absolute calibration of the LIDAR during very clear nights, and the construction of a \textit{calibrated degradation proxy}. The degradation proxy is constructed under the assumption of a two-stage exponential drop of the aerosol extinction coefficient, $\alpha_\mathrm{aer}(h)$, with altitude $h$ during very clear nights and that its average contribution can be extrapolated to altitudes close to ground. Under these assumptions, the \textit{vertical aerosol optical depth (VAOD)} of the ground layer during a clear night can be expressed as:
\begin{align}
\textit{VAOD}_\mathrm{\,clear~night} &= \int_0^\infty \alpha_\mathrm{aer}(h)\,\ud h \quad, \nonumber\\
              &\equiv \int_0^{H_\mathrm{PBL}} \alpha_0\cdot \exp\left(-\ddfrac{H_\mathrm{PBL}} {H_\mathrm{aer}} \right) \cdot \exp\left(- \ddfrac{(h-H_\mathrm{PBL})} {H_\mathrm{Elterman}} \right) \,\ud h + \int_{H_\mathrm{PBL}}^\infty \alpha_0 \cdot \exp\left(- \ddfrac{h} {H_\mathrm{aer}} \right) \,\ud h \quad, \nonumber\\
              & =  \alpha_0 \cdot \exp\left(-\ddfrac{H_\mathrm{PBL}}{H_\mathrm{aer}}\right) 
               \cdot \left\{   H_\mathrm{Elterman} 
               \cdot  \left( \exp\left(\ddfrac{H_\mathrm{PBL}} {H_\mathrm{Elterman}} \right) - 1 \right) +  H_\mathrm{aer} \right\} \quad, \label{eq:vaod_exp}
\end{align} \noindent
where $H_\mathrm{Elterman} = 1.2$\,km is the scale height of the aerosol model used by the default simulations for the aerosol-part of the atmosphere above the MAGIC Telescopes~\citep{Elterman:64,garrido2013}, $H_\mathrm{aer}$ is the (fitted) scale height of the aerosol extinction coefficient above the boundary layer, and  $H_\mathrm{PBL} \approx 800~\mathrm{m} \cdot (\cos\theta)^{0.6}$ is the height at which the transition between the slow exponential drop, characterized by $H_\mathrm{Elterman}$ and the fast drop, characterized by $H_\mathrm{aer}$, takes place. 

\begin{figure}
  \centering
  \includegraphics[width=0.99\columnwidth,trim={0.2cm 0cm 0.2cm 0cm},clip]{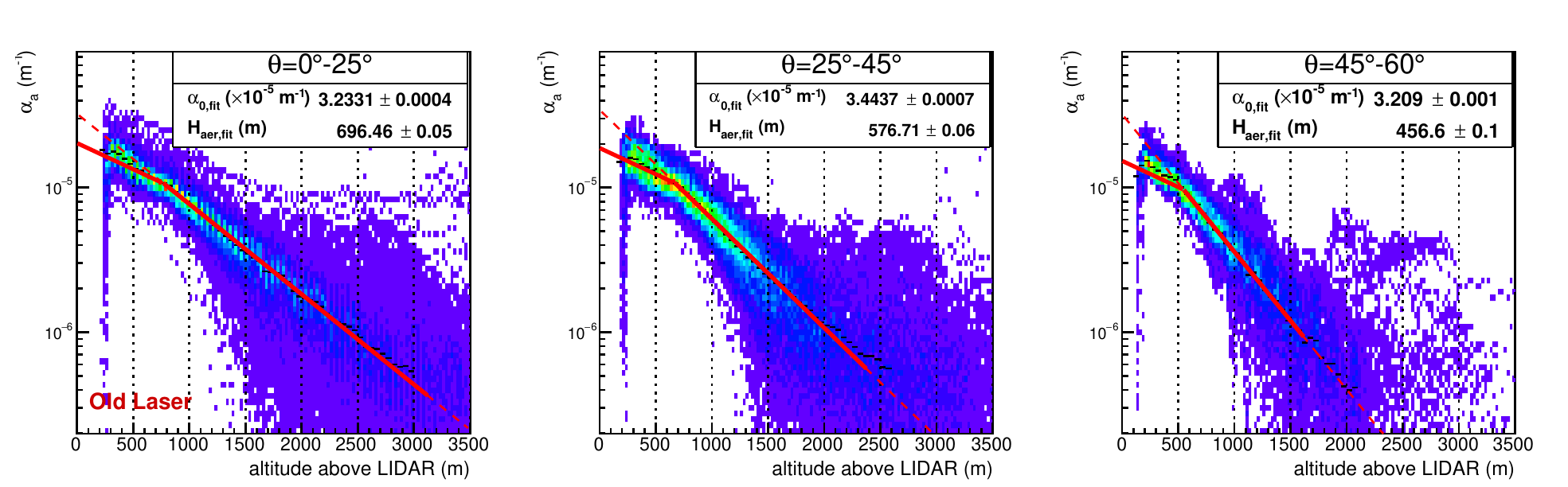}
  \includegraphics[width=0.99\columnwidth,trim={0.2cm 0cm 0.2cm 0cm},clip]{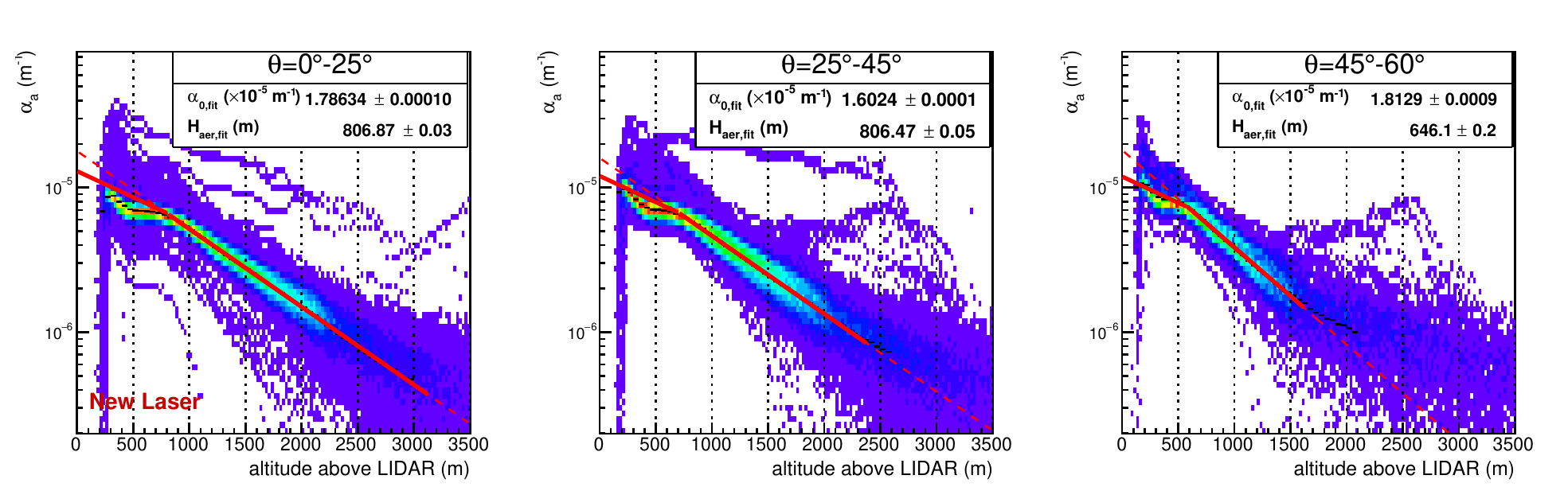}
  \includegraphics[width=0.99\columnwidth,trim={0.2cm 0cm 0.2cm 0cm},clip]{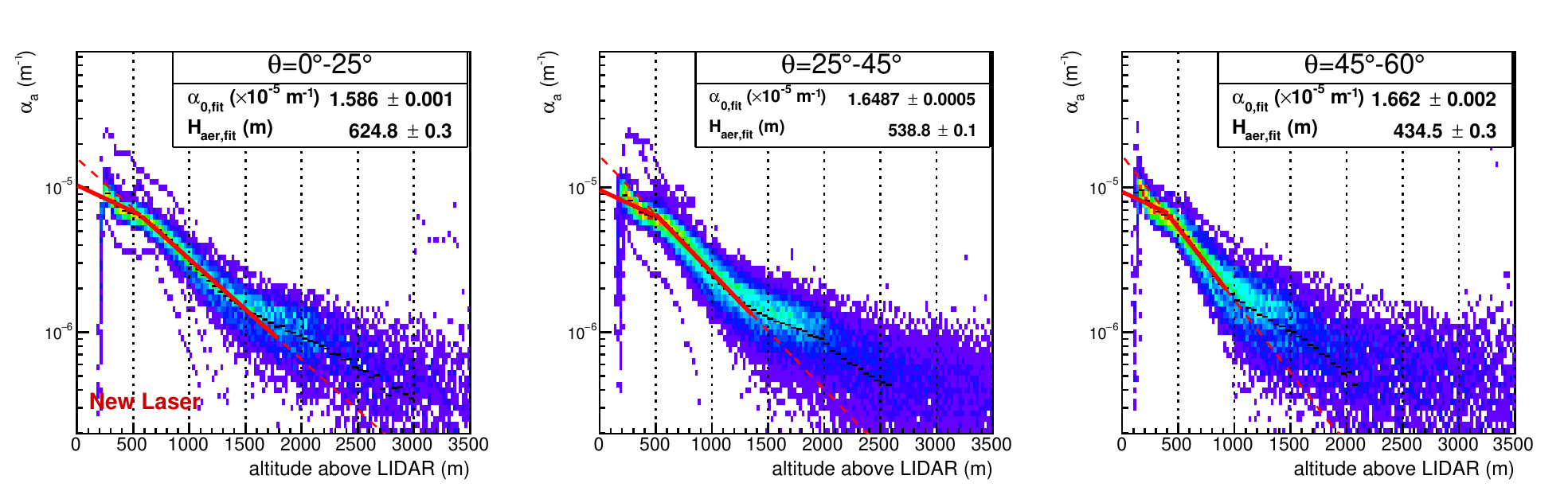}
  \caption[Clear night extinction coefficients]{Reconstructed aerosol extinction coefficients, as a function of atmospheric height above the MAGIC LIDAR, for clear nights during three selected months. Each entry corresponds to one altitude bin of a given LIDAR shot sequence. The median of each altitude bin has been fit to Eq.~\protect\ref{eq:vaod_exp} (full red line), with $H_\mathrm{PBL}$ and $H_\mathrm{Elterman}$ fixed, and $H_\mathrm{aer}$ and $\alpha_0$ as free parameters. The dashed line shows the extrapolation of the function $\alpha_0 \cdot \exp\left(-h/H_\mathrm{aer}\right)$ from ground to $H_\mathrm{PBL}$. The shown data starts from the altitude where the signals are considered free of degradation from FADC saturation or incomplete overlap. The three subsequent plots from left to right show subsequently increasing zenith angles. A LIDAR ratio of 25\,Sr~\protect\citep{Mueller:2011} had been assumed for the inversion. \label{fig:extrapolation}}
\end{figure}\noindent

Fig.~\ref{fig:extrapolation} highlights strengths and limitations of such an assumption. The retrieved extinction coefficients from clear nights follow closely the two exponential drops, except for very low altitudes, where additional structures often become visible. The exponential fits do not take into account such structures, which add to the systematic uncertainty of the method. The exponential drop persists for the different LIDAR pointing zenith angles.

\begin{figure}
  \centering
  \includegraphics[width=0.9\columnwidth,trim={0.6cm 0cm 1.5cm 0cm},clip]{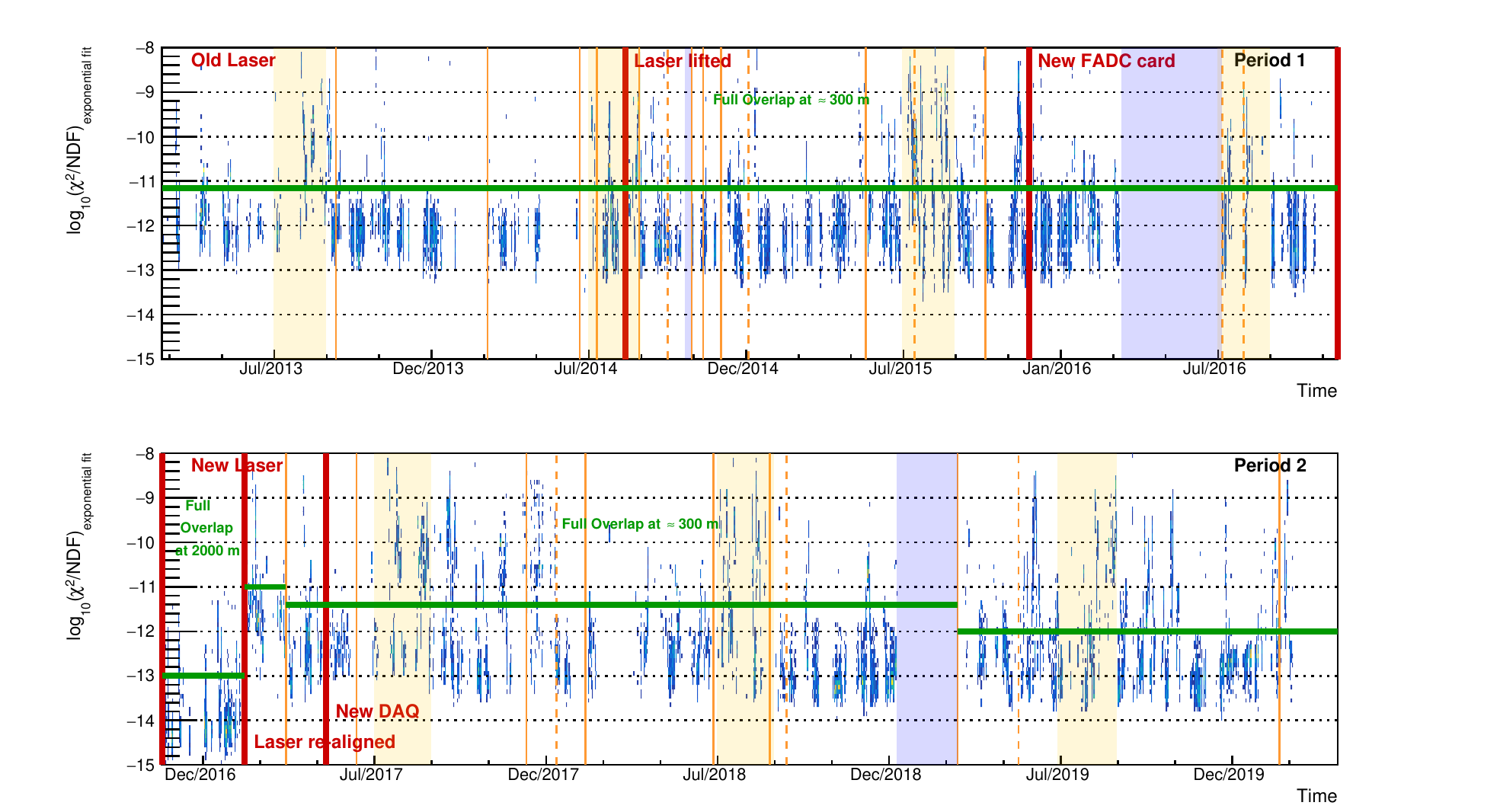}
  \caption[Distribution of chi-squares]{Logarithms of reduced $\chi^2$ from the exponential fits to the ground layer aerosol extinction coefficients $\alpha_\mathrm{aer}(h)$, as a function of time. In green, the selection criteria for the degradation proxy. Vertical red lines indicate substantial changes in hardware, whereas the dimmer full orange lines indicate minor hardware changes or maintenance activities. The dashed orange lines indicate changes in the behaviour of the LIDAR, which could not be traced back to maintenance activities in the technical logs. Blue shaded areas denote times when the LIDAR was not operative, and yellow shaded areas highlight the two summer months July and August, when frequent dust intrusions occur. Note the worse fit qualities during these months, when the aerosol density stops to decay exponentially with altitude.
  \label{fig:chi2}}
\end{figure}

The retrieved values of $\alpha_\mathrm{aer}(h)$ are fit to an exponential decay function above $H_\mathrm{PBL}$, and only those data sets are selected, which yield an acceptable fit quality, expressed in terms of the reduced chi-square. Fig.~\ref{fig:chi2} shows distributions of the logarithm of such reduced fit chi-squares as a function of time. After further cuts removing low-altitude clouds, such data provide then a set of selected \textit{VAOD} values, expressed as:
\begin{align}
    \textit{VAOD}_\mathrm{\,exp.~fit} &= \textit{VAOD}_\mathrm{\,clear~night} (\alpha_{0,\mathrm{fit}}, H_\mathrm{aer,fit}) \quad, \label{eq:vaod_selected}
\end{align}\noindent
where $\alpha_{0,\mathrm{fit}}$ and $H_\mathrm{aer,fit}$ denote the respective fit results. We can also derive the vertical aerosol optical depth  from the "Rayleigh fit" (if successful) above the boundary layer:
\begin{align}
    \textit{VAOD}_\mathrm{\,Rayleigh~fit}(C_0) &=  \ddfrac{(C_0-C)\cos\theta}{2}    \quad, \label{eq:ta_selected}
\end{align}\noindent
where $C$ is the result of the "Rayleigh fit", $C + F(h)$, to the molecular part of the LIDAR return profile (Eq.~\ref{eq:molecular_return}). We construct then a {\bf proxy for the system constant}:
\begin{align}
P(C_0) &:= \ddfrac{2\cdot \textit{VAOD}_\mathrm{\,Rayleigh~fit}(C_0)}{\cos\theta} + C  \quad. \label{eq:def_degrproxy} 
\end{align}
Eq.~\ref{eq:def_degrproxy} does not explicitly rely on the assumption of a horizontally stratified aerosol layer, only indirectly by using the exponential decay model for $\alpha_\mathrm{aer}$. 

Since we are mainly interested in changes of $C_0$ with respect to some reference value $C_{0,\mathrm{initial}}$ -- apart from one absolute calibration at the beginning -- and, moreover, some kind of linear degradation with time of the number of laser photons $N_0$ and effective  area $A$ (e.\,g., due to dust deposit, or fatigue of the laser) is expected, we can define two \textit{related expressions of the degradation proxy}:
\begin{align}
P(\Delta C_0) &:= \ddfrac{2\cdot\textit{VAOD}_\mathrm{\,exp.~fit}(C_0) }{\cos\theta} + C  - C_{0,\mathrm{initial}}\quad \\[0.25cm]
\mathrm{and} \nonumber\\[0.25cm]
P\left(\ddfrac{N_0 \cdot A}{N_{0,\mathrm{initial}} \cdot A_{0,\mathrm{initial}}}\right) &:= 
\exp \big( P \left(\Delta C_0\right)  \big) \quad.
\label{eq:def_deltadegrproxy} 
\end{align}

\noindent

\begin{figure}
  \centering
  \includegraphics[width=0.99\columnwidth,trim={0.7cm 0cm 1.5cm 0cm},clip]{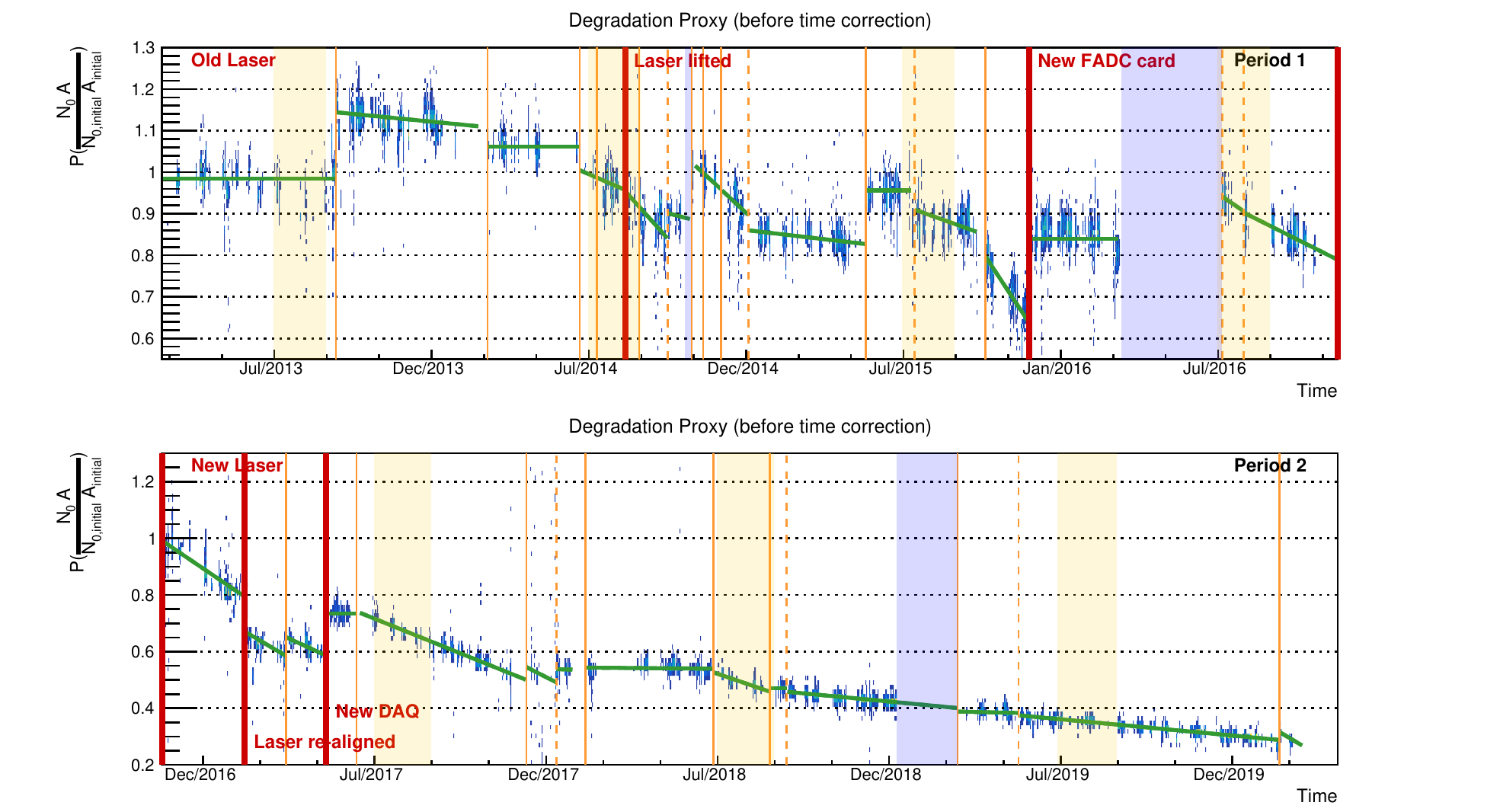}
  \caption[Distribution of degradation proxy, before calibration]{Distribution of the degradation proxy, as a function of time, before correction. Linear fits to each separate hardware period are shown as green lines. Note the different axes scales between both figures. For further details, see also Fig.~\protect\ref{fig:chi2}.
  \label{fig:proxy_before}}
\end{figure}

\begin{figure}
  \centering
  \includegraphics[width=0.99\columnwidth,trim={0.5cm 0cm 0.2cm 0cm},clip]{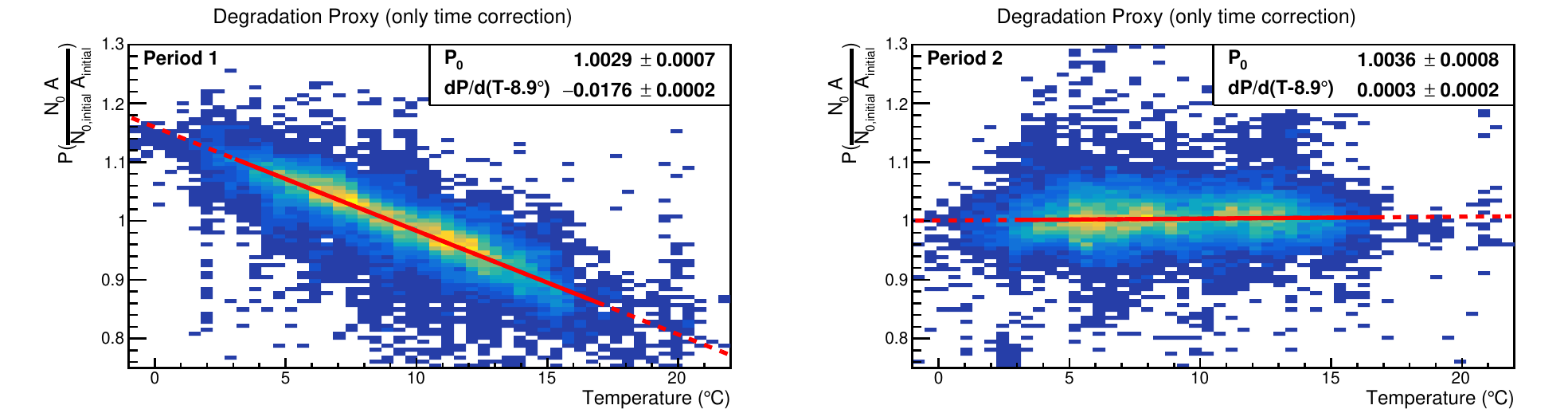}
  \caption[Distribution of degradation proxy, temperature correction]{Distribution of the degradation proxy, as a function of the outside temperature, after correction using the fits to each hardware period (see Fig.~\protect\ref{fig:proxy_before}). The medians of each temperature bin have been fitted to a 1$^\mathrm{st}$-order polynomial.
  \label{fig:proxy_beforetemp}}
\end{figure}

Both quantities depend on time and temperature (see Figs.~\ref{fig:proxy_before} and~\ref{fig:proxy_beforetemp}) and require an iterative correction approach for both. A negative temperature gradient can be explained by the not fully compensated temperature dependency of the gain of the avalanche photo-diode inside the HPD. A software routine has been developed in the new system, where the pulse integrals around the single photo-electron peaks are precisely monitored and used to adjust the APD gain. 

Hence, 
the onsite photon-counting algorithm of the new system 
removes first-order temperature dependencies of the degradation proxy. 

\begin{figure}
  \centering
  \includegraphics[width=0.6\columnwidth]{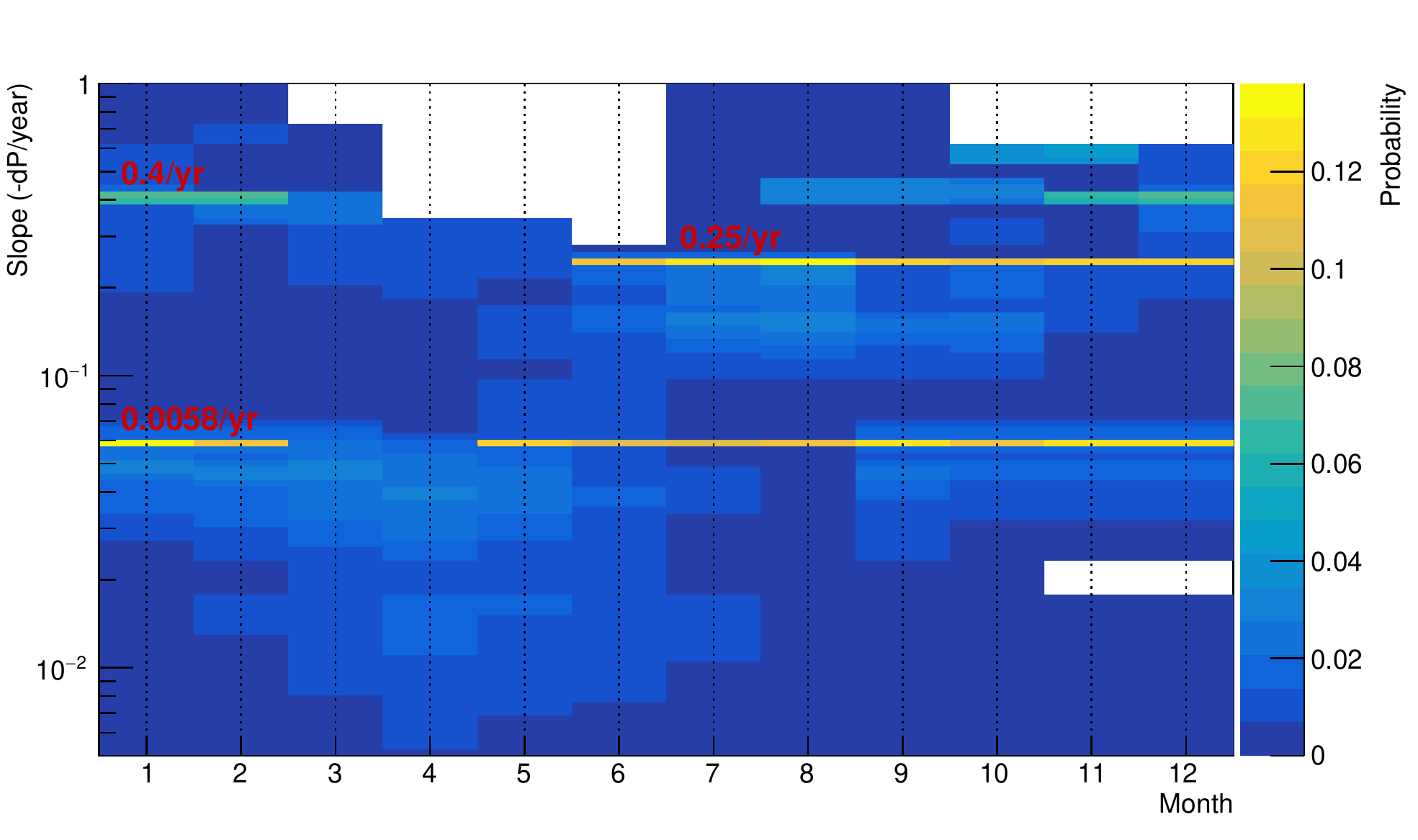}
  \caption[Distribution of fitted degradation slopes]{Distribution of the fitted degradation slopes, as a function of affected month. The probability distributions for each reconstructed slope have been treated as normal distributions centered at the most probable fit value, with sigma equivalent the square root of the corresponding entry of the covariance matrix. The three most probable degradation slopes have been marked with red text.  
  \label{fig:proxy_slopes}}
\end{figure}

Application of this new algorithm has greatly improved the stability of $ P(\Delta C_0)$ in Period~2. 
\collaborationreview{Derived} probability distributions to find a given degradation slope with time are shown in Fig.~\ref{fig:proxy_slopes}. One can see that the system degrades most probably with about $-$0.6\% per year, except for the summer months, where $-$25\%/year 
\collaborationreview{is found to be the}
most probable. The latter is clearly related with the frequent Saharan dust intrusions during the summer months, which cause dust deposit on the mirror. Later during the year, degradation depends on whether the mirror had been cleaned or not (i.e., recovering the $-$0.6\%/year slope) or not. In the latter case, even $-$40\% degradation per year have been observed.

\begin{figure}
  \centering
  \includegraphics[width=0.99\columnwidth,trim={0.6cm 0cm 1.5cm 0cm},clip]{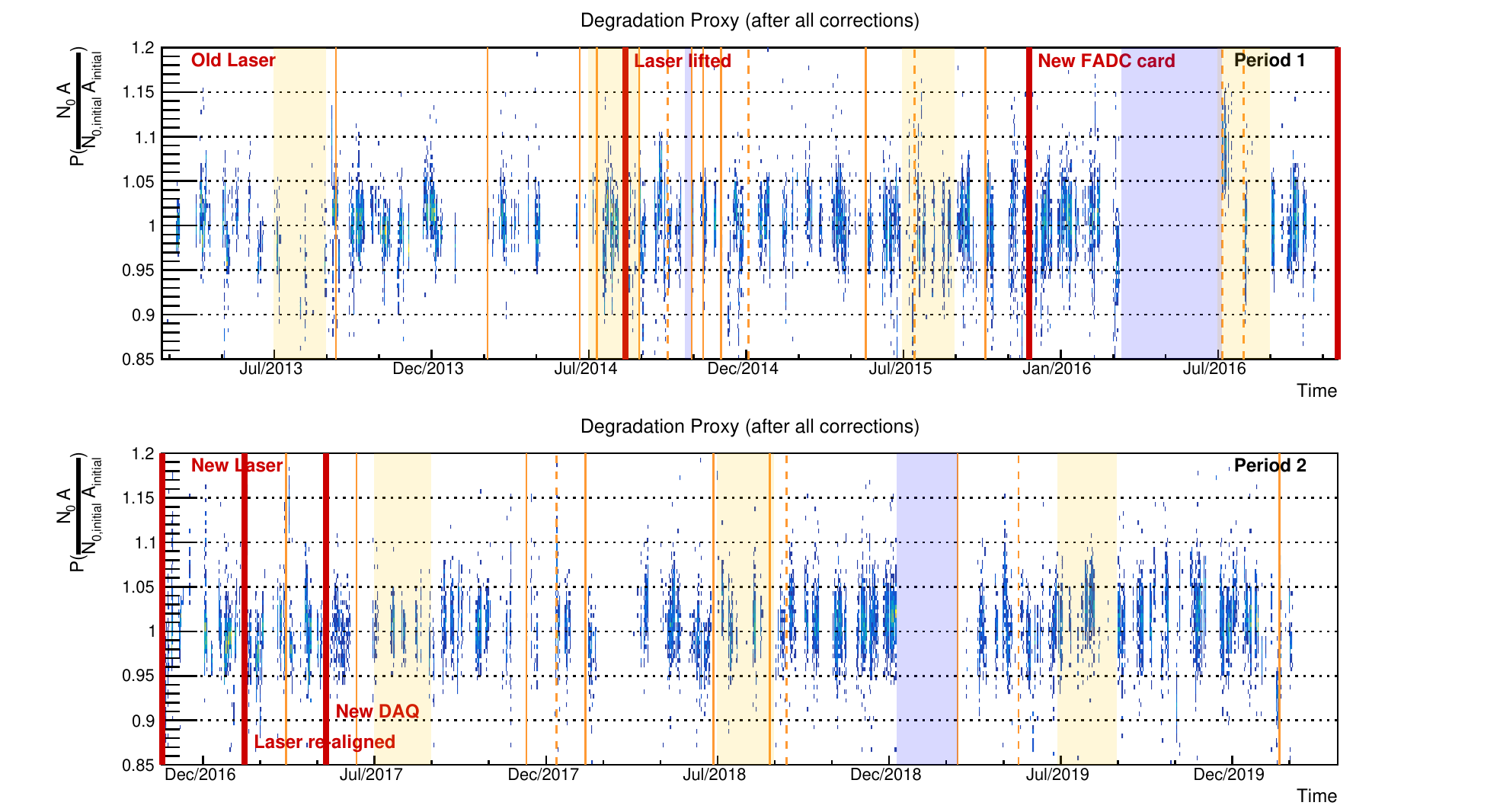}
  \caption[Distribution of degradation proxy, after calibration]{Distribution of the degradation proxy, as a function of time, after correction using the fits to each period (for further details, see Figs.~\protect\ref{fig:chi2} and~\protect\ref{fig:proxy_before}).
  \label{fig:proxy_after}}
\end{figure}

\begin{figure}
  \centering
  \includegraphics[width=0.475\columnwidth]{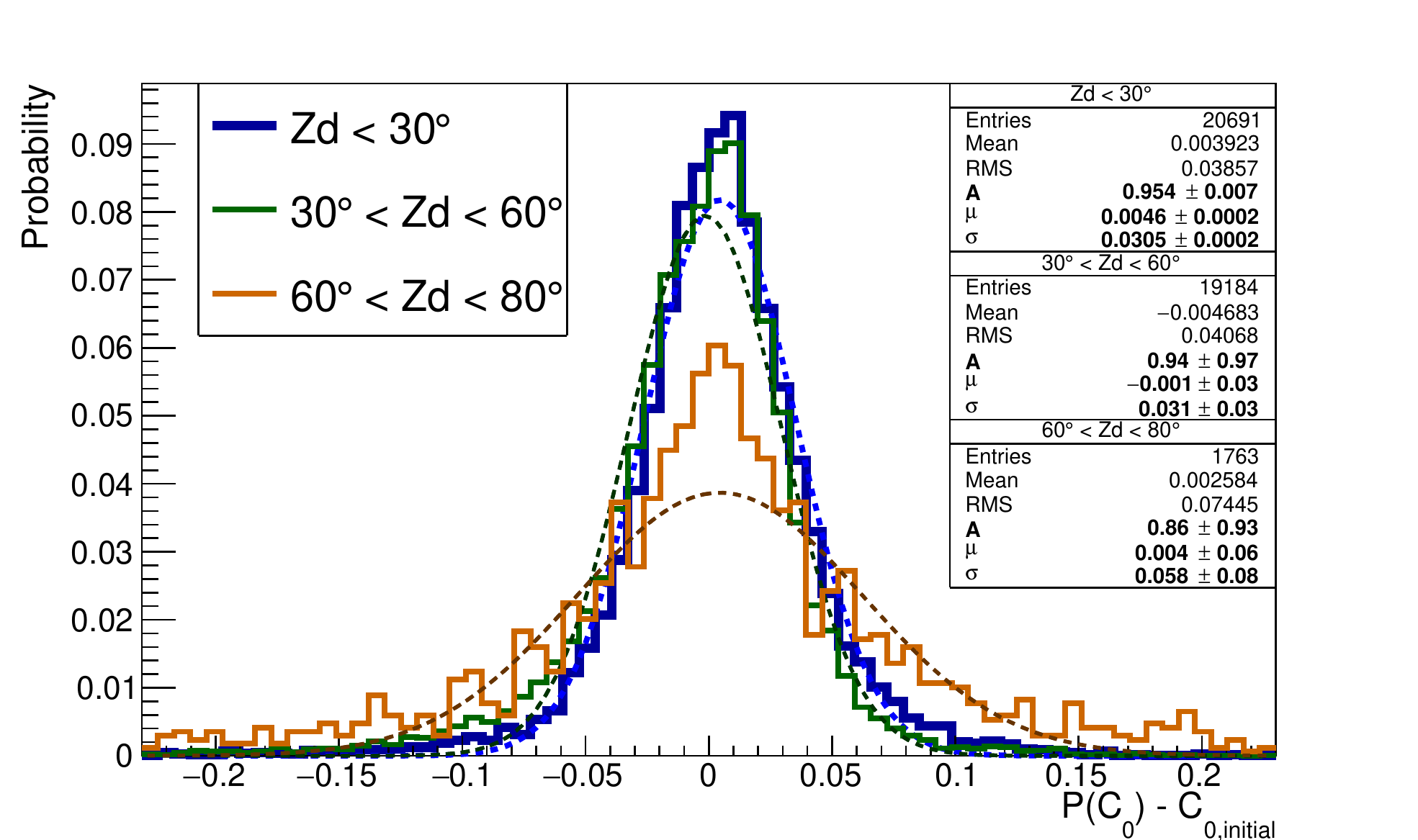}
  \includegraphics[width=0.505\columnwidth]{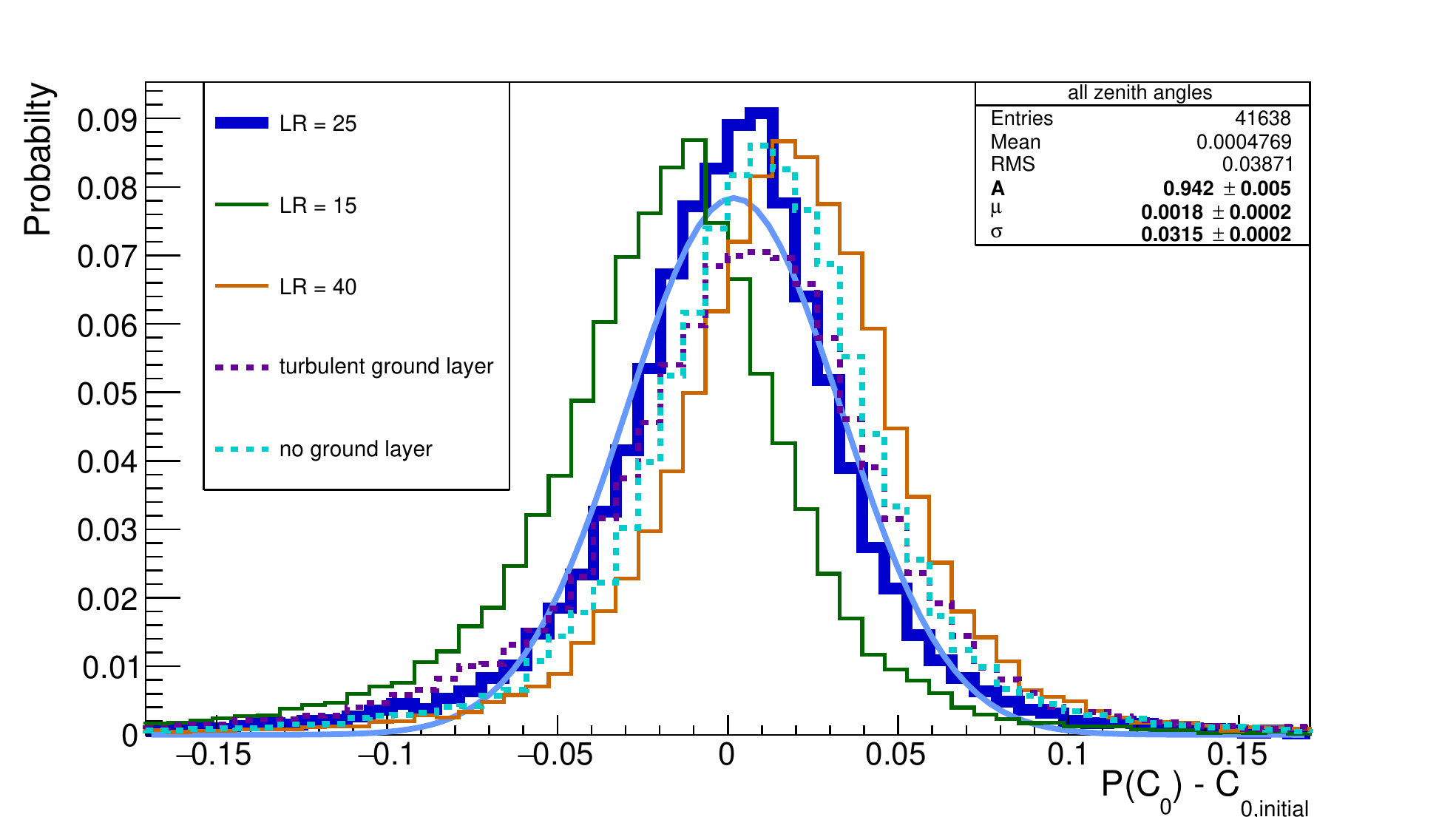}
  \caption[Distribution of degradation proxy after calibration]{Distribution of the degradation proxy after calibration. Left: Distributions are shown for three different zenith angle ranges, all three have been fitted to normal distribution PDF. Right: all zenith angles together are shown (blue). The green and orange histograms show solutions obtained from different, but rather extreme assumptions for the LIDAR ratio (LR) of the clear sky ground layer aerosols. The violet dashed line shows the deviation of the degradation proxy, if instead of an exponentially decreasing aerosol density profile, a turbulently mixed layer of constant aerosol density, up to 800\,m from ground, is assumed, from which the exponential decrease starts. The cyan dashed line, on the contrary, assumes similar exponential decay behavior of the aerosol extinction coefficient, from ground to infinity. Note the different axis ranges, the right image has been magnified for the sake of better visibility.
  \label{fig:proxy_1d}}
\end{figure}

Figs.~\ref{fig:proxy_after} and~\ref{fig:proxy_1d} show the degradation proxy after both temperature and time-wise corrections, as a function of time and the distribution for all data taken which satisfy the clear-night fit criteria. One can see that the resolution of the proxy is about $\pm$3\% (except for the largest pointing zenith angles $>60^\circ$, where it becomes twice as large), with an additional systematic uncertainty of about 2\%. We conclude that we are able to absolutely calibrate our system to better than 4\%.

\noindent

\noindent

\noindent

\noindent

\noindent

%% file: cloudanalysis.tex
\section{Clouds}

Where the atmosphere shows a strong pre-dominance of the free troposphere with practically no aerosols (i.e., well above the boundary layer at the Canary Islands), we can describe the LIDAR return signal by molecular Rayleigh scattering and extinction alone, except for the contribution of regions identified as cloud layers. Under these conditions, it is possible to measure the optical depth of clouds directly with an elastic LIDAR. To achieve this, clouds are first identified by scanning the logarithm of the range-corrected LIDAR return in a sliding window of 20 (6) data points (corresponding to a range window of 1000\,m (300\,m)) for the old (new) laser data, respectively.  In each step, the window is shifted upwards in altitude by one bin, and the shape of the interpolated molecular signal prediction (Eq.~\ref{eq:molecular_return}) is fitted to that interval. The fitted \textit{VAOD} from that part to ground and the reduced fit chi-square per degree of freedom, $\chi^2_\nu$, are then used to recognize clouds.

The cloud base is detected if the following two criteria are fulfilled: \collaborationreview{the}  $\chi^2_\nu$ of the local fit is larger than 5.0 (6.0), and the fitted \textit{VAOD} value from ground to the respective fit range, must \textit{decrease}, compared with fit results from lower altitudes. The second condition follows the assumption that inside a cloud layer, increased aerosol backscattering will increase the signal with respect to the surrounding molecular regions.  Once both conditions are fulfilled, the algorithm moves back to lower altitudes, until the fitted \textit{VAOD} ($\textit{VAOD}_1$) is compatible again with (i.\,e., lying within one standard deviation from) the one obtained from the molecular atmosphere below the cloud. At the same time, the $\chi^2_\nu$ of the local fit must be better than 1.7 (2.0) at that point.

The cloud top, instead, is reached if  $\chi^2_\nu$ reaches again 1.7 (2.0), and the difference between the fitted \textit{VAOD}s from above and below the cloud is \textit{positive}. In that case, the algorithm moves further up to higher altitudes testing a steady further increase of \textit{VAOD}, until a stable \textit{VAOD} ($\textit{VAOD}_2$) is reached again, or a maximum limit altitude of $22\,000 \cdot \cos\theta$\,m (24\,500\,m or $31\,000 \cdot \cos\theta$\,m, whatever is lower) above ground has been reached.

The vertical optical depth of the cloud can then be estimated directly from the difference of \textit{VAOD}s from the molecular fits at the end and start points of each cloud, i.e., $\textit{VOD}_\mathrm{cloud} = \textit{VAOD}_2 - \textit{VAOD}_1$.

\subsection{Signal inversion using an iterative Klett formalism}

We invert the signal region between the estimated cloud top and cloud base with the Klett algorithm, Eq.~\ref{eq:sasano2}, assuming an initial LIDAR ratio of $S_{a,\mathrm{cloud}}= 33$\,Sr. The resulting aerosol extinction coefficients, $\alpha_a(h)$, can then be integrated over the cloud and compared with the cloud's \textit{VOD}, obtained in the previous section. We then re-scale $S_{a,\mathrm{cloud}}$ by the ratio between both optical depth estimates and invert the cloud part of the signal again using Eq.~\ref{eq:sasano2}. This procedure is repeated, until both \textit{VOD} estimates  coincide, providing an indirect measurement of the cloud's LIDAR ratio. The algorithm converges after a few iterations in most cases. Only in case of failure of convergence, or an obtained LIDAR ratio smaller than 8\,Sr or larger than 110\,Sr, an alternative method has been used to approximate the extinction profile of the cloud. Such cases  may be due to the limitations arising from the assumption of one common LIDAR ratio across the cloud (particularly for thick clouds), or from statistical uncertainties in the assumed cloud's \textit{VOD}, particularly for optically very thin clouds. Also the movement of a cloud into or out of the LIDAR field-of-view during data taking may lead to unphysically large LIDAR ratios, and hence failures of this method. 

\subsection{Signal inversion using the Extinction Method}

In that case, we distribute the total extinction  proportionally to the magnitude of the cloud/aerosol back-scattering excess over the mean molecular fit expectation:
%
\begin{align}
	\overline{\left(N\!(r) r^2\right)}_\mathrm{mol.~part}(h) &= 
	\left(\left(N\!(r) r^2\right)_\mathrm{mol.~part}(\textit{VAOD}_1,h) + \left(N\!(r) r^2\right)_\mathrm{mol.~part}(\textit{VAOD}_2,h)\right)~/~2  \nonumber\\
	\alpha_a(h) & \approx (\textit{VAOD}_2 - \textit{VAOD}_1) \cdot \ddfrac{ \left(N\!(r) r^2\right)(h)-\overline{\left(N\!(r) r^2\right)}_\mathrm{mol.~part}(h)}{ \displaystyle\int_{h_1}^{h_2}{  \left(N\!(r) r^2\right)(h)-\overline{\left(N\!(r) r^2\right)}_\mathrm{mol.~part}(h) \, \ud h } }\, . \label{eq:distrExtinction}
\end{align}

This approximation neglects the fact that the signal excess over the molecular expectation gets modified by attenuation within the cloud itself and is good only for optically thin clouds~\citep{fruck:phd,Fruck:2015}. However, Eq.~\ref{eq:distrExtinction} does not modify the \textit{VOD} of the cloud, only the spatial distribution of $\alpha_a$ within the cloud is affected.

\subsection{Signal inversion using the LIDAR Ratio Method}

In a third approximation, from now on referred to as ``LIDAR ratio method'', the LIDAR ratio is assumed to be known and constant inside a given cloud. Also the total transmission of the cloud is assumed to be close to unity, such that back-scattered light from the far end of the cloud does not get attenuated much by the rest of the cloud. The extinction coefficient $\alpha_a(h)$ can then be approximated from the back-scattering excess on top of the molecular scattering pedestal directly, assuming that the LIDAR-ratio of cloud scattering $S_a$ is known.
\begin{equation}
	\alpha_a(h) = S_a \cdot \beta_\t{mol}(h) \cdot \ddfrac{  \left(N\!(r) r^2\right)(h)-\overline{\left(N\!(r) r^2\right)}_\mathrm{mol.~part}(h) }{\overline{\left(N\!(r) r^2\right)}_\mathrm{mol.~part}(h)}\, .
\end{equation}

The LIDAR ratio method only makes sense if the assumption of a known and stable value of the LIDAR ratio is valid. Therefore, a typical value for the LIDAR ratio was chosen from a test sample and the resulting total cloud transmission for both methods over a large sample of LIDAR measurements has been compared (see~\citet[][for further details]{fruck:phd,Fruck:2015}).

The extinction method performs reasonably well for clouds of moderate extinction. Instead, as the cloud \textit{VOD} approaches zero, the relative error on the measurement becomes very large. For clouds absorbing less than $\np[\%]{10}$ of the light, it is more sensible to use the LIDAR ratio approximation, if the direct (iterative) signal inversion using Klett's algorithm fails.

\begin{figure}
  \centering
  \includegraphics[width=0.69\columnwidth,trim={0.9cm 0 1.9cm 0},clip]{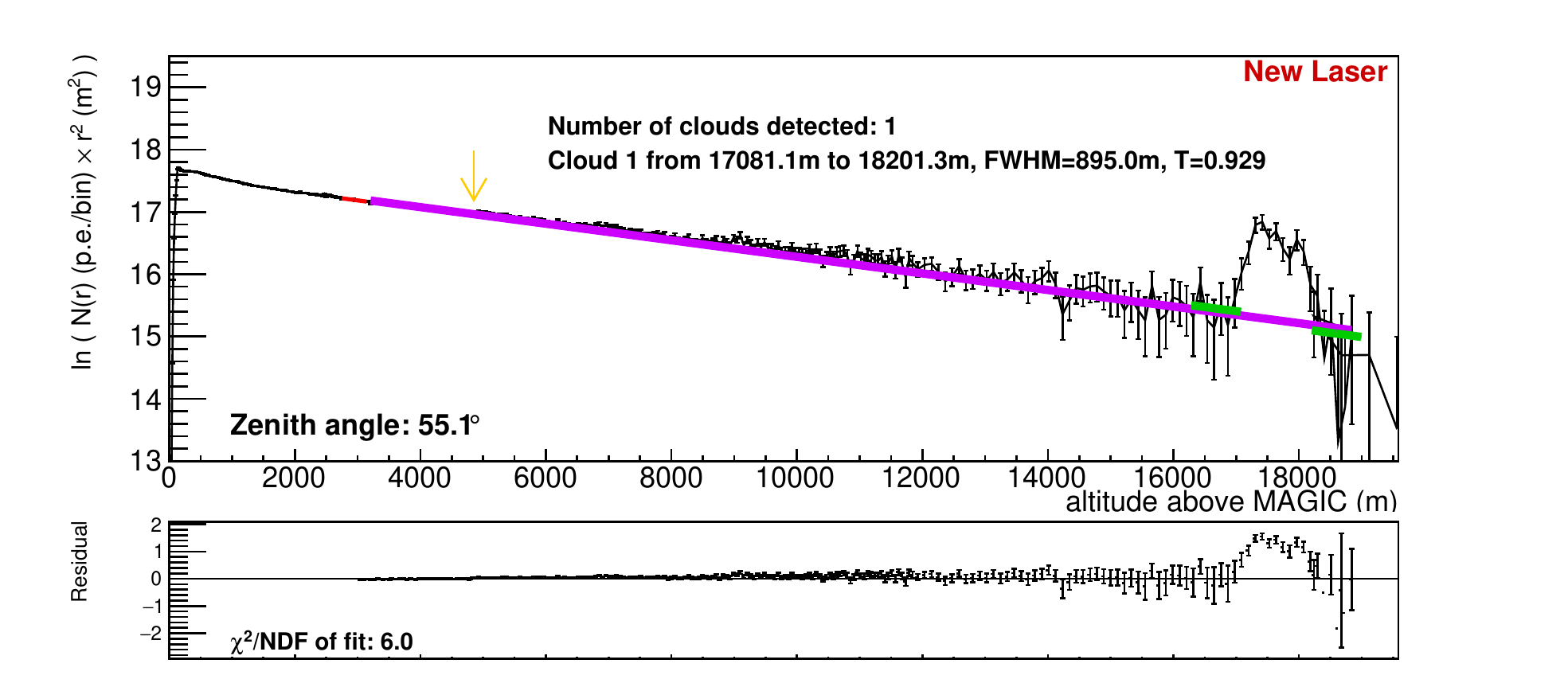}
  \caption[Example of high cloud]{An example of a high cloud observed under a large zenith angle. Green  lines show the fits to $F(h)$ right below and above the detected cloud. The small yellow arrow indicates the transition from the amplitude to the photo-electron counting regime. Below each signal plot, the residuals between data and $F(h)$ are displayed and the corresponding reduced $\chi^{2}$. The latter is used to determine whether a cloud search is initialized at all.
  \label{fig:highcloudexample}}
\end{figure}

Fig.~\ref{fig:highcloudexample} shows an extreme case of a high cloud observed under a large zenith angle, during August when the tropopause can reach altitudes of up to 20\,km a.s.l.\ above the Canary Islands~\citep{rodriguez2013}. Detection of an optically thin cloud at a distance of more than 30\,km from the LIDAR reaches the edge of sensitivity, but also demonstrates the capability of our system.

\noindent

%% file: sixyears.tex
\section{Analysis of seven years of LIDAR data}
\label{sec:sixyears}

\begin{figure}
  \centering
  \includegraphics[width=0.99\columnwidth,trim={0.5cm 0cm 0.5cm 0cm},clip]{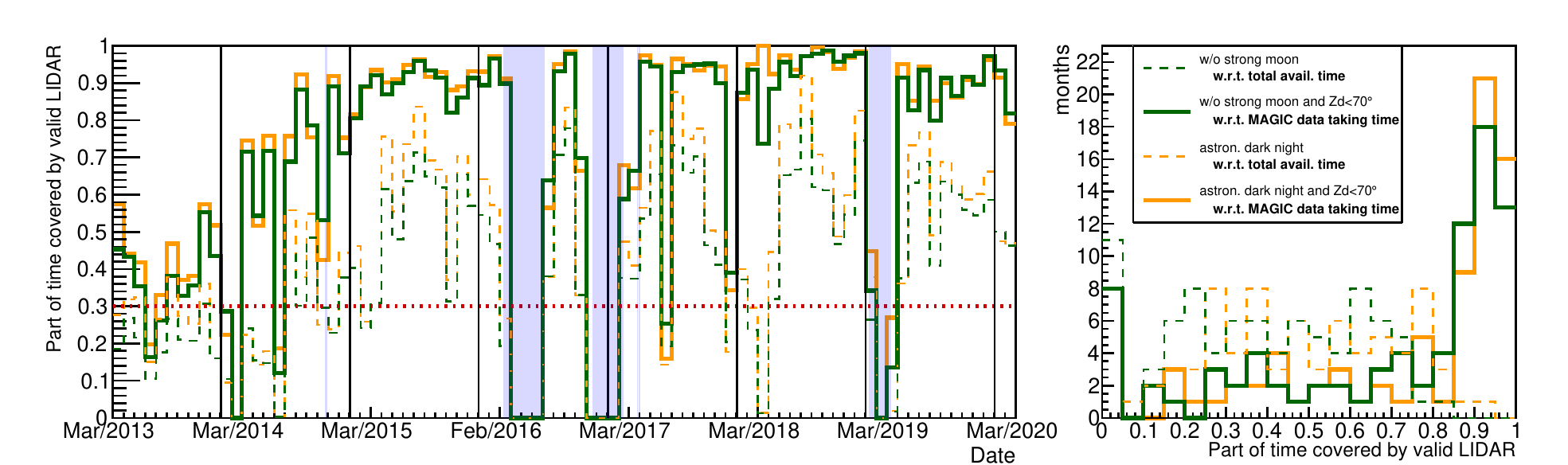}
  \caption[Coverage of LIDAR data]{Coverage of the MAGIC LIDAR data as a fraction of the time available (dashed lines) or time actually used by the MAGIC Telescopes  (full lines). Left: evolution of valid LIDAR coverage with time, every bin corresponds to one month. Orange: part of astronomical dark night covered by valid LIDAR, green: part of the full night including astronomical twilight, after excluding strong moon and very high zenith angle ($\theta>70^\circ$) observations. Blue shaded areas denote times when the LIDAR was not operative. The dotted red line displays the minimum full time coverage required for a month to enter the statistical analysis of Section~\ref{sec:sixyears}. Right: distribution of monthly LIDAR data coverage.
  \label{fig:coverage}}
\end{figure}

In the following, we present results from data taken with the MAGIC LIDAR during seven years, from March 2013 until March 2020. The LIDAR was operated in semi-continuous mode during night, very closely following the observation schedule of the MAGIC Telescopes and has collected $\sim$10$^5$ atmospheric profiles. Fig.~\ref{fig:coverage} shows the time coverage of these LIDAR data, the time covered by LIDAR, once compared to the observation up-time of the telescopes (full lines), and then compared to a total of night time available. The differences reside in the uptime of the MAGIC Telescopes, which stop observing if humidity on ground is higher than 90\%, the speed of wind gusts becomes larger than 40\,km/h, the aerosol transmission drops to values well below 50\% during an extended time interval (and if technical observations are not feasible), or for technical maintenance of the telescopes. Unfortunately, some of these criteria have changed over time. 

The LIDAR is operated when the MAGIC Telescopes observe targets at zenith angles below 70$^\circ$ (above 20$^\circ$ elevation). The so-called ``very-large zenith angle'' (VLZA) observation mode~\citep{VLZA:2019}, which has gained more importance over the past years, is usually \textit{not} accompanied by LIDAR. On the one hand, the range of the LIDAR does not reach the part of the atmosphere where VLZA air showers develop, on the other hand the edge of the LIDAR dome presents a hardware limit to the zenith angles the LIDAR can reach. Observations during moon~\citep{moonlight} can lack LIDAR coverage if the background light caused by the moon becomes too strong for the instrument. For this reason, Fig.~\ref{fig:coverage} shows two separated cases, one for astronomical dark night time and data taken at zenith angles below 70$^\circ$, when the LIDAR is supposed to be operative always, and another comparison using data without very strong moon. Those MAGIC data, which ought to be covered by contemporaneous LIDAR data reach regularly coverage well above 90\% from 2015 on, after a first year of low performance and a second year of continuous improvement. Three periods of system upgrades are marked by blue shades. The overall night-time coverage of LIDAR data (compared to the total available night time) fluctuates considerably, very dependent on the weather conditions. A general trend to higher coverage is found over the years, partly due to continuous improvements of the overall data taking efficiency of the MAGIC Telescopes.

\begin{figure}
  \centering
  \includegraphics[width=0.5\linewidth]{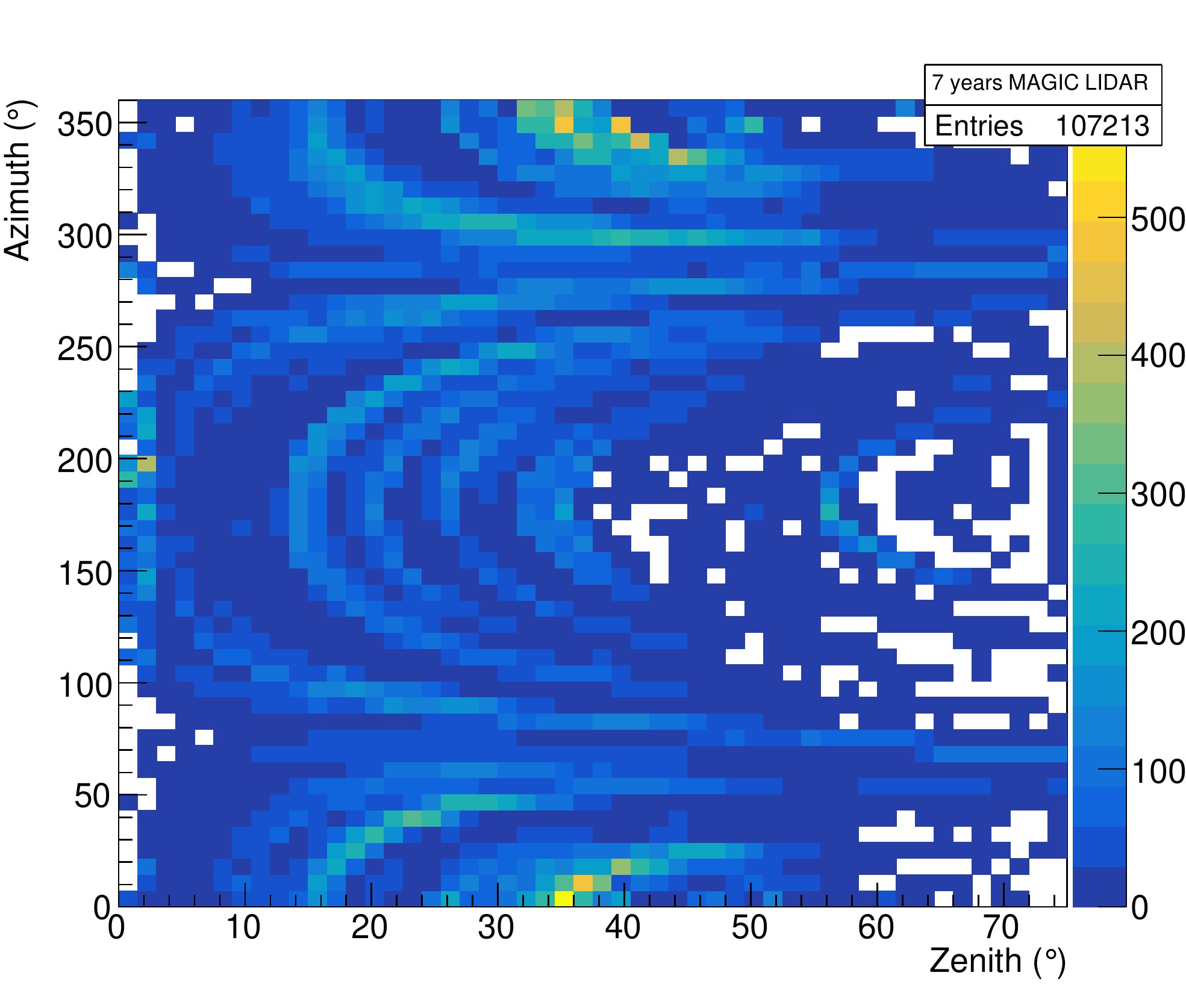} 
  \caption{Distribution of LIDAR pointing angles of the analyzed data set. \collaborationreview{0$^\circ$} azimuth corresponds to North, \collaborationreview{90$^\circ$} to East.
  \label{fig:zdaz}}
\end{figure}

Fig.~\ref{fig:zdaz} shows the distribution of LIDAR pointing angles of the whole sample. A clear signature of the movement of several primary MAGIC observation targets across the sky is visible, which the LIDAR has followed. Nevertheless, a large fraction of the sky is covered by data, except for the Southern and Northern directions at large zenith angles.

In the following sections we will use either the full data set when general atmospheric properties are highlighted, or a selection of data when statistical properties are derived. The selection criteria \collaborationreview{have} been chosen such that only months with a minimum of 30\% night-time coverage enter statistical analyses as this is considered sufficiently representative for the behaviour of a full month. We emphasize that this criterion has been chosen {\textit after} looking at the data, and to ensure sufficient overall statistics. Related possible systematic biases will be discussed in Section~\ref{sec:discussion}.

\subsection{Aerosol altitude profiles of clear nights}

About 60\% of the LIDAR data sample ($\sim 6\times 10^4$ profiles) follow a ground layer aerosol extinction profile in the form of that introduced in Eq.~\ref{eq:vaod_exp}, with acceptable fit quality (see Fig.~\ref{fig:chi2}). This sample was taken during time periods where the LIDAR reached full overlap at less than 350\,m. The profiles have been classified as representative for \textit{clear nights} and  further analyzed. Aerosol extinction scale heights $H_\mathrm{aer}$ ranging from 250\,m to more than 4000\,m have then been obtained for clear nights during the seven years analyzed. 

\begin{figure}
  \centering
  \includegraphics[width=0.5\linewidth]{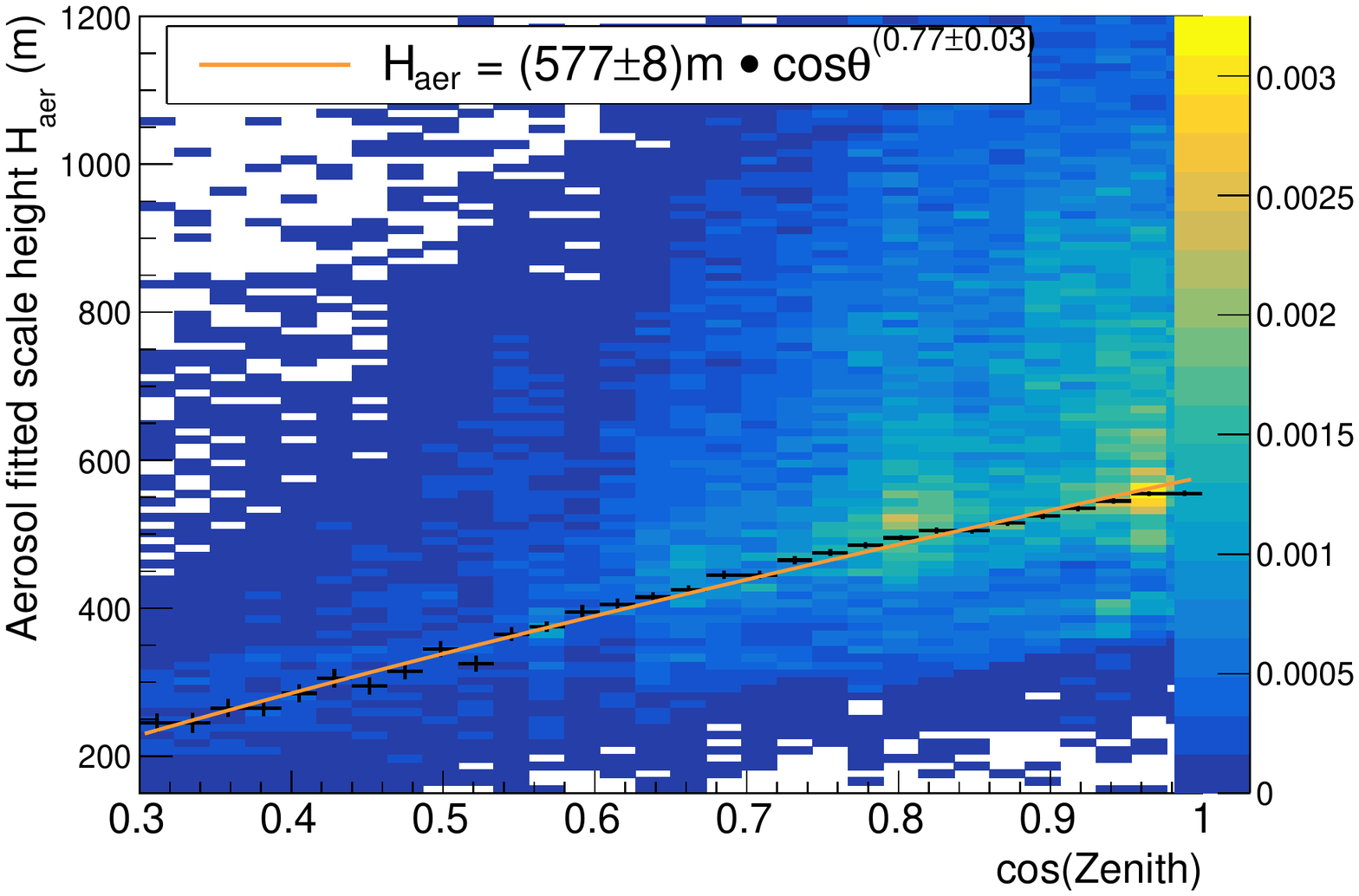} 
  \caption{Distribution of fitted aerosol extinction coefficient scale heights $H_{aer}$ (see Eq.~\protect\ref{eq:cosfit}). The black crosses are located at the position of the weighted median at each bin and fitted to a linear dependency of the cosine of zenith angle.
  \label{fig:alphascalezd}}
\end{figure}

In a stratified atmosphere, the height dependency of aerosol extinction does not depend on the direction to which the LIDAR points. This is, however, not the case for the ORM. Figure~\ref{fig:alphascalezd} shows the distribution of fitted aerosol extinction coefficient scale heights, $H_\mathrm{aer}$, as a function of LIDAR pointing zenith angle. Contrary to the stratification case, the scale heights do depend on observation zenith angle, in a form that can be approximated by
\begin{align}
  H_\mathrm{aer} \approx H_\mathrm{aer,0} \cdot \left(\cos\theta\right)^\gamma~,
  \label{eq:cosfit}
\end{align}
%
where the exponent $\gamma$ parameterizes the curvature of the aerosol layer. A completely stratified atmosphere would reproduce $\gamma \approx 0$, whereas a round convex-shaped layer produces $\gamma \approx 1$. 

The aerosol ground layer at the ORM has an intermediate shape, represented by $\gamma \approx 0.77$ on average. We investigated further possible dependencies of $H_\mathrm{aer}$ on azimuthal pointing directions and seasonal variations, shown in Table~\ref{tab:scaleheightszd}. Southern directions

show a shallower drop of scale height with zenith angle on average, corresponding to a smaller curvature of the aerosol layer. This is in agreement with expectation because of the elongated shape of the island of La Palma, where the ORM is located towards its Northern edge. Nevertheless, this result should be taken with some caution, due to the much lower angular LIDAR coverage towards Southern directions, as discussed in the previous subsection. Besides that, there is a hint of a small seasonal dependency of the ground layer during clear nights, with a higher overall scale height, but less stratification during summer. A similar trend has been observed in the background AOD by~\citet{Laken:2016} (see, e.\,g.\, their Fig.~9b).

\begin{table}
  \centering
  \begin{tabular}{ccc}
  & $H_\mathrm{aer,0}$ (m) & $\gamma$ \\ \addlinespace
  \toprule
  North & (581$\pm$ 12)& (0.83 $\pm$ 0.04) \\ \addlinespace[1mm]
  South & (552$\pm$ 12)& (0.60 $\pm$ 0.04) \\\addlinespace[1mm]
  East & (553$\pm$ 12) & (0.72 $\pm$ 0.04) \\\addlinespace[1mm]
  West & (578$\pm$ 12) & (0.79 $\pm$ 0.04) \\\addlinespace[1mm]
  \midrule
  Winter (S,O,N,D,J,F) & (568$\pm$ 12)& (0.75 $\pm$ 0.04) \\\addlinespace[1mm]
  Spring (M,A,M,J) & (580$\pm$ 12)& (0.79 $\pm$ 0.04) \\\addlinespace[1mm]
  Summer (J,A) & (606$\pm$ 12)& (0.88 $\pm$ 0.04) \\\addlinespace[1mm]
  \bottomrule \addlinespace[1mm]
  All data & (577$\pm$ 8) & (0.77 $\pm$ 0.03)
  \end{tabular}
  \caption{Results of the fit Eq.~\protect\ref{eq:cosfit} to aerosol scale heights $H_\mathrm{aer}$ for different azimuth pointing angles and separated by seasons. At the bottom, the total of all data are shown. Uncertainties are statistical only and do \textit{not} take into account the effect of the assumption of a uniform LIDAR ratio across the ground layer. 
  The months have been grouped in those of similar behaviour. Note that September sometimes behaves rather like a summer month, whereas in other years it is more winter-like. In order to keep the small summer data sample clean, we have opted to subsume September within the winter case. 
  \label{tab:scaleheightszd}}
\end{table}

\begin{figure}
  \centering
  \includegraphics[width=0.55\linewidth]{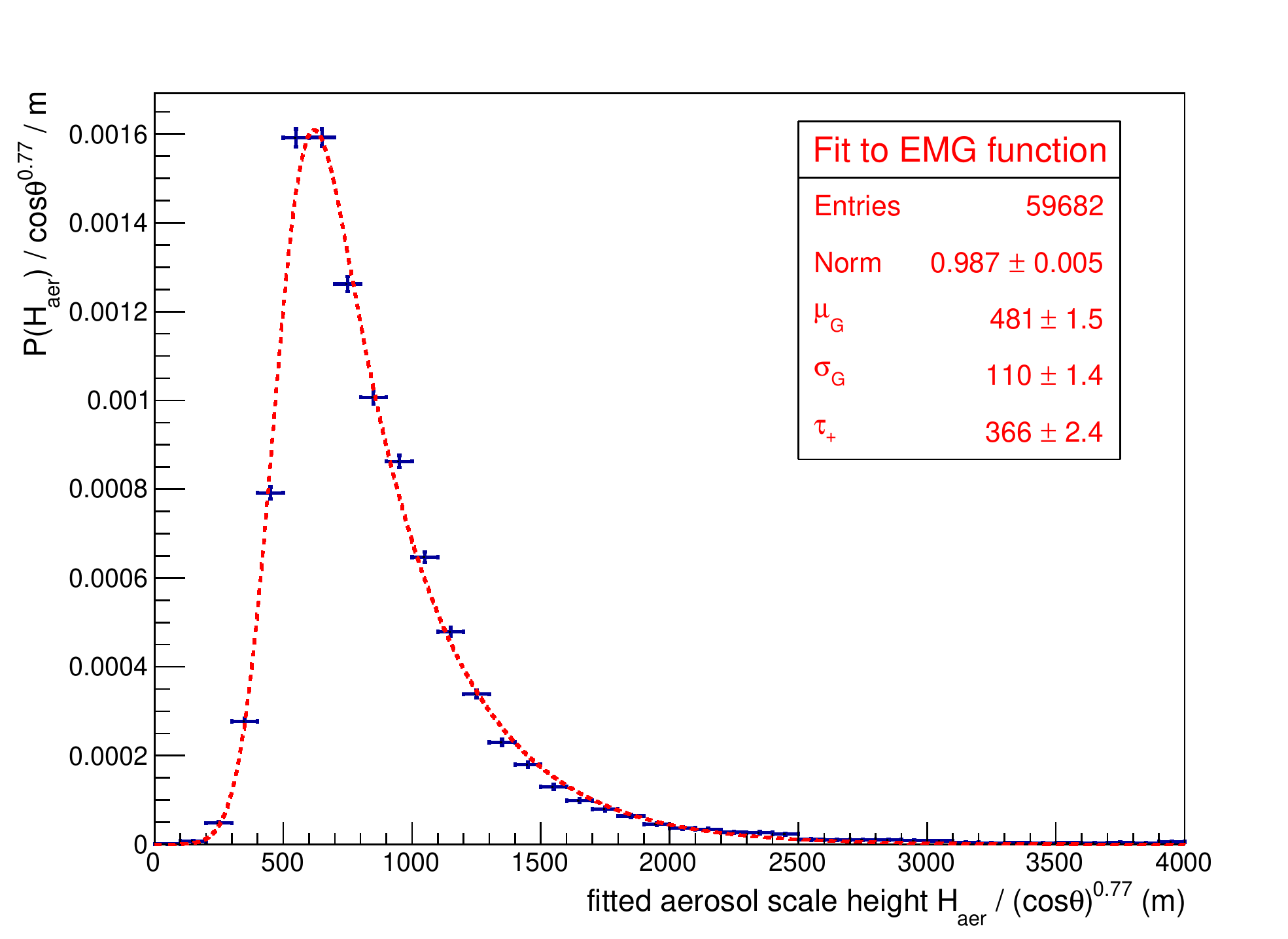} 
  \caption{Distribution of zenith-corrected reconstructed aerosol extinction coefficient scale heights $H_\mathrm{aer}$. The asymmetric distribution has been fitted to an exponentially-modified Gaussian probability distribution, with a mean of $(\mu_G+\tau_+)=590\pm 2$\,m and a standard deviation of $\sqrt{\sigma_G^2+\tau_+^2} = 388\pm3$\,m. 
  \label{fig:alphascale}}
\end{figure}

The vertical scale heights, $H_{\mathrm{aer,0}}$, show an asymmetric distribution with and a mode of 620\,m. It can be fit with an exponentially-modified Gaussian distribution (see, e.\,g.,~\citet{EMG:2017}) and is shown in Fig.~\ref{fig:alphascale}.

\begin{figure}
  \centering
  \includegraphics[width=0.65\linewidth]{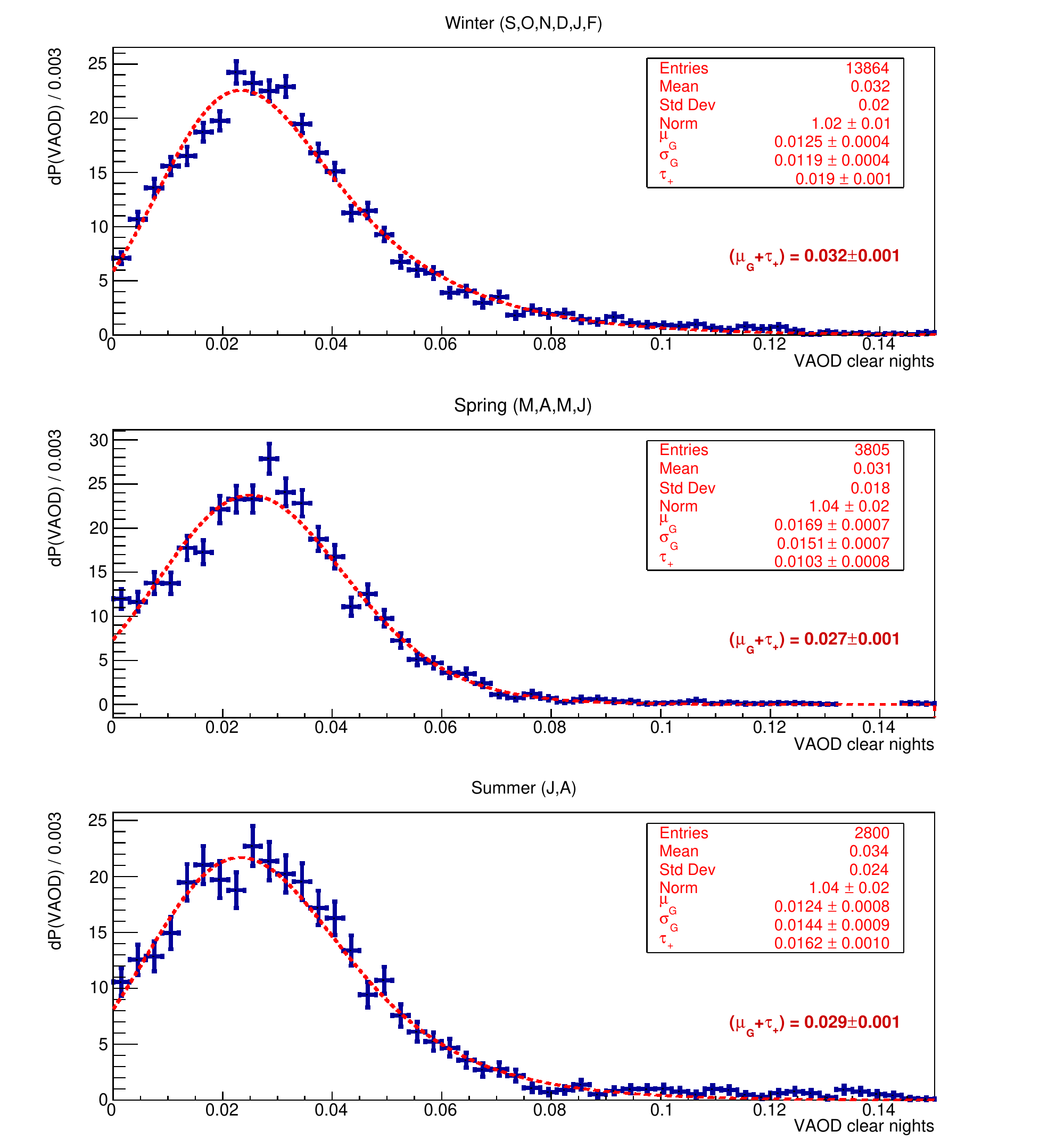} 
  \caption{Distributions of ground layer vertical aerosol optical depths, reconstructed by calibrated ``Rayleigh fits'' (Eq.~\protect\ref{eq:molecular_return}), for selected clear nights at low zenith angles for different seasons. The distributions have been fitted to an exponentially-modified Gaussian probability distribution. 
  \label{fig:vaodseasons}}
\end{figure}

The reconstructed ground layer \textit{VAODs} (now using the calibrated ``Rayleigh fit'' Eq.~\ref{eq:molecular_return}) for those selected clear nights that can be described by Eq.~\ref{eq:vaod_exp}, seem to be very stable throughput the year and do not show any clear seasonal dependency (see Fig.~\ref{fig:vaodseasons}). This result may seem in contradiction with the findings of \citet{Laken:2016}, but can be easily explained by the selection criteria (Fig.~\ref{fig:chi2}) applied to this subsample of data, which have a much higher selection efficiency for the winter months than the two summer  months.

The resulting mean VAOD for clear nights yields $\textit{VAOD} = 0.030^{+0.021}_{-0.013}$, where the uncertainty represents the (asymmetric) standard deviation. An additional systematic uncertainty of $\Delta \textit{VAOD} = 0.01$ stems from the assumptions made for the LIDAR ratio. The corresponding ground layer transmission for clear nights at 532\,nm is then $T_\mathrm{aer} = 0.970^{+0.016}_{-0.023}$. This result is compatible with the winter 50\% percentile background optical extinction ($\tau$-values) of~\citet{Laken:2016} ($\tau = 0.042 \pm 0.001$ mag/airmass), retrieved from extinction measurements at 770\,nm, by two optical telescopes at 2400\,m a.s.l. These measurements contain a contribution from molecular extinction and stratospheric aerosol extinction. Subtracting the molecular extinction due to Rayleigh scattering ($\tau_{\mathrm{Rayleigh},770\,\mathrm{nm}} \approx 0.020$) and assuming a background stratospheric extinction of $\tau_{\mathrm{aer,strat.}} \approx 0.005$ and converting to optical depth, we obtain $\textit{VAOD}_\mathrm{ref,770\,nm} \approx 0.016$. In the same reference, baseline aerosol optical depths at 675\,nm ($\tau_a = 0.017 \pm 0.01$ systematic uncertainty~\citep{Holben:1998}) are provided from AERONET Sun photometers installed at the Iza\~na Atmospheric Research Center (IARC) at about 150\,km distance on the neighbouring island of Tenerife, also located at 2400\,m altitude, i.\,e., 200\,m higher than the MAGIC~LIDAR. Such low \collaborationreview{\textit{VAOD}s} of the ground layer at 2400\,m~a.s.l.\ at the ORM have also been found by~\citet{Sicard:2010} in a LIDAR measurement campaign lasting several days during May, June and July, 2007. At 532\,nm, two out of three measurement campaigns yielded values of $\textit{VAOD}_\mathrm{ref,532\,nm} \approx 0.011$ and $0.018$, respectively, close to their actual detection limit.

Using Eq.~\ref{eq:vaod_exp}, $\Delta \textit{VAOD} \approx 0.004$ can be attributed to the 200\,m difference in altitude between the locations used by \citet{Laken:2016} and \citet{Sicard:2010}, and our LIDAR, respectively. Another correction should be made to account for the different wavelengths, \citet{maring2000} found \AA ngstr\"om exponents ranging~\citep{angstrom:1929} from 1.5 to 2.0 for non-dusty periods at Iza\~na. This would result in wavelength-corrected mean aerosol optical depths of $\textit{VAOD}_\mathrm{corr,532\,nm} \approx (0.025-0.033)$, translated  from \citet{Laken:2016}, and compatible with our results.

\subsection{Aerosol Transmission Statistics}
\label{sec:aertransmission_lidar}

\begin{figure}
  \centering
  \includegraphics[width=0.99\columnwidth,trim={0.6cm 0cm 1.5cm 0cm},clip]{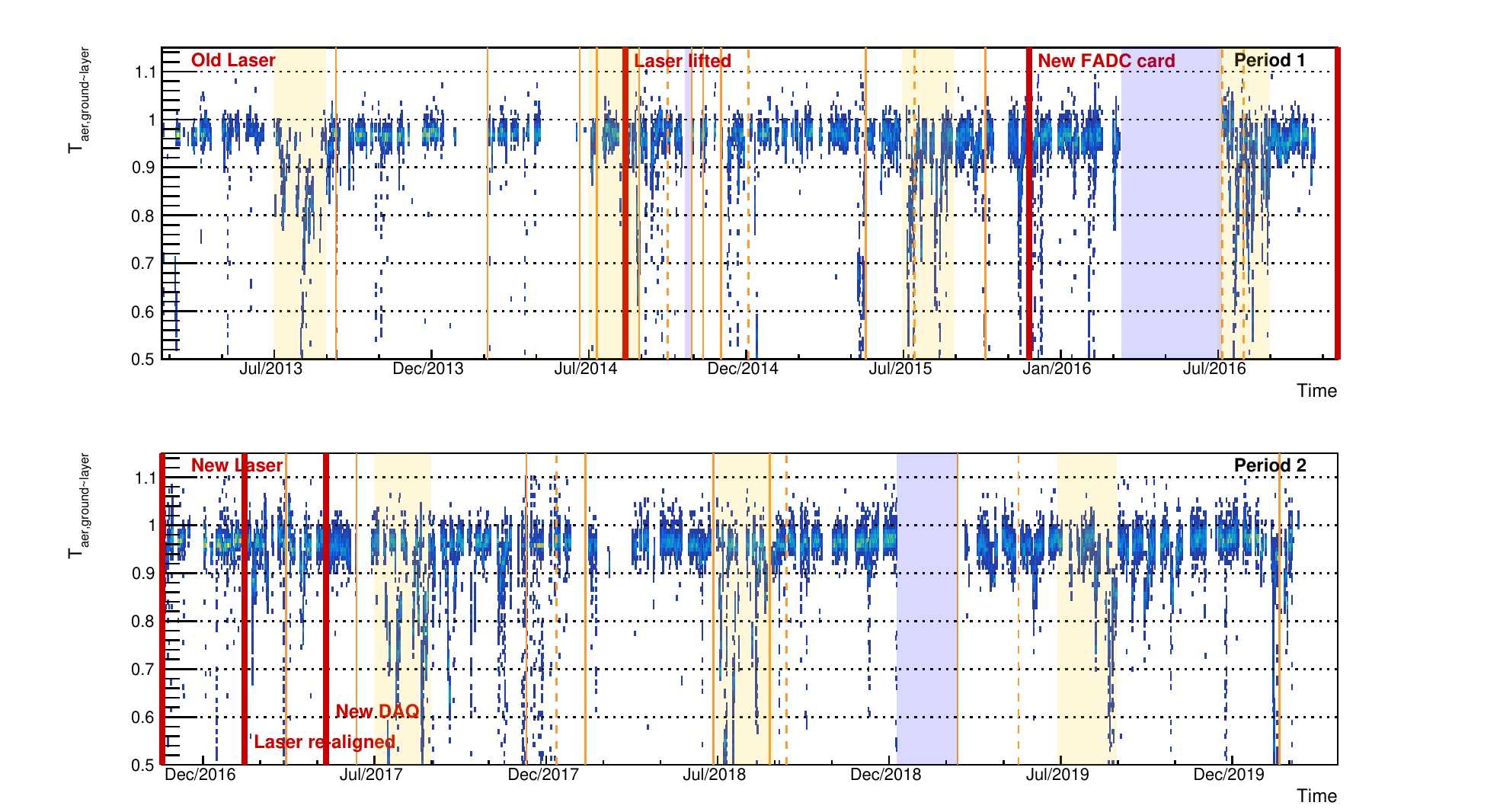}
  \caption[Distribution of aerosol transmission]{Distribution of ground layer aerosol transmission retrievals as a function of time. See Fig.~\protect\ref{fig:chi2} for details. 
  \label{fig:ta}}
\end{figure}\noindent

\begin{figure}
  \centering
  \includegraphics[width=0.485\linewidth, clip, trim=5.6cm 5.8cm 6cm 5.8cm]{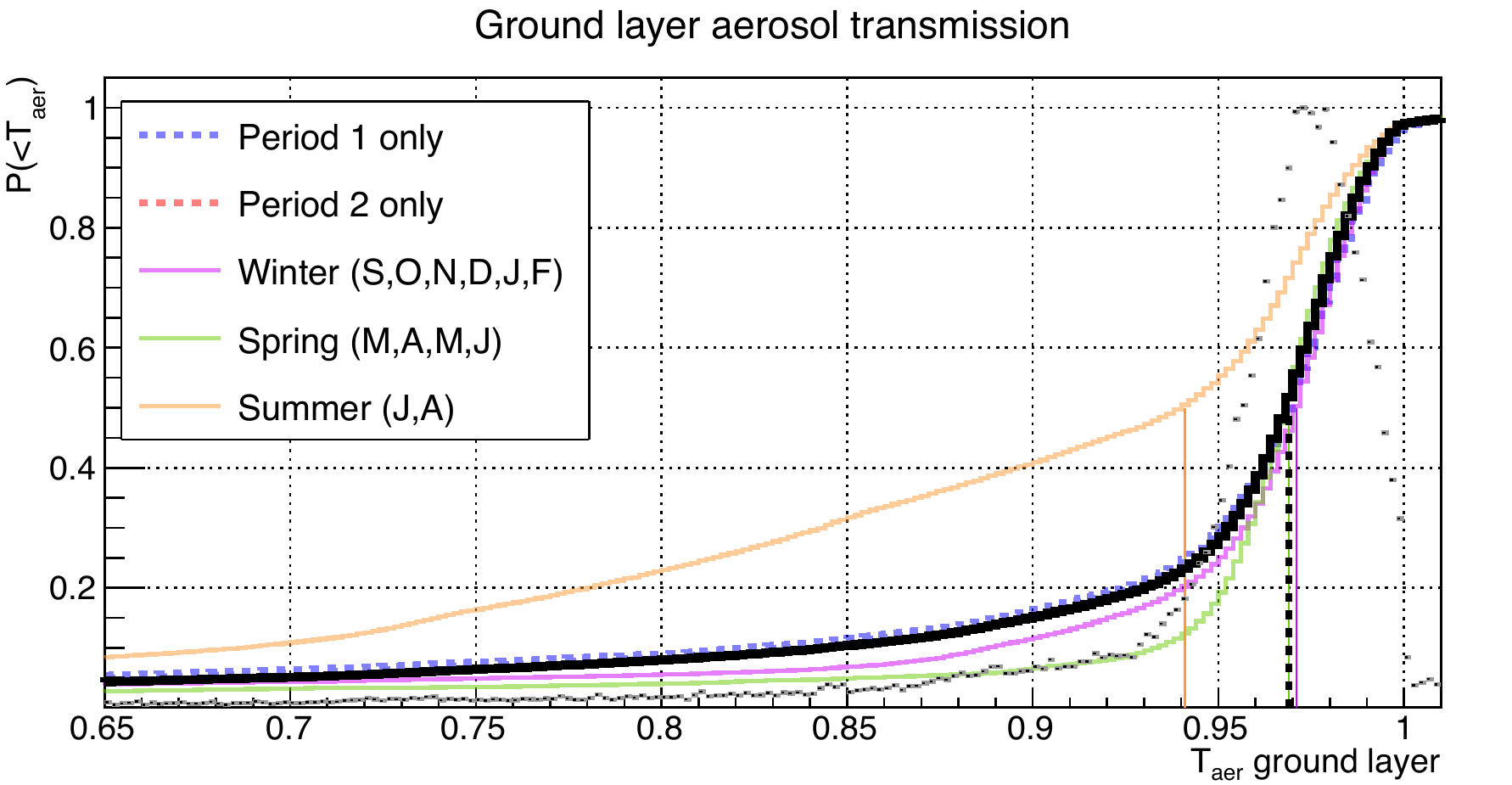}
  \includegraphics[width=0.485\linewidth,clip,trim=0.1cm 0 1.5cm 0]{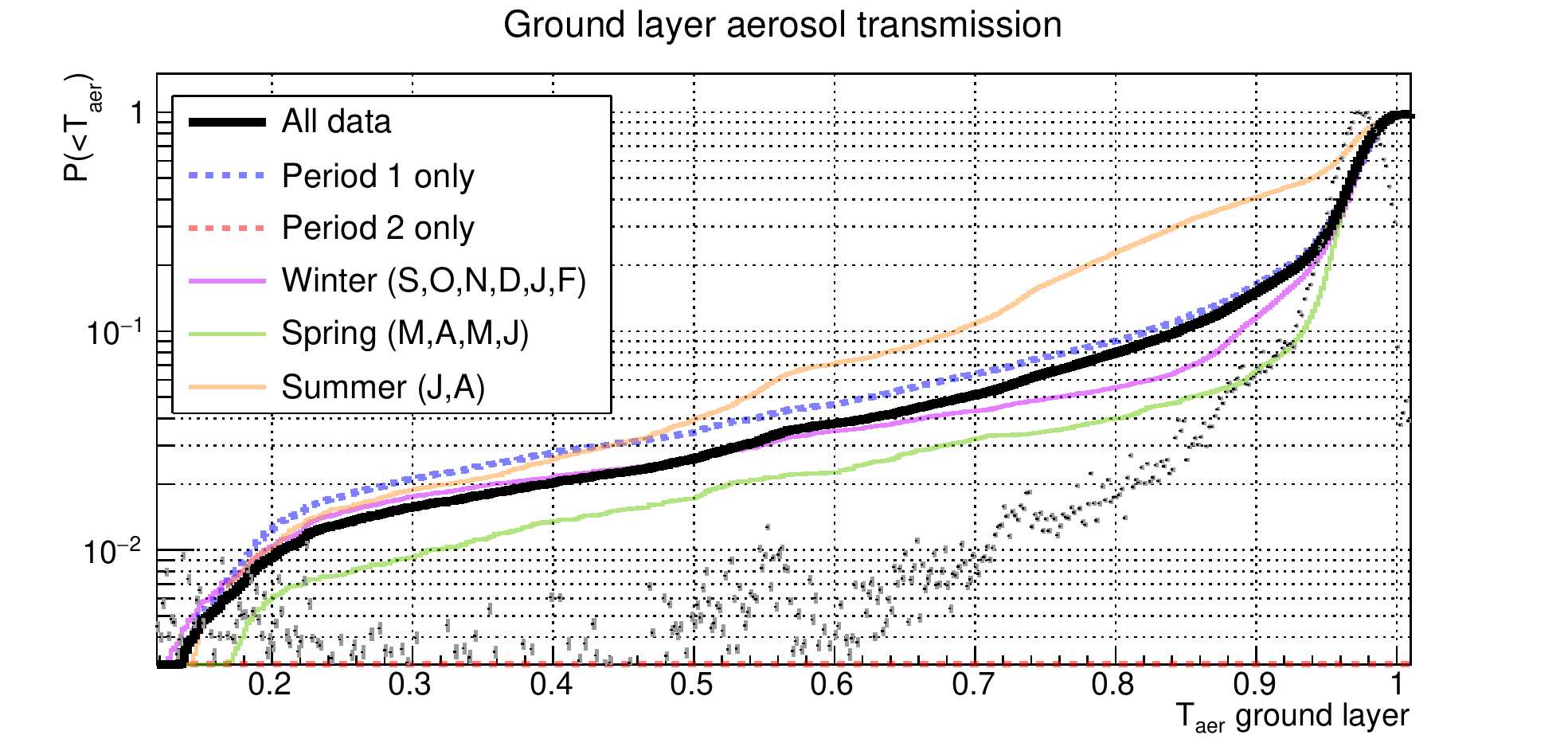}
  \caption{\label{fig:MAGIC_calima} Cumulative vertical aerosol transmission probabilities of green light for ground-layer aerosols during MAGIC data taking conditions (RH~$<$~90\%, wind gust $<$~40\,km/h) and weighted according to Eq.~\protect\ref{eq:weights}. The vertical lines refer to the medians of the distributions. Gray dots show the differential probability of the full data set, scaled to reach a maximum of 1 for better comparability. Left: in linear scale, right: in logarithmic scale. Note the different axis ranges. 
  }
\end{figure}

\begin{figure}
  \centering
  \includegraphics[width=0.485\linewidth, clip, trim=0.3cm 0 1.5cm 0]{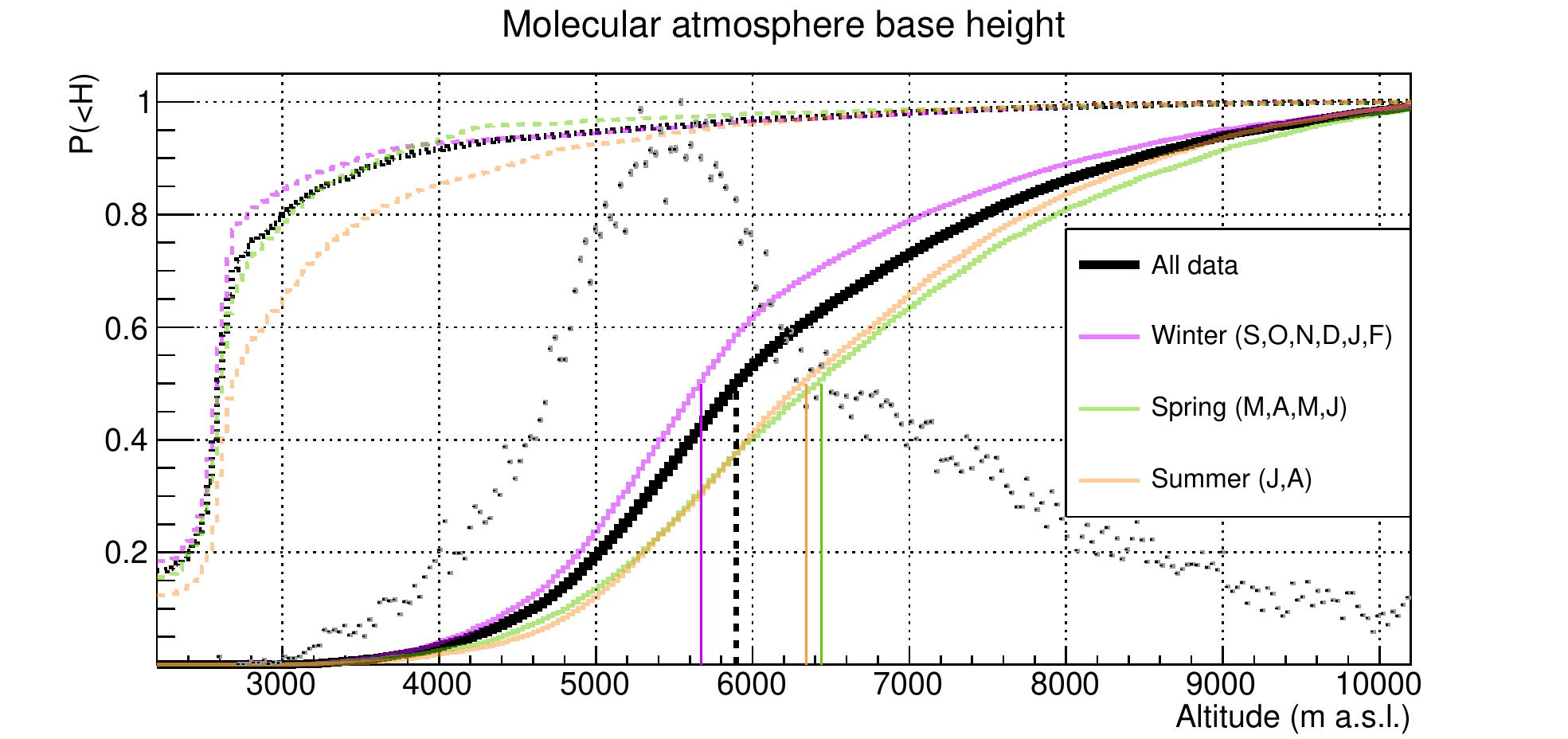}
  \includegraphics[width=0.485\linewidth,clip,trim=0.1cm 0 1.5cm 0]{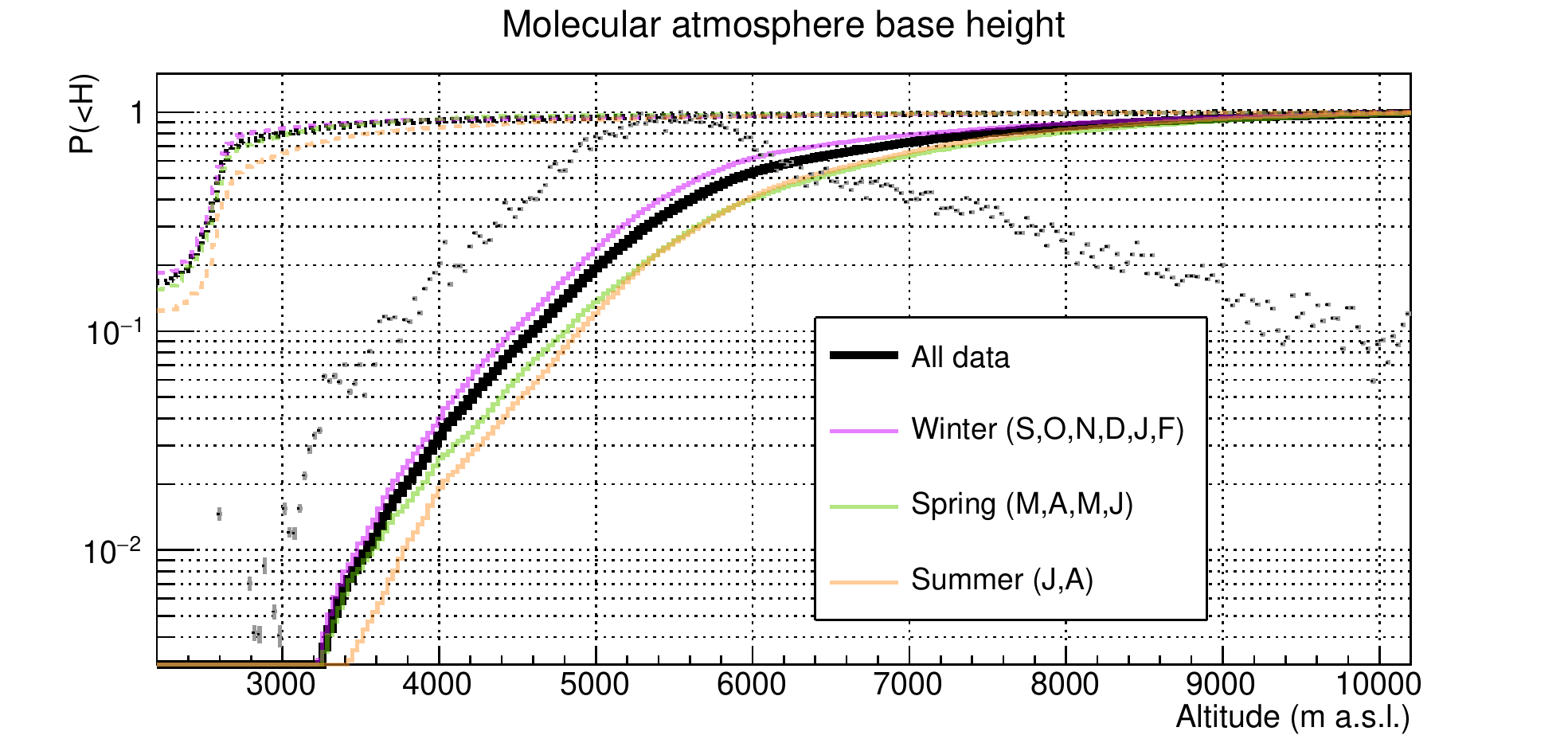}
  \caption{\label{fig:MAGIC_molheight} Cumulative probabilities for pure molecular atmosphere base heights (full lines) and the PBL heights using the gradient method~\protect\citep[][dashed lines]{Sicard:2006} during MAGIC data taking conditions (RH~$<$~90\%, wind gust $<$~40\,km/h) and weighted according to Eq.~\protect\ref{eq:weights}. Where the signal gradient coincides with the start range of full overlap of the LIDAR, a value of 2200~m has been artificially attributed to PBL height. The vertical lines refer to the medians of the molecular base height distributions. Gray dots show the differential probability of the full data set, scaled to reach a maximum of 1 for better comparability. Left: in linear scale, right: in logarithmic scale.
  }
\end{figure}

With an absolutely calibrated elastic LIDAR, a ground layer aerosol transmission statistics for observable night times can be established.  Figure~\ref{fig:ta} shows the estimated ground layer aerosol transmission as a function of time, for transmission values above 0.5. The two summer months July and August show clearly lower transmission on average, although low transmission due to dust intrusion is occasionally also possible during other months~\citep{lombardi2008}. At the same time,  high aerosol transmission, typical for clear nights, also frequently occurs during the summer months.

In order to quantify a probability of occurrence of a given aerosol transmission, averaged over the seven years of our data set, we \collaborationreview{first calculated} the month-wise (normalized) ground-layer probability of occurrence $P_{<T_\t{aer}}(m)$, obtained from those months with at least 30\% absolute dark-night coverage (Fig.~\ref{fig:coverage}).  This threshold has been chosen ad-hoc, in order to guarantee sufficient statistics for all months. 

The data from these months have been considered sufficiently representative for the entire month. The month-wise probability distributions were then weighted and added, according to:
\begin{align}
    P_{<T_\mathrm{aer}} &= \sum_{m=1}^{12} w(m)\,  P_{<T_\t{aer}}(m) ~ {\Big / } ~ \sum_{m=1}^{12}
    w(m) \quad \mathrm{with:} \nonumber\\
    w(m) &= \epsilon(m) D(m)~,
    \label{eq:weights}
\end{align}
where $D_m$ is the average night-time available for a given month and $\epsilon(m)$ the average monthly data taking efficiency of the MAGIC Telescopes. The former provides higher numbers for the winter months, whereas the latter gives a somewhat higher weight to the summer months (see also Fig.~9 of~\cite{garciagil2010}).

\begin{figure}
  \centering
  \includegraphics[width=0.65\linewidth, clip, trim=0cm 0 1cm 0]{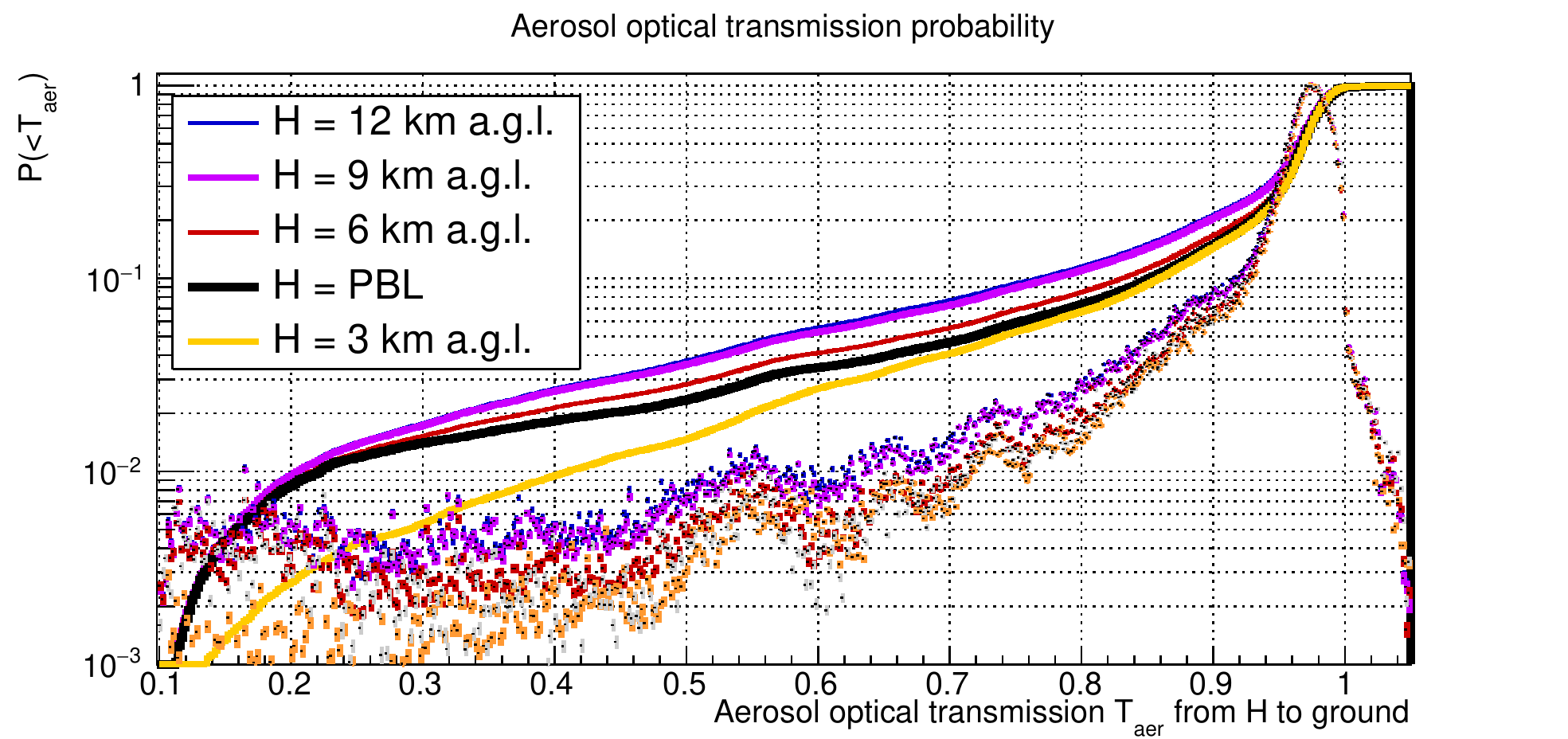}
  \caption{\label{fig:trans_full} Cumulative     probabilities for vertical aerosol transmission from different altitudes above ground, during MAGIC data taking conditions (RH~$<$~90\%, wind gust $<$~40\,km/h) and weighted according to Eq.~\protect\ref{eq:weights}.  The  dots show the differential probability of the full data set, scaled to reach a maximum of 1 for better comparability. \collaborationreview{Note that the blue curve (corresponding to 12~km above ground) is almost hidden behind the purple one (9~km above ground).} }
\end{figure}

Figure~\ref{fig:MAGIC_calima} shows the results for $P_{<T_\mathrm{aer}}$ for different sub-samples of the data: separated for Period~1 and~2 only, separated by seasons, and for the entire data set. At the level of occurrence probabilities of less than about 10\%, visible differences between the old and new laser periods appear, which are probably due to different criteria for aborting MAGIC Telescope data taking, which became stricter over time.  Such criteria include, among others, the use of the VLZA observation types, which has been introduced and become popular during the past years and entail abortion of the LIDAR data taking. Also aerosol transmission  $T_\mathrm{aer}<0.4$ has lead more often to abortion of  MAGIC Telescope science data taking, because of unacceptable loss of sensitivity.   

Hence the old laser data set shows higher contamination of very bad ground-layer transmission. Also statistical fluctuations may explain part of the difference. All in all, the observed spread of occurrence probabilities below 0.85~$T_\mathrm{aer}$ (ground layer) between Period~1 and~2 may serve to indicate the size of the systematic uncertainties on this statistical assessment. Furthermore, Fig.~\ref{fig:MAGIC_calima} makes apparent the worse atmosphere obtained during the summer months of July and August, where a lower median ground layer transmission is obtained, and sometimes transmission of green light drops below 70\%, due to the recurrent phenomenon of calima~\citep{Laken:2016}. The best conditions are observed, on average, during the Spring months of March, April, May and June.
 
Considering $T_\mathrm{aer} \approx 0.93$ as the rough transition  between the normally-distributed clear nights and others, we derive an overall occurrence of $\sim 80$\% for the clear night. This number is slightly lower than, but compatible with, the 84\%  occurrence of the TIL  for the Western Canary Islands, found by~\cite{Carrillo2016}.

Finally, Figure~\ref{fig:trans_full} displays the occurrence frequency of vertical aerosol transmission probabilities from different altitudes to ground. These numbers are needed to estimate the impact of aerosols and clouds on the Cherenkov light emitted by gamma-ray induced air showers, which ranges between the shown altitudes. The differences between 6\,km above ground and 9\,km stem from clouds in the field of view. One can also see that above 9\,km (11.2\,km a.s.l.), the effect of clouds becomes negligible.

Figure~\ref{fig:MAGIC_molheight} shows the occurrence frequency of molecular atmosphere base heights $h_t$ (see Eq.~\ref{eq:ht}), and, for comparison, the Planetary Boundary Layer, using the gradient method as used in~\citet{Sicard:2006}. The difference between both can be explained by the large exponential tails of the clear-night ground-layer aerosol density with altitude. One can also see that in about 25\% of the cases, the PBL could not even be determined with the gradient method, because the minimum of the first derivative coincides with the height at which full LIDAR overlap is reached, and is hence artificially determined by the latter. The distributions shown in Fig.~\ref{fig:MAGIC_molheight} are compatible with, but much more detailed than those obtained by~\citet[][see, e.g., their figure 5]{Sicard:2006}.

\subsection{Clouds}

The effect of clouds above the ORM \collaborationreview{has} been investigated in the past mainly by means of photometry measuring atmospheric extinction~\citep{garciagil2010}. Most of these (narrow-band filter) observations do not allow to distinguish between dust and clouds, however. Clouds found at altitudes higher than about 6000\,m a.s.l.\ 
are most often optically thin Cirrus, i.\,e., clouds formed of ice crystals. Lower, optically thick clouds above the ORM are normally of type Cumulonimbus and Altostratus.

\begin{figure}
  \centering
  \includegraphics[width=0.485\linewidth, clip, trim=0.3cm 0 1.5cm 0]{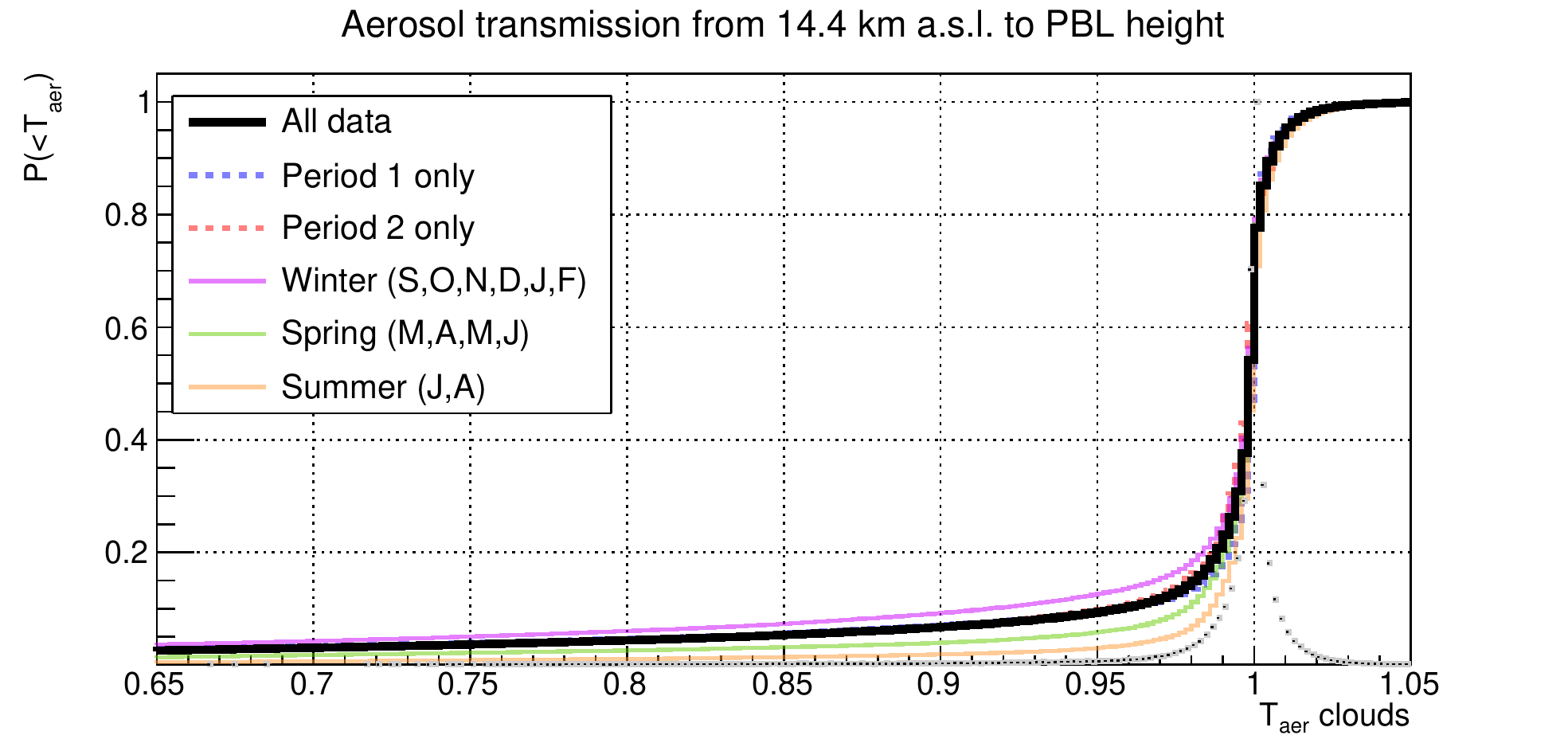}
  \includegraphics[width=0.485\linewidth, clip, trim=0.1cm 0 1.5cm 0]{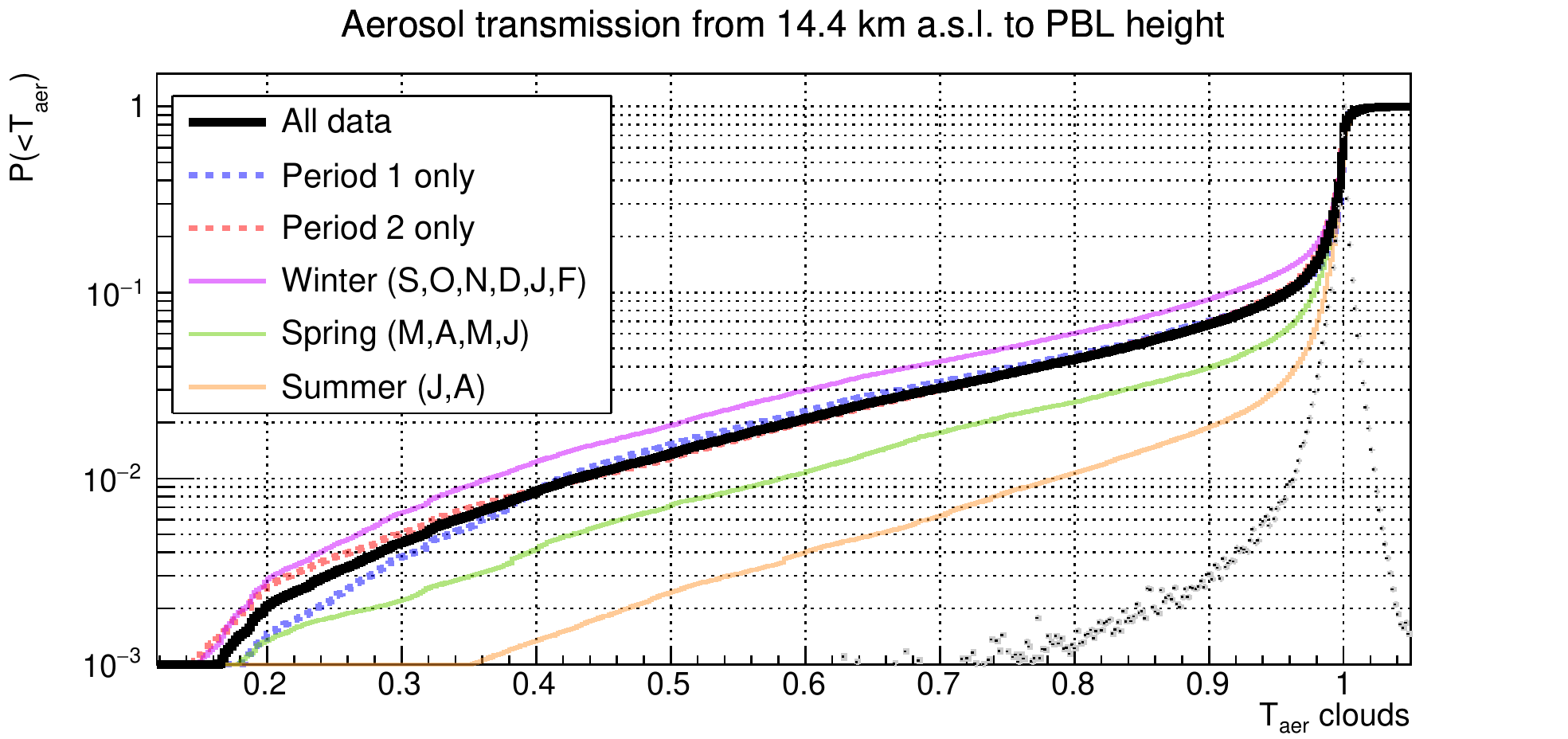}
  \caption{\label{fig:MAGIC_clouds} 
  Lines: cumulative vertical aerosol transmission probability of green light for  aerosols/clouds between 12~km above ground (14.4~km a.s.l.) and the molecular base height  during MAGIC data taking conditions (RH~$<$~90\%, wind gust $<$~40\,km/h) and weighted according to Eq.~\protect\ref{eq:weights}. Gray dots show the differential probability of the full data set, scaled to reach a maximum of 1 for better comparability. Left: in linear scale, right: in logarithmic scale. Note the different horizontal axis ranges.}
\end{figure}

Our 7-years data sample contains $\sim$16500 clouds, of which 92\% are composed of only one single cloud layer, 8\% of two layers, separated by a pure molecular atmosphere part. More than two layers have been found, but are very rare ($<$0.4\%) and all concentrated in the month December. Our sample is, however, strongly biased towards nights free of Cumulonimbus, because the MAGIC Telescopes and its auxiliary LIDAR normally stop operations under these conditions.

Figure~\ref{fig:MAGIC_clouds} shows the cumulative vertical aerosol transmission probabilities from 12\,km above ground to the base height of the pure molecular atmosphere, i.e. the part of the troposphere not affected by the ground layer. Roughly one eighth of the data taken would then need a correction for clouds, a value compatible with the relative amount of photometric nights of 84\%, claimed by~\cite{Erasmus:2006} from satellite observations. The occurrence probability of clouds is higher in winter and lowest in summer, with intermediate values for the Spring months. The MAGIC LIDAR does not operate when MAGIC observations are aborted due to a sky completely covered by optically thick clouds or during rain. For that reason, the statistics shown in Fig.~\ref{fig:MAGIC_clouds} is biased towards lower cloud occurrence probabilities and is only representative for standard astronomical observation conditions.  On the other hand, a comparison between old laser Period~1 and new laser Period~2 shows excellent agreement for this parameter, an indication  that the conditions for aborting MAGIC data taking have remained stable over time.

\begin{figure}
  \centering
  \includegraphics[width=0.485\linewidth,clip,trim=0.3cm 0 1.5cm 0]{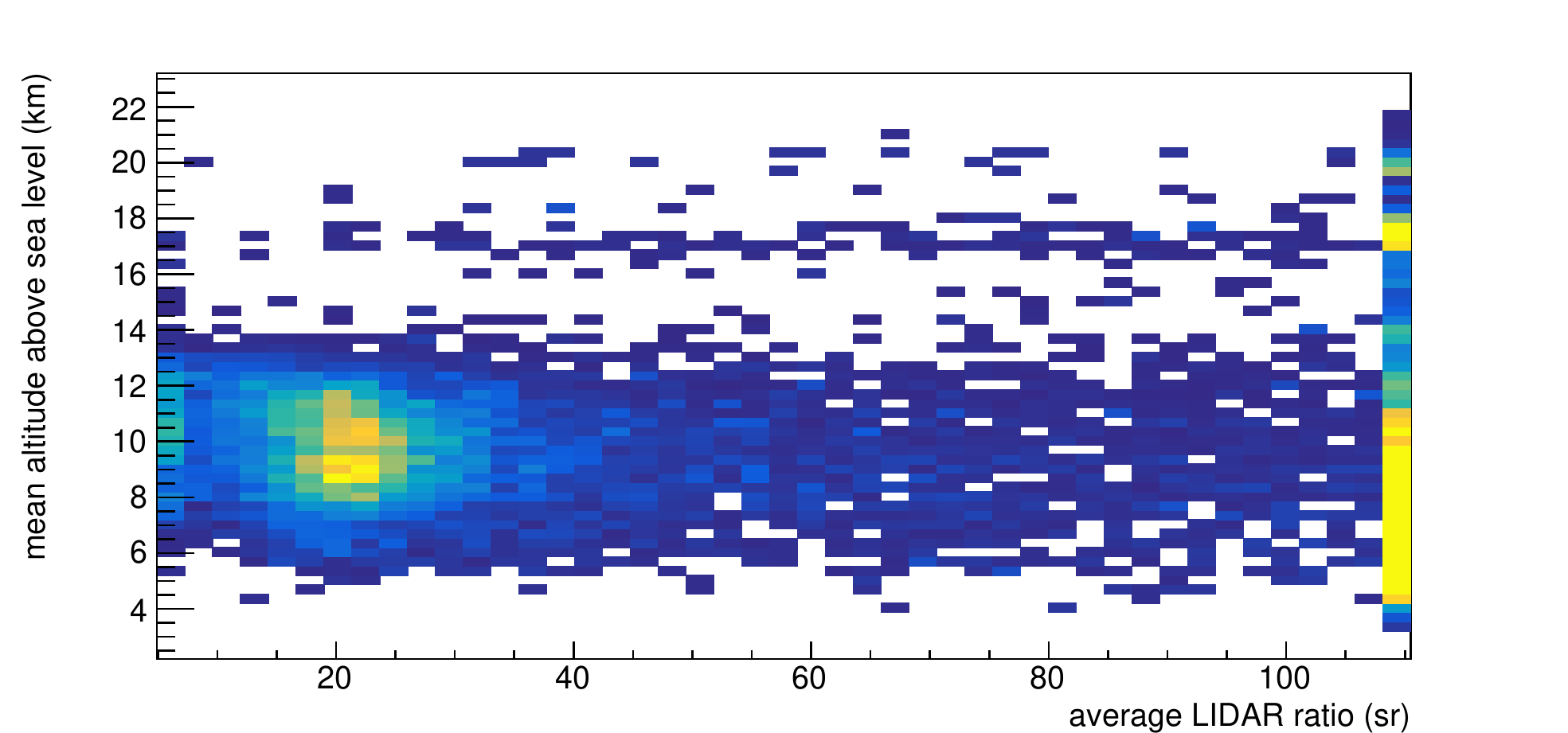}
  \includegraphics[width=0.485\linewidth,clip,trim=0.1cm 0 1.5cm 0]{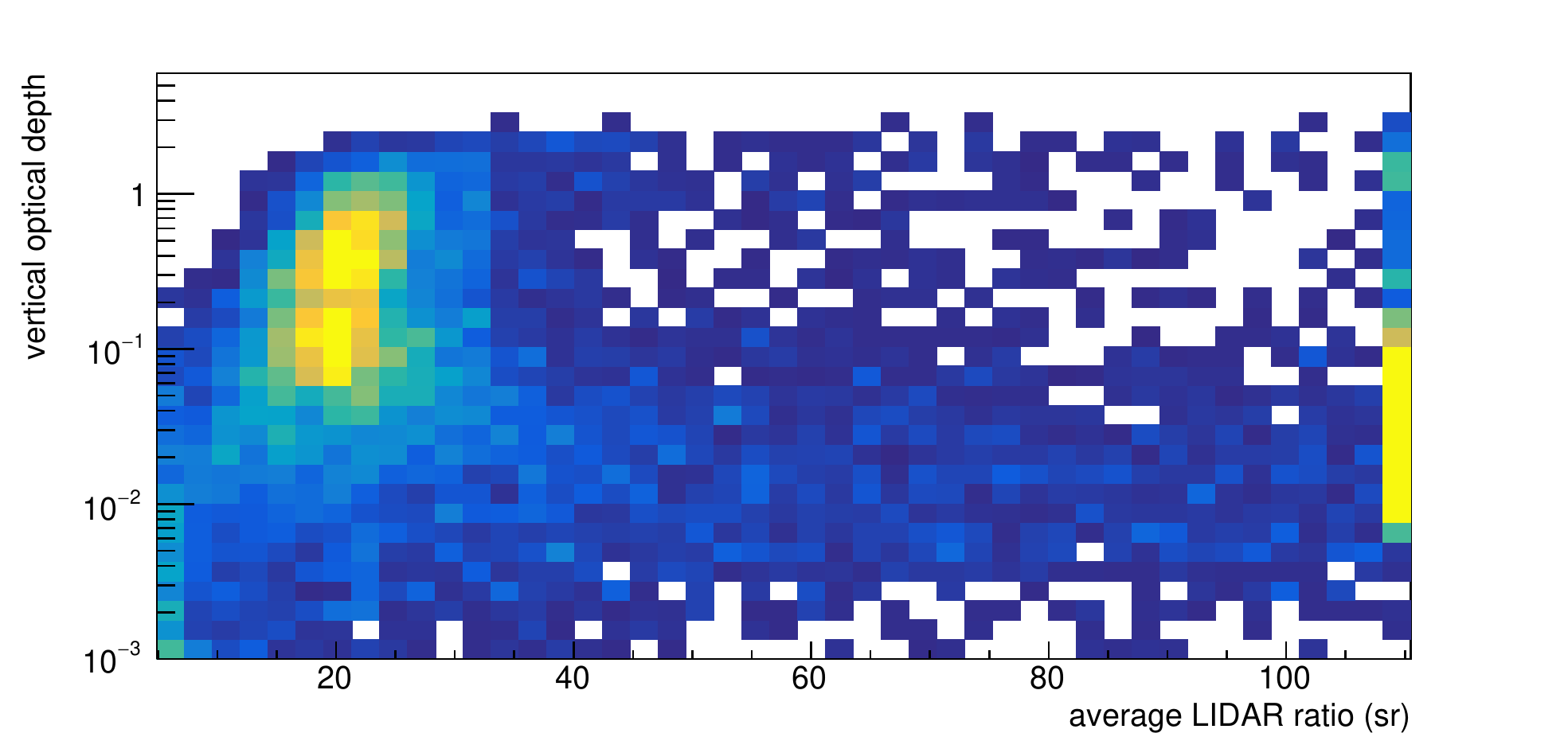}
  \caption{\label{fig:MAGIC_clouds_LR} Distribution of cloud mean altitudes (left) and vertical optical depths (right) vs. reconstructed average LIDAR ratios. LIDAR ratios smaller than 5\,sr have been collapsed into the first bin and larger than 110\,sr into the last bin.}
\end{figure}

The retrieved cloud is contaminated, on one hand, by small temperature fluctuations~\citep[see, e.\,g.,][]{Carrillo2016}, which have not been included in the molecular profile model and may be misinterpreted as a cloud of very small optical depth, and, on the other hand, by clouds only partially illuminated by the laser. This happens if the LIDAR moves away from the cloud during its 100~seconds long data taking sequence. Visual inspection of many LIDAR profile sequences have revealed that a cloud entering the field-of-view produces such artificially increased LIDAR ratios, which then remain constant at a physically possible value, and finally increase again, as the cloud slowly disappears. Finally, thin clouds with fine substructures may cover only parts of the field-of-view of the LIDAR. All these cases lead to artificially increased reconstructed LIDAR ratios. Figure~\ref{fig:MAGIC_clouds_LR} shows mean cloud altitude and optical depth  as a function of the reconstructed average cloud LIDAR ratio. As one can observe, the altitude of clouds does not show any dependency on the reconstructed LIDAR ratio, but instead the optical depth decreases as the LIDAR ratio increases. There is a transition from one behaviour to the other at a LIDAR ratio of about 40\,sr, above which vertical optical depths above 0.1 are only rarely found. For these reasons, we will characterize, in the following, only those parts of our data sample, which have a reconstructed LIDAR ratio below 40\,sr. Finally, LIDAR ratios below 7\,sr are accompanied with predominantly very low optical depths and interpreted as temperature fluctuations. These data have also been removed from the sample.

\begin{figure}
  \centering
  \includegraphics[width=0.485\linewidth,clip,trim=0.3cm 0 1.5cm 0]{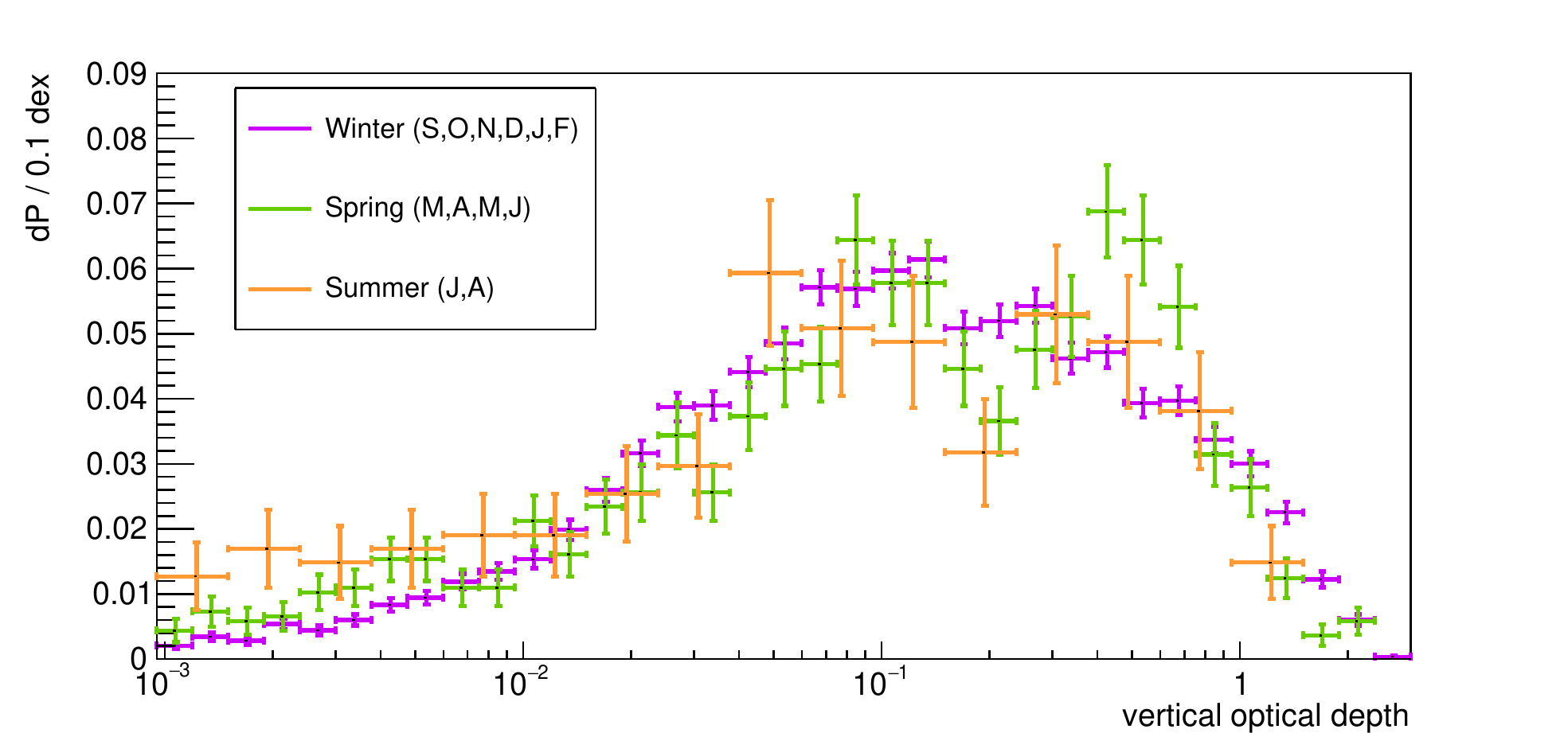}
  \includegraphics[width=0.485\linewidth,clip,trim=0.1cm 0 1.5cm 0]{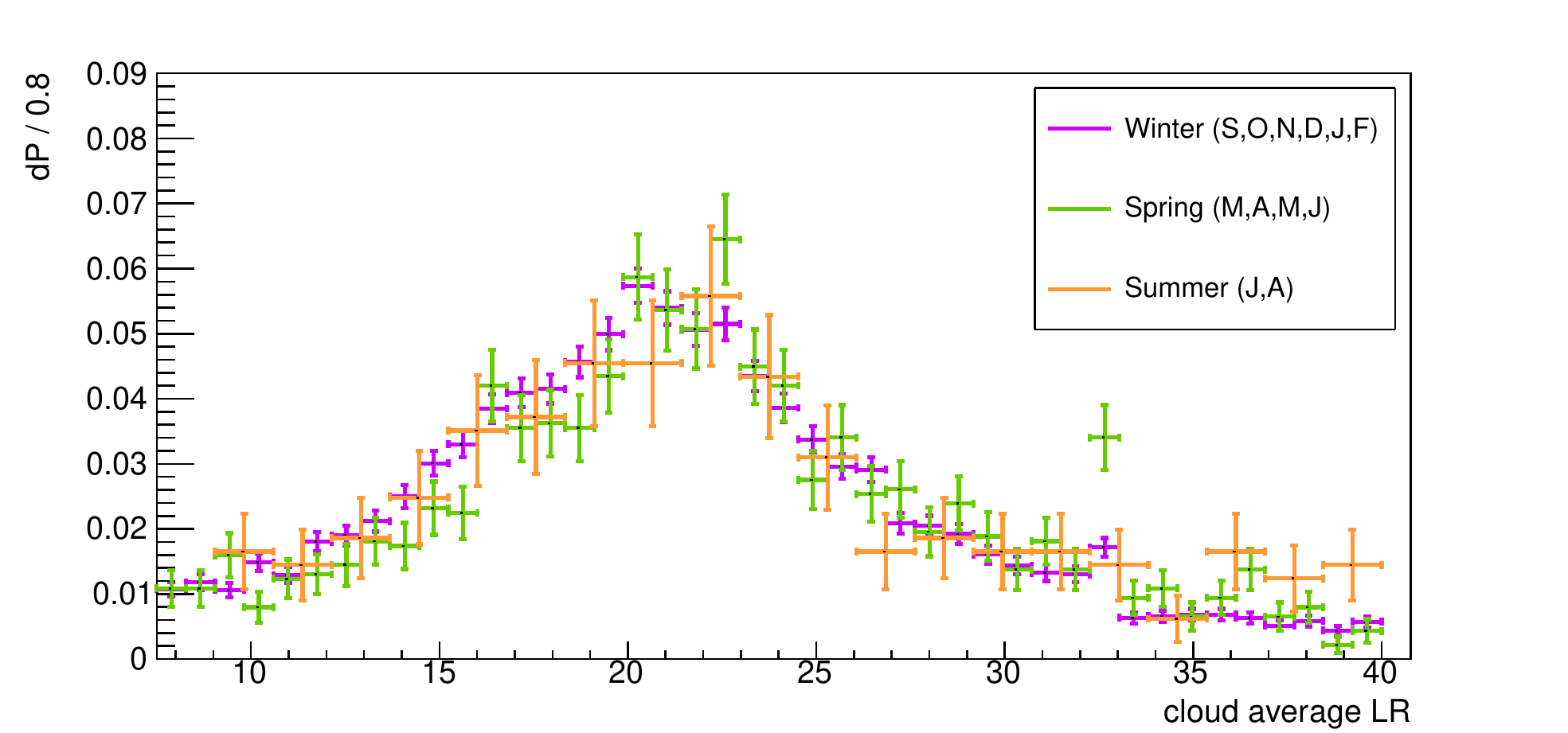}
  \includegraphics[width=0.485\linewidth,clip,trim=0.3cm 0 1.5cm 0]{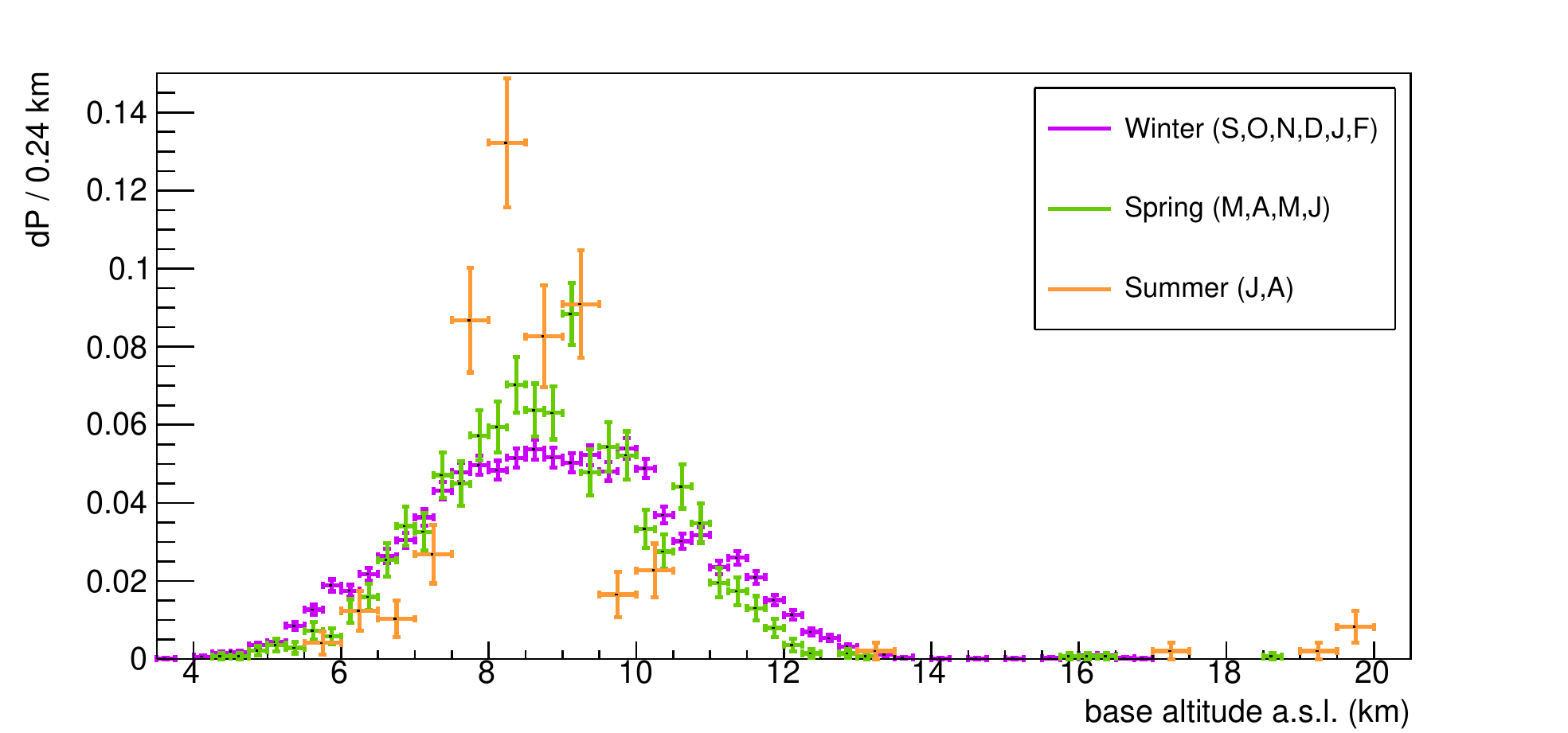}
  \includegraphics[width=0.485\linewidth,clip,trim=0.1cm 0 1.5cm 0]{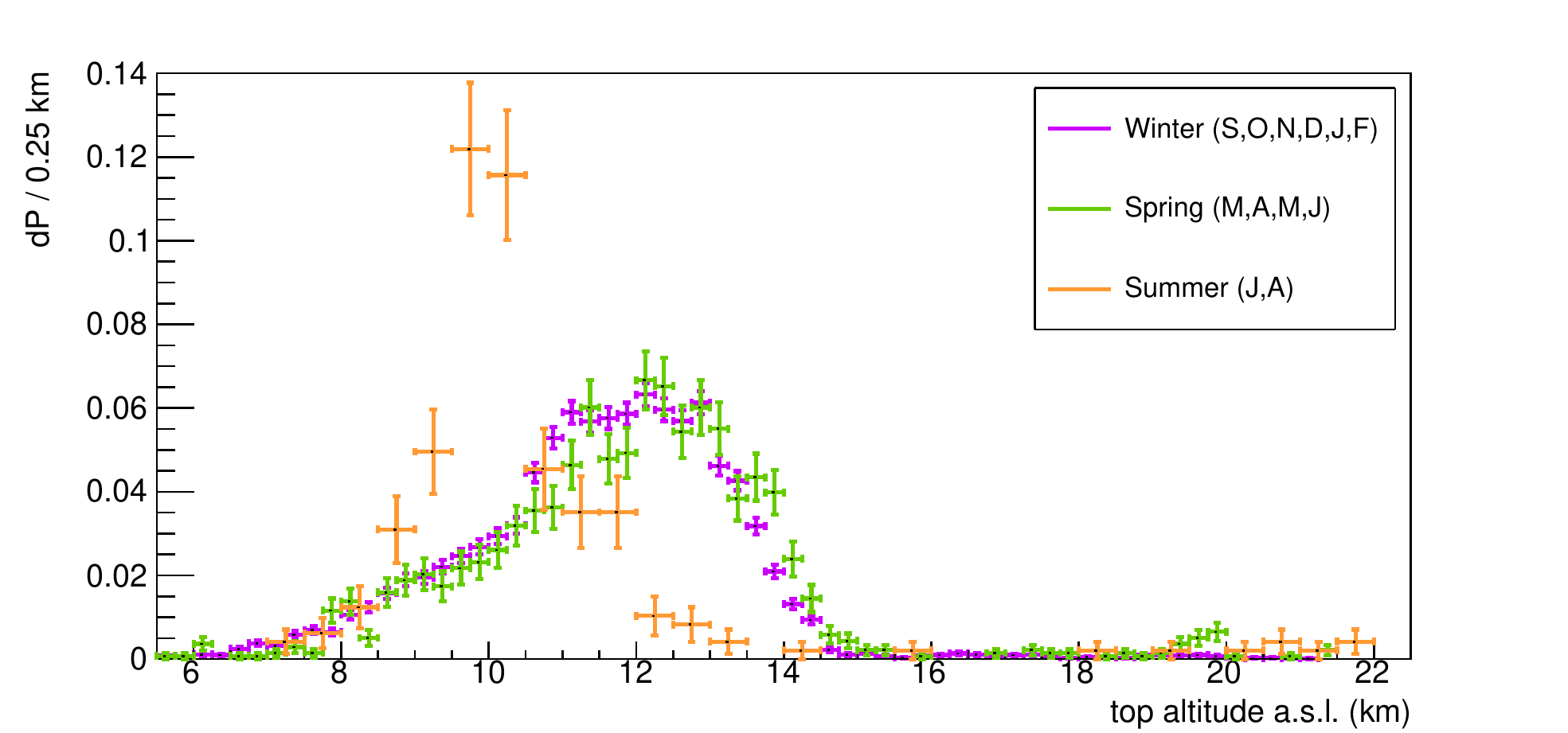} \caption{\label{fig:MAGIC_clouds_OD} Distributions of cloud parameters for the sample with reconstructed LIDAR ratios larger than 7\,sr and smaller than 40\,sr. Upper left: vertical optical depths, upper right: LIDAR ratios, lower left: cloud base altitudes and lower right: cloud top  altitudes.}
\end{figure}

Figure~\ref{fig:MAGIC_clouds_OD} (top) shows the distribution of vertical optical depths and LIDAR ratios from the remaining cloud sample. The LIDAR ratios are centered at 21\,sr, with a width of 6\,sr, whereas the vertical optical depths (VOD) show a multi-modal distribution, with centers at 0.09 and 0.5. These values are consistent with expectations for C3 and C2 type Cirrus~\citep{gouveia,keckhut2013}, with a higher relative prevalence of C2-type cirrus in spring. Our system is, though, not particularly sensitive to subvisible cirrus~\citep[SVC,][]{reverdy}, which form part of the sample with VOD$<$0.01 and are detected particularly in summer, but are certainly under-represented here. Cloud base and top altitudes are shown in the lower graphs of Fig.~\ref{fig:MAGIC_clouds_OD}. The summer data show clouds with base altitudes stronger concentrated around 8\,km a.s.l.\ and top altitudes around 10\,km a.s.l. Such summer clouds display smaller geometric thicknesses than the winter and spring clouds, which reach higher up into the tropopause, on average. Spring and winter data show cloud top heights reaching up to altitudes between 8\,km and 14\,km, reproducing very closely the distribution of the lower part of (multiple) thermal tropopause event heights, as found in Fig.~5 of~\citet{rodriguez2013}.

\begin{figure}
  \centering
  \includegraphics[width=0.99\linewidth]{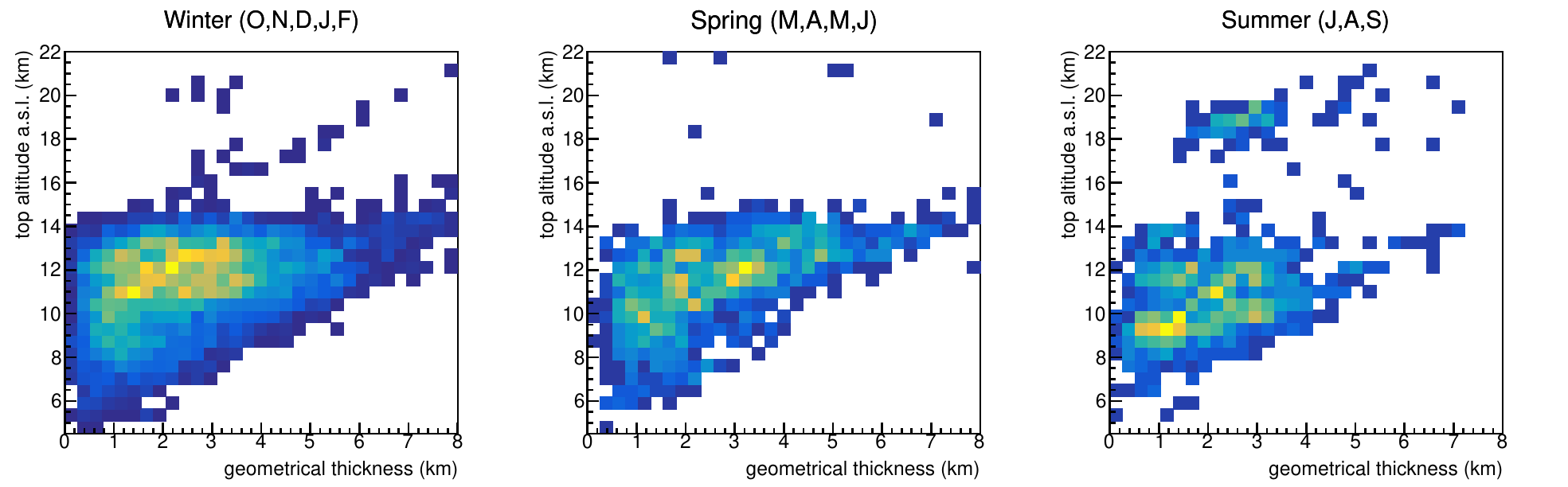}
  \caption{\label{fig:MAGIC_cloudseasons} Distribution of reconstructed cloud top altitudes vs.\ geometric thickness. The figures have been separated into Winter (left, comprising October, November, December, January, February), Spring (March, April, May, June), and Summer (July, August, September). Note that in this case and unlike previously, September has been included in the summer data sample, because of its clear preference for high cloud top altitudes (see also Fig.~\protect\ref{fig:MAGIC_clouds_BaseTop}, left).}
\end{figure}

\begin{figure}
  \centering
  \includegraphics[width=0.485\linewidth,clip,trim=0.3cm 0 1.5cm 0]{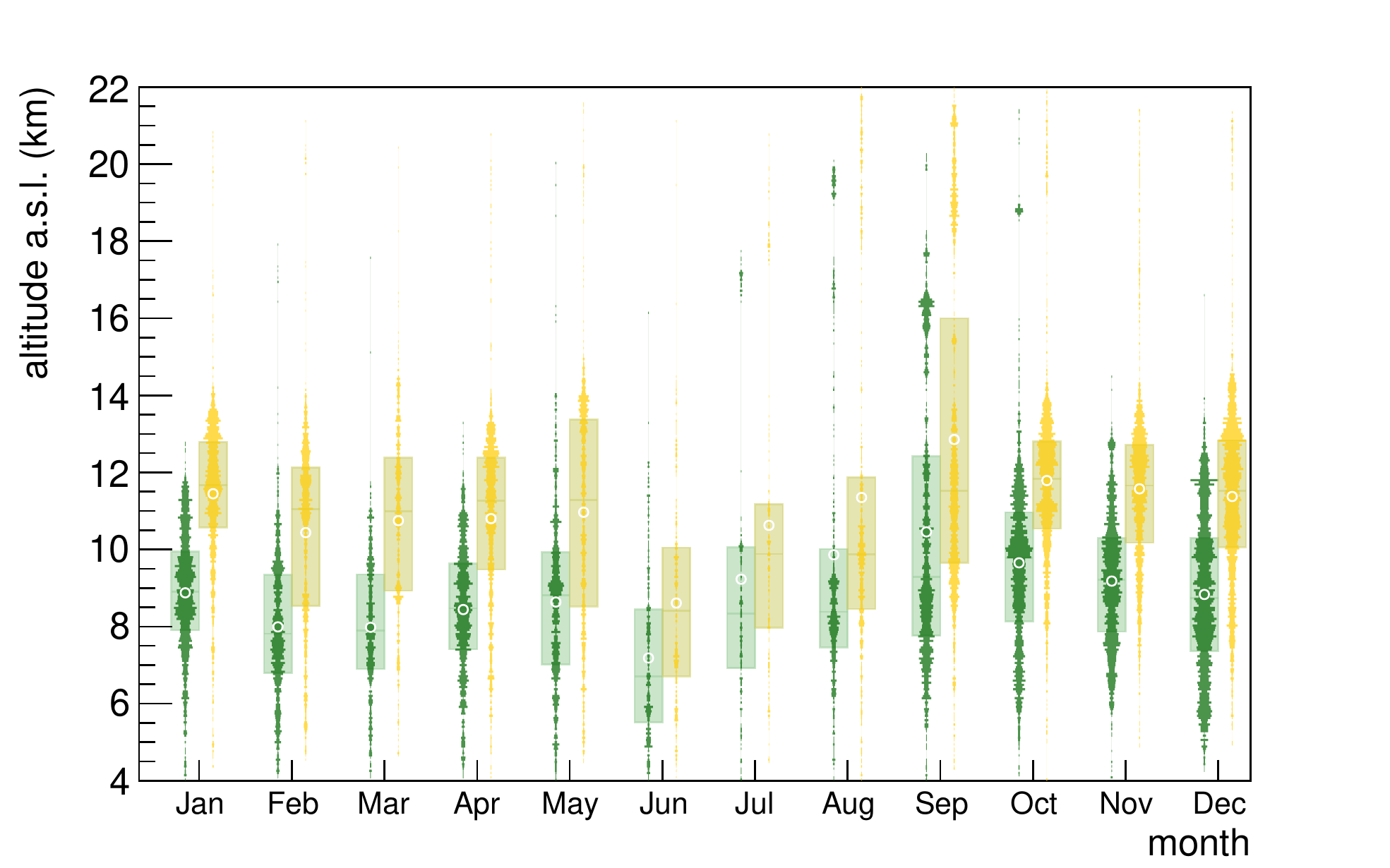}
  \includegraphics[width=0.485\linewidth,clip,trim=0.1cm 0 1.5cm 0]{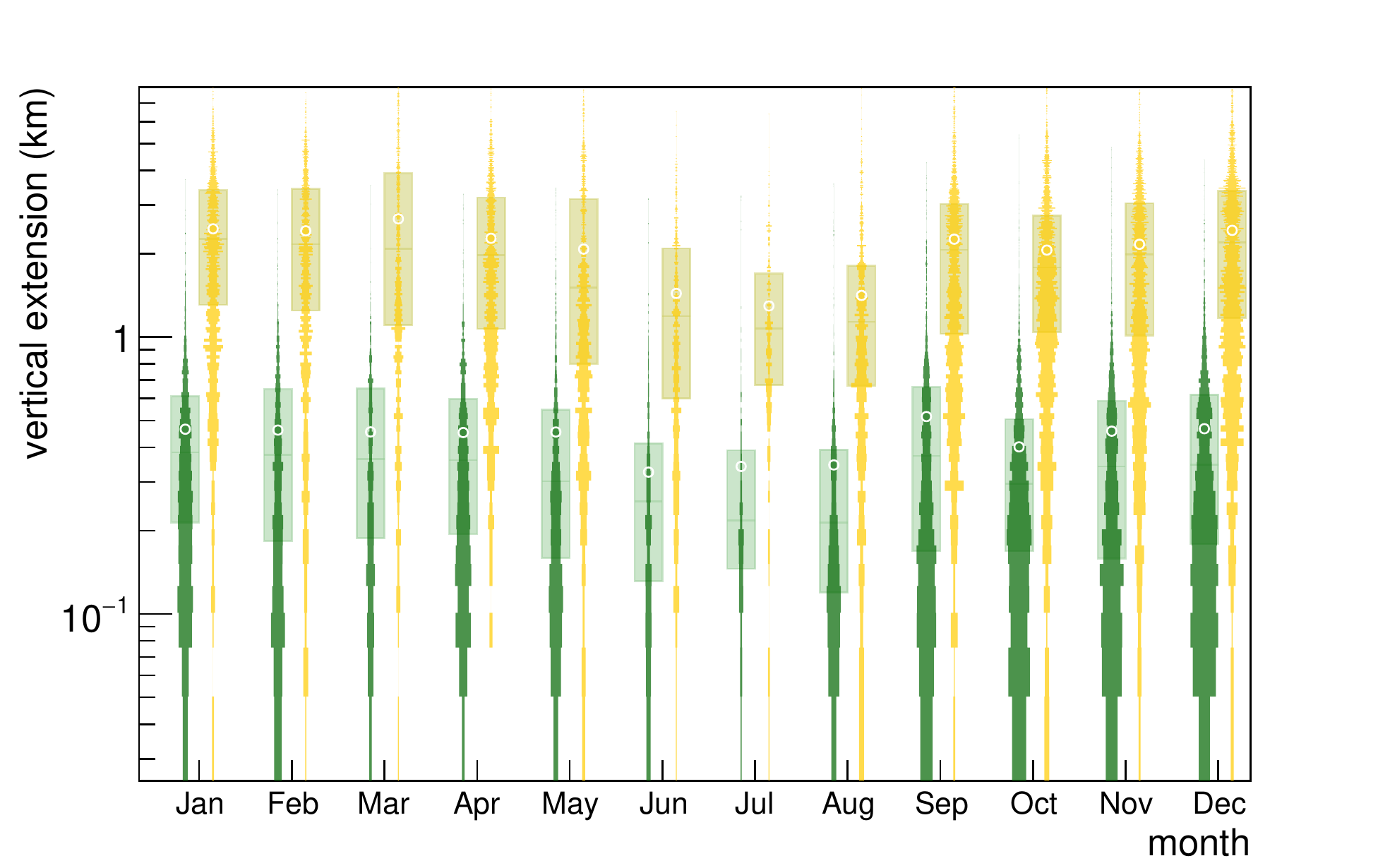}
  \caption{\label{fig:MAGIC_clouds_BaseTop} Left: seasonal cycle of cloud base heights (green) and top heights (yellow), right: seasonal cycle of standard deviations of clouds extinction-weighted vertical profile (green), and geometric thickness (cloud top minus cloud base height, yellow). The overlaid bars show the 25\% to 75\% percentiles of the distributions. Note the logarithmic scale of the right-hand graph.}
\end{figure}

Extremely high clouds above 14~km base height are occasionally found across all seasons and compatible with the thermal tropopause height distribution for single tropopause events found in \citet{rodriguez2013}. Single tropopause events occur at a  frequency of around 20\% in winter, rising to almost 100\% in summer \citep[see Fig.~3 of][]{rodriguez2013}, with intermediate occurrence frequencies in spring and autumn. Such large prevalence is not reproduced by our cloud top height distribution. On the contrary, summer clouds normally reach up to \textit{lower} altitudes between 8\,km and 12\,km, despite of the existence of a single tropopause well above 14\,km, but on rare occasions do reach up to extreme altitudes of up to 22~km~a.s.l. It is not surprising that the high cloud example of Fig.~\ref{fig:highcloudexample} was observed during August. The influence of a very high single tropopause becomes more evident in Figure~\ref{fig:MAGIC_cloudseasons}, which shows the reconstructed mean altitudes of clouds against the vertical optical depth for the full cloud sample (i.\,e., including those reconstructed with unphysical LIDAR ratios), separated into three seasons. Here,  clouds moving up to the single thermal tropopause above 14\,km a.s.l.\ occurs predominantly during the summer months, whereas winter and spring clouds are normally trapped by the lower part of a double or triple \collaborationreview{tropopause} temperature inversion.

Figure~\ref{fig:MAGIC_clouds_BaseTop} highlights the seasonal cycle of the basic cloud parameters, base and top heights and geometric thickness. Also the standard deviation of the vertical cloud extinction profile is shown.  One can see that lower and thinner clouds occur during the months June, July and August, whereas the extremely high altitude clouds concentrate in September and August. Geometric thicknesses are larger during winter and spring (median thicknesses between two and three kilometers) and lower during the months June, July and August (with medians between 1 and 1.5~km). At the same time, the spread between geometric thickness and  standard deviation of the vertical cloud extinction profile diminishes during these three summer months, indicating a more homogeneous distribution of extinction along the vertical profile. On the contrary,  winter and spring clouds can show a stronger concentration of extinction within the cloud. This is particularly striking for the winter months September, October, November, December and January, where very small standard deviations of the vertical cloud extinction profile are possible, down to less than several tens of meter, accompanied by geometrical thicknesses of several hundreds of meters to one kilometer.

%% file: nsb.tex
\subsection{Night-sky brightness}


Any LIDAR is able to measure, and subtract from its laser return signals, a corresponding contribution from the night-sky brightness. In the case of the MAGIC LIDAR, the 
background light is measured as the median photo-electron rate from the region before the laser return signal becomes visible
(see the number of ADC slices used in the 6$^\mathrm{th}$ column of Table~\ref{tab:hardware_components}). Our LIDAR is perfectly suited to measure photo-electron rates falling through a well-defined diaphragm and from a known solid angle, due to the use of an HPD with high quantum efficiency and charge resolution. This function is decoupled from the laser shooting of the LIDAR.

The obtained photo-electron background rate depends on the light of night sky (LoNS), influenced by the presence of the moon, star fields, zodiacal light and possible anthropogenic contributions. The measured rate depends on the mirror reflectance and HPD photon detection efficiency. Also the size of the diaphragm in front of the receiver optics has changed with time and causes changes in the observed background. In order to use the background rates for a characterization of the night sky background conditions at the site, the instrumental contributions need to be corrected for ageing and hardware changes. We did so using the degradation proxy, separately for the two laser periods. Here we have implicitly assumed that the laser power within both periods has remained stable over time. 
 We checked that the median NSB has remained constant after correction with the degradation proxy justifying the previous assumption.

In the following, we show the effect of each contribution, after removing all data taken during twilight, i.e. keeping only data taken under astronomical darkness (solar zenith angle $\theta_\odot > 108^\circ$). Also Galactic observation fields have been excluded requiring Galactic latitudes $l >10^\circ$.

\begin{figure}
  \centering
  \includegraphics[width=0.45\linewidth]{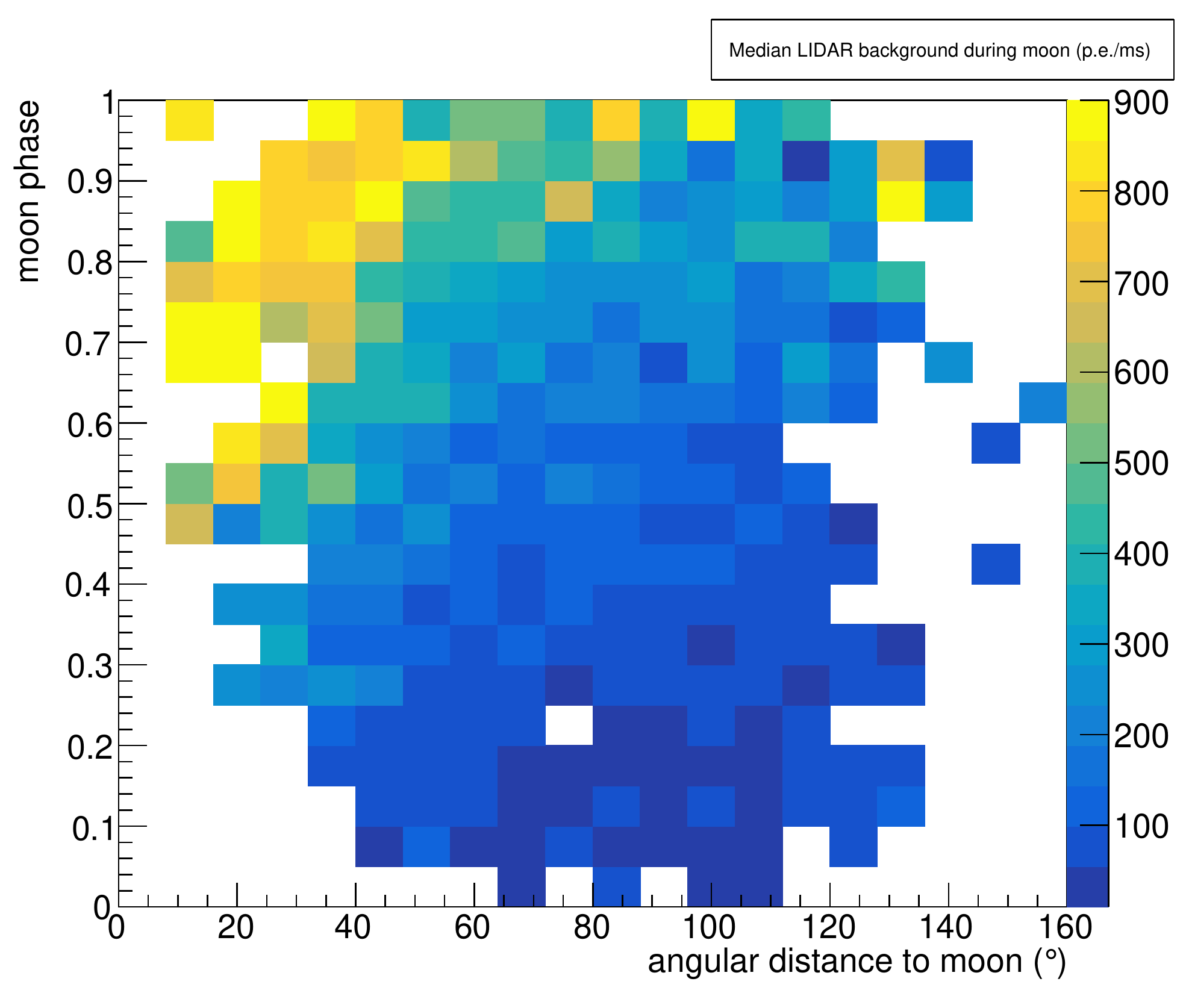}
  \hspace{0.5cm}
  \includegraphics[width=0.45\linewidth]{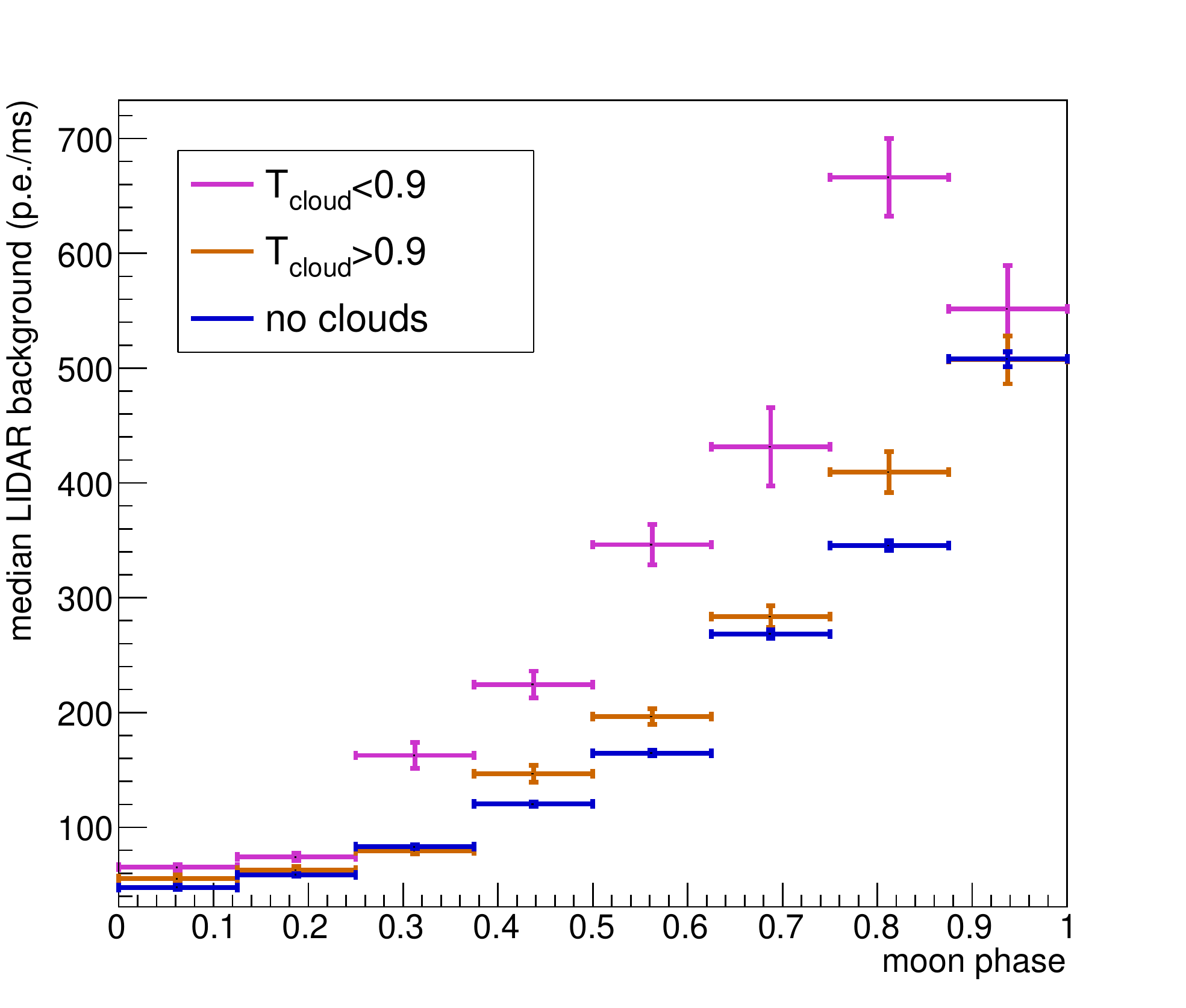}
  \caption{\label{fig:moonbg} Left: Median LIDAR background rates observed as a function of angular distance to the moon and the moon phase for cloudless nights. Right: median background rates as a function of moon phase, for data with cloud base higher than 2~km and cloud top lower than 12~km above ground. The cloud data has been divided into those with cloud transmissions higher and lower than 0.9. For comparison, also the cloudless data are shown. Only angular distances between 40$^\circ$ and 110$^\circ$ from the moon have been used here, in order to ensure a similar coverage of all angular distance and moon phase bins. Galactic star fields and those affected by zodiacal light had been excluded previously. }
\end{figure}

Figure~\ref{fig:moonbg} shows the median background rates in the presence of moon: \collaborationreview{as} the LIDAR points closer to the moon and the moon becomes brighter, background rates increase by up to a factor of 20. Even larger increases have been observed, but led to invalid LIDAR data. Fig.~\ref{fig:moonbg} also highlights the role of clouds on the background photo-electron rates: As clouds cover the LIDAR field-of-view, the background rates \textit{increase}, with optically thicker clouds leading to more background light. This behaviour can be explained by back-reflection of moon light from the clouds.  

\begin{figure}
  \centering
  \includegraphics[width=0.45\linewidth]{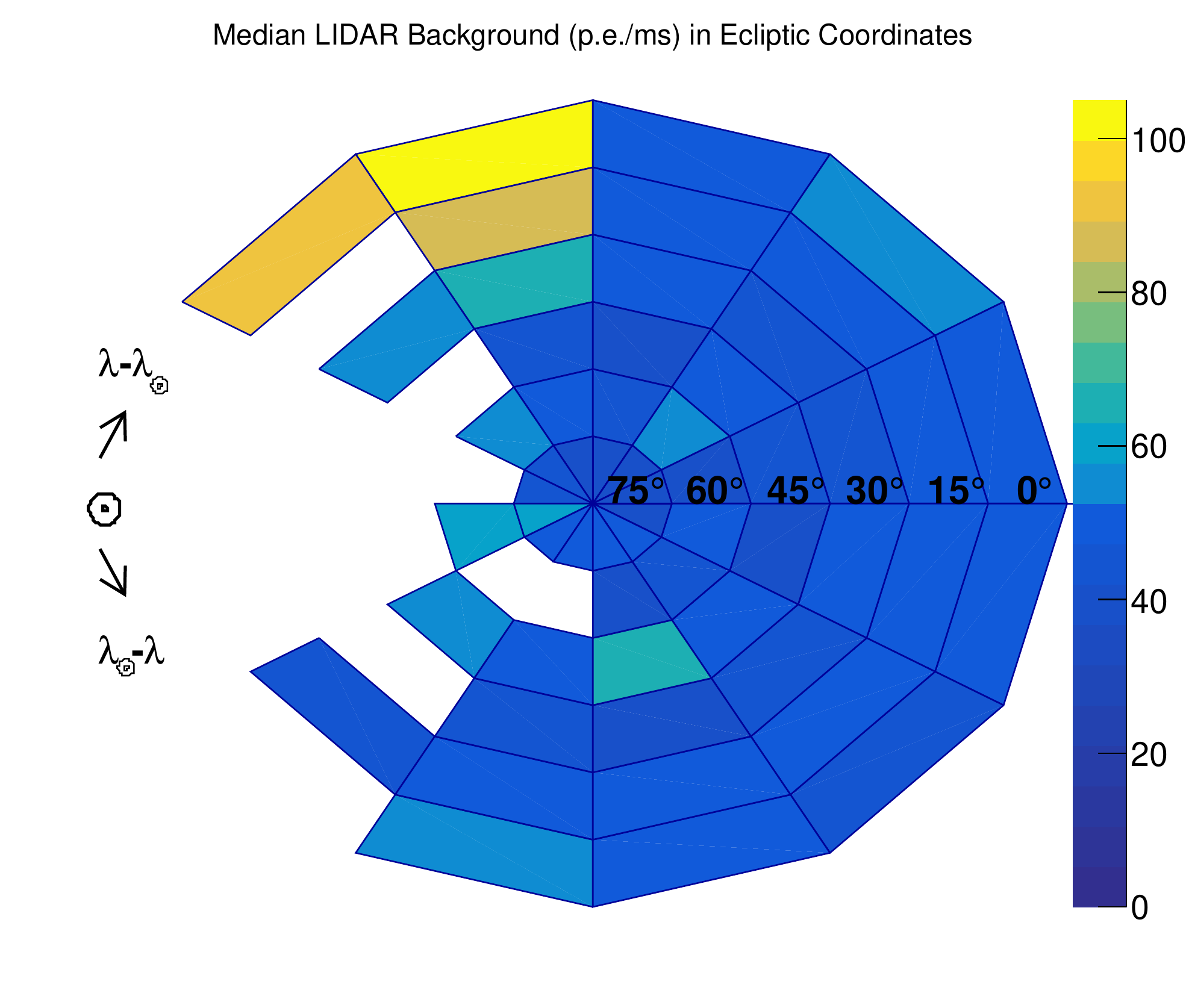}
  \includegraphics[width=0.45\linewidth]{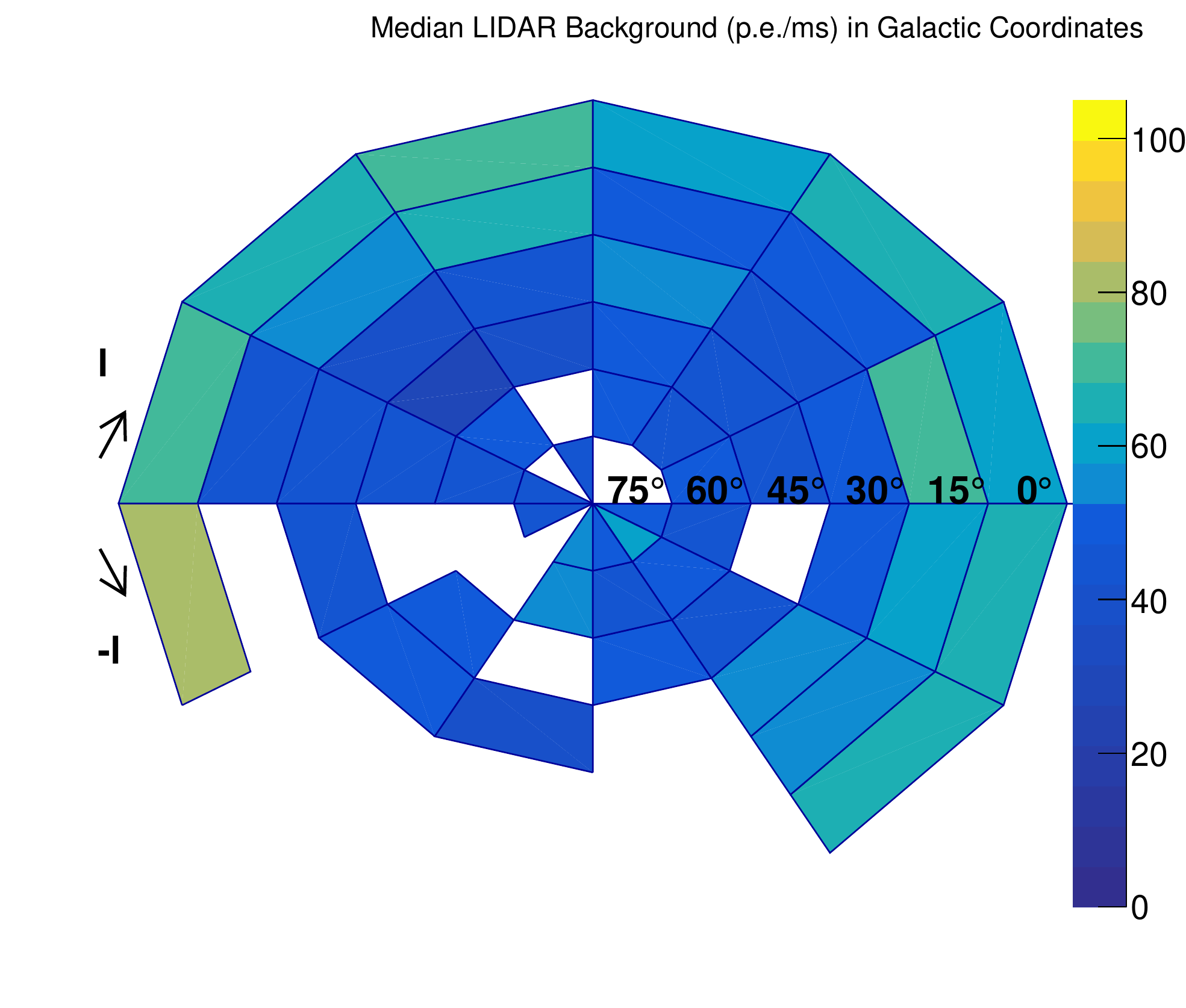}
  \caption{\label{fig:hrbg}
  Observed median LIDAR background rates in ecliptic (left) and galactic (right) coordinates. The Sun (Galactic Center) is located at the left, the ecliptic (galactic plane) is represented by the outer circle, whereas the ecliptic (galactic) pole is found in the center. The numbers show absolute ecliptic (galactic) latitudes. Moon, twilight data  have been excluded previously, as well as data taken before 23~h UTC time. In the left plot, also galactic fields have been excluded, whereas in the right plot zodiacal light has  been removed. Empty fields denote lack of data.
  }
\end{figure}

After removing data contaminated by moon, we display the impact of zodiacal light~\citep{LevasseurZodiacalLight:1980,leinert,May:PhD} on the background rates. Figure~\ref{fig:hrbg} (left) shows the median background rates in ecliptic coordinates. An increase of up to 150\% background light is observed, predominantly for the evening zodiacal light. In the following, we remove data with ecliptic latitudes smaller than 30$^\circ$ and longitude differences w.r.t. the sun smaller than 90$^\circ$. The right side of Fig.~\ref{fig:hrbg} shows the median background rates in galactic coordinates. The galactic plane is seen clearly brighter, particularly the Galactic Center, where an increase of background light of the order of 100\% w.r.t. to the extra-galactic fields is observed. For the following studies, data with absolute galactic latitudes smaller than 10$^\circ$ have been excluded from the data.

\begin{figure}
  \centering
  \includegraphics[width=0.45\linewidth]{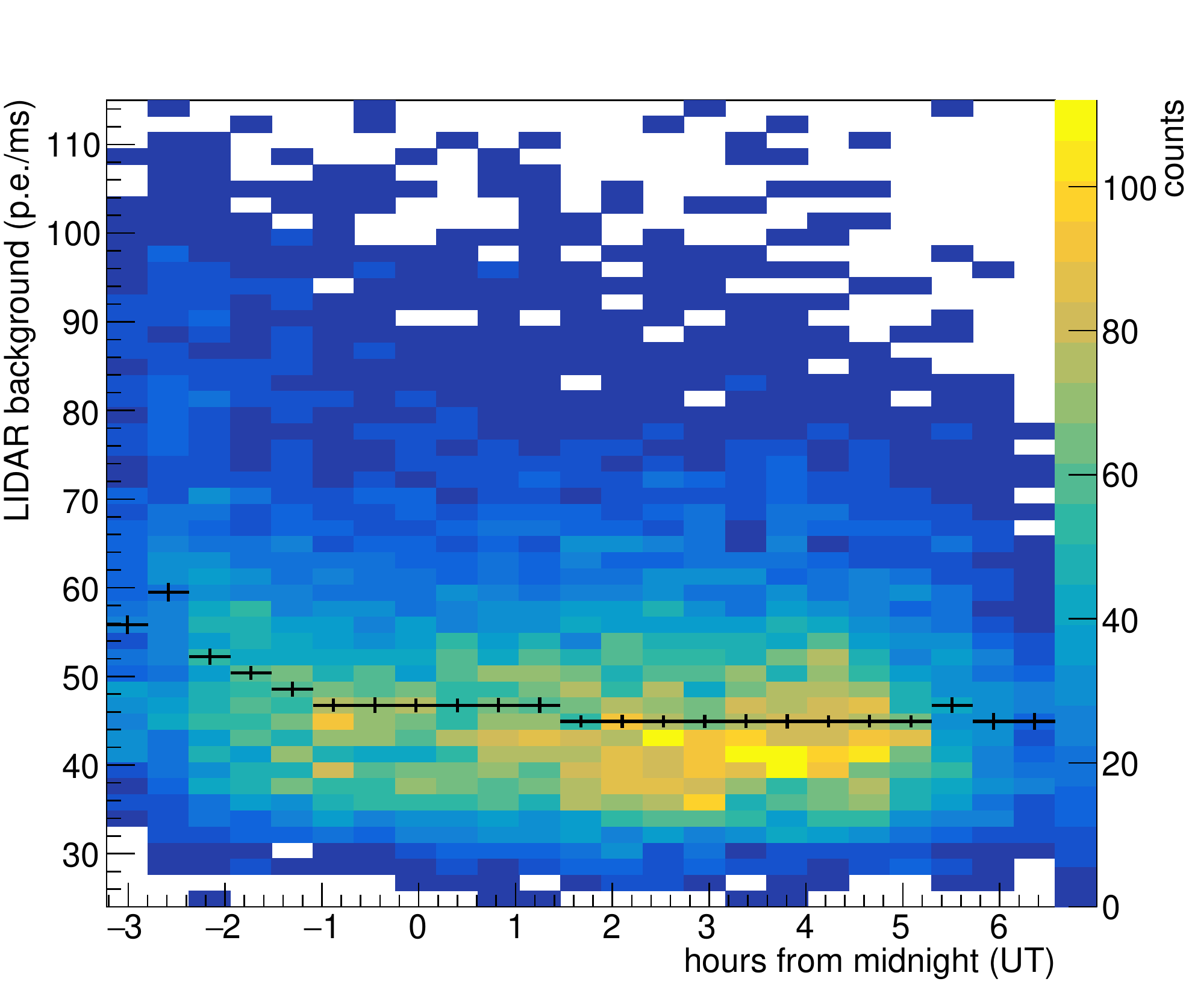}
  \caption{\label{fig:hrbg2}
  Observed LIDAR background rates, as a function of local time. Moon, twilight, zodiacal light and galactic fields have been excluded previously.}
\end{figure}

\begin{figure}
  \centering
  \includegraphics[width=0.99\linewidth]{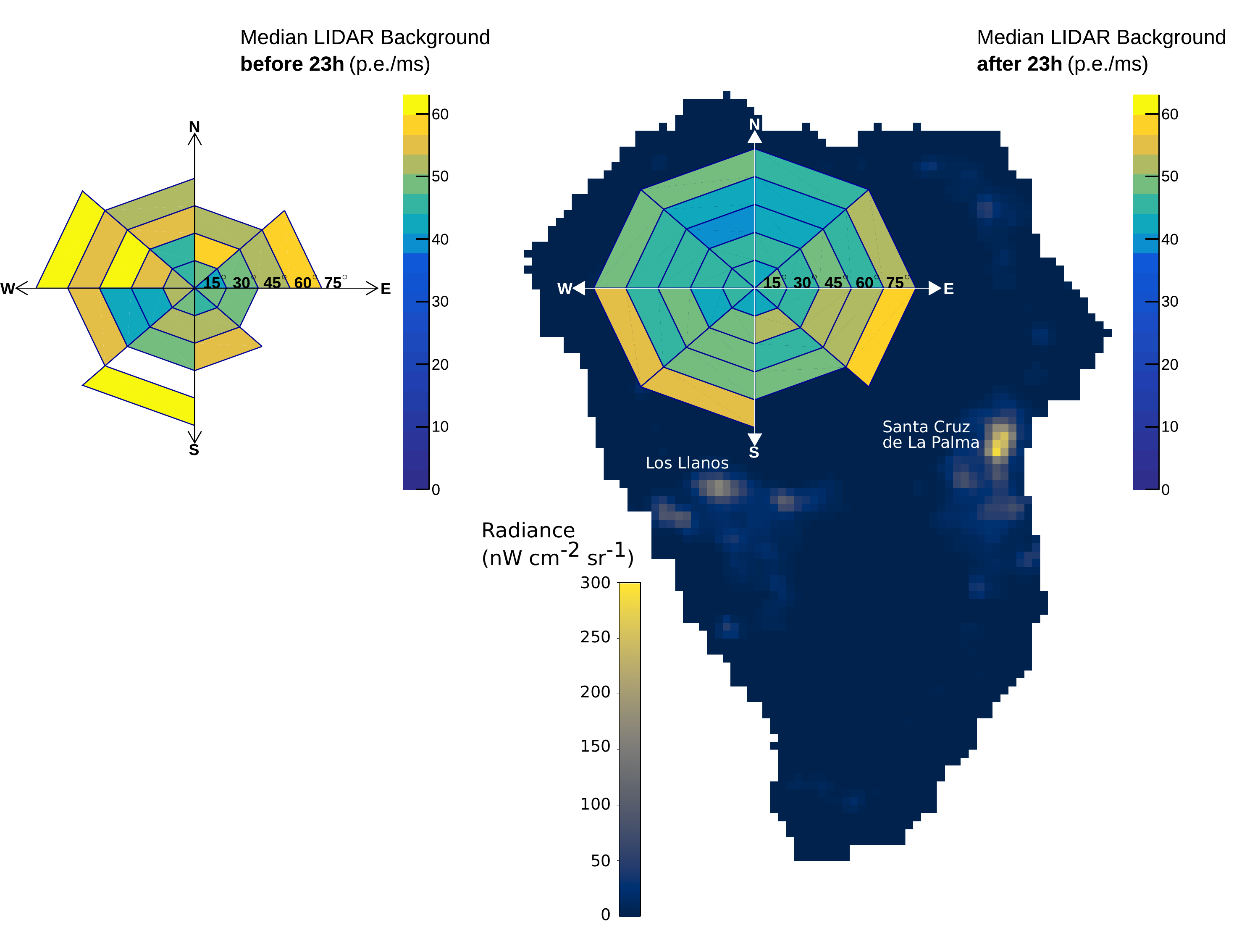}
  \caption{\label{fig:bg}
  Observed median LIDAR background rates in local coordinates. Left: only data before 23h UT, right: after 23h UT. Moon, twilight, zodiacal light and galactic fields have been excluded previously. Empty fields contain no data. \collaborationreview{In addition, the averaged and bidirectional reflectance distribution function (BRDF) adjusted temporal radiance during snow-free periods in 2020 is shown \citep{roman_mo_viirsnpp_2021}. These satellite observations were performed using the panchromatic ($500$\,nm -- $900$\,nm) Day-Night-Band (DNB) of the Visible Infrared Imaging Radiometer Suite (VIIRS) onboard the Suomi-National Polar-orbiting Partnership (Suomi-NPP). All Suomi-NPP data were gathered after 23h UT as a consequence of its Sun-synchronous orbit (see \citeauthor[ Historical Predications of NPP Trajectories]{noaastarncc_historical_nodate}).
  } }
\end{figure}

The remaining sample is left with spurious bright stars in the field-of-view, airglow~\citep{leinert}, diffuse star light and anthropogenic contributions. The latter stem mainly from the major two cities of La Palma: Santa Cruz, located south-east of the ORM, and Los Llanos, located south-south-west. Also the airport and its surroundings, located south-south-west of the ORM, provide artificial light backgrounds. In that direction, also the northern coast of the 130~km distant, but much denser populated, island Tenerife is found. Street lighting on La Palma is regulated by law~\citep{skyprotectionlaw1, skyprotectionlaw2}, and any type of light installations are getting advice by the IAC's Sky Protection Unit~\citep{otpr}. These measures have greatly reduced light pollution on the Canarian observatories down to values typical for world-class observatories. Nevertheless, the effect of residual anthropogenic lights becomes visible, when the LIDAR photo-electron rates are plotted against local time, see Figure~\ref{fig:hrbg2}. A clear increase of light pollution is visible before 23h~UT (24h local time), when the street lights turn off and only sodium lamps, undetectable at 532~nm wavelength, remain switched on.   

In Figure~\ref{fig:bg} we study the arrival directions of these residual light backgrounds, separated into two samples taken before 23h~UT and after. In all cases, light pollution gets stronger towards the horizon, as expected from airglow~\citep{leinert}. Nevertheless, certain azimuthal directions show a clear increase of residual background, by up to 50\% (before midnight) and 25\% (after midnight). Light pollution from the direction of Los Llanos and Santa~Cruz/Tenerife is detected, but also from west-north-west before midnight, where the small town of Puntagorda is located. Another \collaborationreview{explanation} could be residual light from the observatory residence, located north-west of the MAGIC LIDAR.  After midnight, an $\sim$5--10\% increase of LIDAR background rates is observed at very large zenith angles, in the directions of Santa~Cruz/Tenerife and Los Lllanos.

\clearpage

%% file: discussion.tex
\section{Discussion}
\label{sec:discussion}

A pure elastic LIDAR system is by design limited in its capabilities to fully characterize optical properties of aerosols, particularly the relation between extinction and backscatter coefficients, expressed by the LIDAR ratio. \collaborationreview{A Raman LIDAR~\citep{Sicard:2010,Ballester:2019} is free of those limitations, although more caution needs to be taken not to disturb gamma-ray observations with the neighboring IATCs, due to its much higher laser power required.} 

The very particular atmospheric conditions found at the ORM have allowed us to mitigate several of the limitations of an elastic LIDAR: the existence of a stable clear night with similar aerosol properties throughout the year has made possible the absolute calibration of our system, to a precision, which is not limited anymore by the assumed LIDAR ratio of the ground layer aerosols. This has permitted us to assess the optical transmission of the full ground layer, in any condition (i.e., not only \collaborationreview{on} clear nights), with an accuracy of 4\%. Our assessment of average properties of the atmosphere is even better and limited only by the assumed \textit{average} ground layer LIDAR ratio or a possible evolution of it with altitude (2\%) and residual errors of the assumed molecular density profile (1\%). At very large zenith angles, the neglected curvature of the Earth introduces an additional small bias. For typical values of the free troposphere base height of 3.3\,km above ground and a zenith angle of 70$^\circ$, the effect is of the order of 0.5\%, growing to 1.5\% at 80$^\circ$ zenith.

%

So, although our system is not able to accurately reconstruct the internal structure of both backscatter and extinction coefficients within any type of ground layer -- as a Raman LIDAR may do -- and \collaborationreview{as} such provide even further information on the size and shape properties of the aerosol, we have been able to make a relative comparison of the clear night ground layer structure, e.g.\ under different viewing directions and \collaborationreview{during different} seasons. Given the seasonal stability found, it is improbable, but not impossible, that a seasonal change in aerosol LIDAR ratios cancels out a similar effect on scale heights and \textit{VAOD}s of same magnitude. Since our LIDAR pointing directions are distributed more or less randomly throughout the year, it is also improbable that (undetected) seasonal variations may mimic the average directional dependencies of the aerosol scale height found.

Our analysis of clouds does not depend on the absolute calibration of the LIDAR, but instead on the assumption of a free troposphere below and above each cloud detected. Such an assumption may not always be correct for low stratocumulus clouds, below which residual aerosols of the ground layer may not yet have vanished. Due to the operation mode of the MAGIC \collaborationreview{telescopes} (the LIDAR is operated in slave mode), such clouds are found very rarely in our sample and hardly influence the statistical properties of the overall sample. Only in winter, about 5\% of our sample might be attributed to such cases. For very high altitude cirrus, our sample is, however, limited by the laser power of our system, leading to a strong underdetection of type SVC clouds of very small optical depths, because the statistical fluctuations of the signal at the far end \collaborationreview{are becoming}   dominant.

%% file: conclusions.tex
\section{Conclusions}

The presented micro-power elastic LIDAR with HPD photon detection was originally designed to provide the MAGIC \collaborationreview{telescopes} with a simple and eye-safe system to characterize the optical properties of the part of the atmosphere 
which gets illuminated by the observed Cherenkov light from gamma-ray induced air showers. \collaborationreview{A follow-up publication will describe how this knowledge is used to correct the MAGIC science data.} 

The innovation of an HPD-based LIDAR photon detector -- facilitated by previous work on these devices carried out at the Max Planck Institute for Physics (Munich, Germany)~\citep{Mirzoyan:2000,Orito:HPD:2009,Saito:2011yp} -- has allowed us to exploit its exceptional charge resolution and digitize the pulse waveform produced by each individual photo-electron, or by ion feedback, and to distinguish both by means of waveform analysis. This has made it possible to \collaborationreview{obtain virtually} afterpulse-free return signals used for further   offline analysis. A laser upgrade in December 2016 has allowed us to further extend the maximum accessible range (in the absence of calima or optically thick clouds) from 22\,km to about 31\,km. 

Exploiting the very particular and stable properties of the clear night atmosphere above the ORM, we have been able to absolutely calibrate the system constant of our LIDAR for 24 adjacent time intervals, separated by hardware interventions. The absolute calibration has a statistical uncertainty of only 3\% and a systematic uncertainty of 2\%.

With this calibrated system and data set of $\sim 10^5$ atmospheric profiles taken from March 2013 to March 2020, we have been able to extensively characterize the nocturnal ground layer and clouds above the ORM.

For the clear (photometric) night, we have found a nocturnal boundary layer characterized by at least two substructures: an initial exponential decrease of aerosol extinction characterized by an Elterman scale height of $\sim 1.2$\,km, and valid up to 800\,m if pointing to zenith; and a subsequent stronger fall-off characterized by a scale height of $H_\mathrm{aer}\sim580$\,m on average. A closer look at the details of $H_\mathrm{aer}$ has revealed that the clear night nocturnal boundary layer is non-stratified, but curved around the Roque~de~los~Muchachos mountain. The curvature can be parameterized as $H_\mathrm{aer} = H_\mathrm{aer,0} \cdot (\cos\theta)^\gamma$, with the dependency on the zenith angle $\theta$ characterized by an exponent of $\gamma \approx 0.77$ on average. The curvature shows, however, a slightly asymmetric form, with strongest curvature towards the North and least towards the South. Curvature of the layer is also more pronounced during the summer months July and August.

For the full data sample, we have derived probability distributions for ground layer aerosol transmission and the base heights of the free troposphere, for winter, spring and summer atmospheres. We confirm previous findings that degraded optical transmission is primarily found during the summer months, but also possible during winter. The cleanest atmosphere is found in spring from May to June. The aerosol-free troposphere typically starts between 5\,km and 6\,km a.s.l., although also lower values down to 3\,km are possible, and higher altitudes up to 10\,km a.s.l., confirming previous results from few individual nights by~\citet{Sicard:2010}. These distributions can be used to plan observations, particularly in view of the forthcoming Cherenkov Telescope Array Observatory~\citep{cta}\,\footnote{see also \url{www.cta-observatory.org}}.

We have developed a cloud search and analysis algorithm, which permits retrieval of an cloud-averaged LIDAR ratio in most cases. Due to the operation scheme of our LIDAR working as a slave of the MAGIC Telescopes, our sample of clouds is, however, strongly biased towards optically-thin clouds. We find a typical LIDAR ratio of 21$\pm$6\,sr for our clouds, compatible with expectations for cirrus of type C2 and C3. Because of the long maximum range achievable by our system, we have been able to assess the interplay of the single, double and triple temperature inversions in the tropopause, discovered by~\citet{rodriguez2013}, and cloud top heights. We have found that during winter and spring, where double and triple tropopauses dominate, clouds are usually trapped by the lowest of these, whereas in summer, clouds usually extend up to about four kilometers below the (single) tropopause, and only sometimes move up to the thermal tropopause itself. In these cases, though, extremely high clouds reaching up to 22\,km a.s.l.\ can be found. Furthermore, summer clouds show smaller geometric thicknesses than spring and winter clouds. Winter clouds are, however, often vertically structured with optical extinction concentrated within hundred or less meters.

Finally, we have exploited the absolute calibration of our LIDAR system to produce corrected photo-electron background rates independent of hardware changes and representative \collaborationreview{for} the night sky background photon rates. We detect a clear increase of photon background rates due to \collaborationreview{back-scattering of night sky background light fro}, and increasing with optical depth of clouds during moon. Our analysis clearly detects evening zodiacal light and the galactic plane. Residual anthropogenic contributions to the 532\,nm light background is detected, particularly before local midnight, after which decorative lighting, halogen, mercury and high pressure sodium street lighting has to switch off, leaving low pressure sodium lamps only. 
The residual background light increases with zenith angle and correlates with the directions towards more densely populated areas of the island, or the neighbouring island of Tenerife.